%% file: wall_laws_letterv9.tex
\pdfoutput=1
\documentclass[aps,prl,twocolumn,superscriptaddress,nofootinbib,floatfix,longbibliography]{revtex4-2}

\usepackage{amsmath,amssymb}
\usepackage{booktabs}
\usepackage{microtype}
\usepackage{placeins}
\usepackage{tikz}
\usetikzlibrary{arrows.meta}
\usepackage{capt-of}   %
\usepackage{xcolor}
\usepackage[colorlinks=true,linkcolor=blue,citecolor=blue,urlcolor=blue]{hyperref}

\makeatletter
\@booleantrue\acknowledgments@sw
\makeatother

\newcommand{\mindeg}{\operatorname{mindeg}}

\newtheorem{theorem}{Theorem}
\input{figures_v9_palette}
\input{figures_v9_chamber_components}

\begin{document}
\setcounter{secnumdepth}{2}

\title{Gluons and Knots}

\author{Alfredo Guevara}
\affiliation{School of Natural Sciences, Institute for Advanced Study, Princeton, NJ 08540, USA}

\begin{abstract}

We argue that the scattering amplitude of a single minus helicity gluon with arbitrarily many positive helicity gluons is topological. The amplitude is constant in chambers of kinematic space, its resonances are located at walls and its factorization is determined by wall-crossing. Yangian invariance dictates the amplitude to be determined by a 2d null Wilson loop that can be lifted into a positive and transverse 3d link%
. By reinterpreting wall-crossing and the Weinberg soft theorem as skein relations and Reidemeister moves of the subknots, we construct a state-counting description consisting of adding the `corners' of their HOMFLY-PT polynomials. Beyond the classical limit, we argue that this provides a conjectured but precise connection to topological field and string theory.

The new formula is rigorously derived via a Lean-formalized dependency graph (DAG) as detailed in a companion paper, laying out the precise assumptions including Reidemeister III invariance. It is further tested exhaustively, including novel `star' kinematics where the amplitude and state count are shown to become Catalan numbers, in agreement with knot theory literature.

\end{abstract}

\maketitle

The answer is gluon scattering in four dimensions.  The search for the question has driven tremendous progress in our understanding of quantum field theory, from AdS/CFT \cite{Maldacena}, through celestial holography \cite{PasterskiPateRaclariu} and on to the modern scattering-amplitudes program \cite{WittenTwistor,ElvangHuang}. The integrability of $\mathcal{N}=4$ Super Yang-Mills \cite{BeisertReview,ADMSTV} paved the way for several unexpected developments, notably, a strong-coupling duality between null Wilson loops and amplitudes \cite{Bern:2005iz,AldayMaldacena07,AldayMaldacena}.

MHV scattering, involving two negative helicity gluons, is widely regarded as the simplest possible non-trivial scattering in this realm. However, recently \cite{GLSSW} presented a new configuration involving instead a single minus helicity gluon, later extended to single-minus graviton amplitudes as well \cite{GLSSWgrav}. The configuration becomes non-trivial only when the particles are all fully collinear, which is best understood in $(2,2)$ signature.   

The single-minus amplitudes were first presented as sums over an exponentially large number of terms, akin to all-valence Feynman diagrams. However, very recently the gravity result has been reframed conceptually \cite{GLMS}: It was argued, through the lens of M-theory and the BFSS model, that the single-minus graviton amplitude is in fact a non-perturbative count over BPS states in a gauge theory. This formulation then allows us to find simplified expressions that do not rely on Feynman diagrams.

We are led to ask: If the gluon scattering question could also be rephrased as a state count%
, which states would such an amplitude count? Does this provide a reformulation which simplifies the gauge dependent Feynman diagrams and Berends-Giele currents into more fundamental constituent blocks?

The amplitude that concerns us here is an integer
function $A$ of the 2d momentum spinors $\{\widetilde\lambda_{\dot{\alpha}}\}_{i=1}^n$, which is constant in domains we call \textit{chambers}. This topological nature is what makes the
states countable by simple rules. We rely on the notion of unitarity which takes a particularly enlightening form here. Recall scattering theory produces physical states at threshold, via the well known decomposition of e.g. a tree level amplitude,
\begin{equation}
\mathcal{A}_n = \frac{\mathcal{A}_L \mathcal{A}_R}{(p_L+p_R)^2} + \text{regular},
\end{equation}
so singularities appear at poles. It turns out that in the degenerate, collinear locus $(p_L+p_R)^2=0$, the singularities become discontinuity walls, and the unitarity statement is
\begin{equation}\label{eq:spuni}
A_n
=\pm\Theta_+\!\bigl([\widetilde\lambda_L\,\widetilde\lambda_R]\bigr)
 A_L A_R+\text{continuous},
\end{equation}
with $\Theta_+$ the unit step across the collinear component on which
$\widetilde\lambda_L$ and $\widetilde\lambda_R$ point in the same direction.
\input{figures_v9_intro}
\noindent Here, as in gravity \cite{GLMS}, $\widetilde\lambda_L+\widetilde\lambda_R$ involves all the particles but one. This shows in particular that many of the walls in the diagrammatic formula \cite{GLSSW}, not of the above type, are nothing but spurious singularities!

What is the mathematical origin of this jump?  In this paper, we will trace it to a state-sum description of the amplitude in terms of knots and links. To understand this, we start our discussion from a null Wilson polygon, Fig.~\ref{fig:intro}(b), that factorizes both at cusps and self intersections, as imprints of unitarity.\footnote{Cusp factorization is well explored in the usual WL/amplitude correspondence \cite{AGMSV}; self-crossing Wilson loops are studied in \cite{DixonEsterlis}.} In the first half of this letter we study the symmetries and classifications of such Wilson polygons. We lay out the wall-crossing properties leading to \eqref{eq:spuni} and show that they fully constrain the amplitude. We will prove that all the chambers are connected through smooth deformations, due to two fundamental facts coming from different realms: wall-crossing and the Weinberg soft theorem. 

A priori, reproducing the gluon amplitude from knots may be counterintuitive. The latter are insensitive to cusps and do not self intersect. In the second half of the letter we resolve this by introducing the states: we prescribe lifts of the 2d polygon to 3d links, using its orientation to construct \textit{positive} and \textit{transverse} knot diagrams, Fig.~\ref{fig:intro}(c). The transverse lift provides a self-linking number and, through the Morton-Franks-Williams (MFW) inequality, we use it to design a novel `corner' quantity that satisfies the wall-crossing laws. This provides a new representation of the single-minus amplitude, formalized as a Lean-verified proof in \cite{CornerLawsLean}. In addition, the lift induces a particular framing which allows us to interpret the polygon Wilson loop as an expectation value in analytically continued 3d Chern-Simons theory  \cite{WittenJones}, see Fig.~\ref{fig:intro}(d).

At this level, as we will see, the fundamental relation \eqref{eq:spuni} is an avatar of the more fundamental algebra of knots invariants, i.e. the skein relation of the HOMFLY--PT polynomial.

\section{The null Wilson loop in twistor space}

We will employ two key insights about the single-minus amplitude provided in Ref.~\cite{BPPTW}.  First, the scattering admits a direct $\mathcal N=4$
supersymmetric extension.  Supersymmetry protects non-collinear corrections that can emerge at loop level.

Second, it can be checked that the amplitude is a Yangian invariant, with
dual conformal symmetry directly acting on the stripped coefficient.  More precisely, the superamplitude factorizes as
\begin{equation}
 \mathcal A_n^{(-)}(\lambda_i,\widetilde\lambda_i,\eta_i)
 =i^{\,2-n}\,\delta^{(2)}\!\left(\sum_i\widetilde\lambda_i\right)
 A(\widetilde\lambda_i)\,
 \mathcal K_n(\lambda_i,\eta_i).
 \label{eq:superfactor}
\end{equation}
The index $i$ runs over the $n$ color-ordered legs, with $k_i=\lambda_i \widetilde\lambda_i$.  The delta function
imposes momentum conservation on half the momenta $\widetilde\lambda_i$. The factor
$\mathcal K_n$ is a universal supersymmetric factor, here fulfilled by choosing the very convenient frame $\lambda=(0,1)$. We refer to \cite{GLSSW} for further conventions. All the dynamical information resides in the piecewise-constant coefficient
$A(\widetilde\lambda_i)$, the main object of this paper. Because \eqref{eq:superfactor} is supersymmetric we know that $A(\widetilde\lambda_i)$ is cyclic invariant.

We employ the color ordering to introduce dual points $x_i$,
\begin{equation}
 k_i=x_{i+1}-x_i=\lambda_i\widetilde\lambda_i,
 \qquad x_{n+1}=x_1 .
 \label{eq:dualpoints}
\end{equation}
Recall we work on signature $(2,2)$; we choose real coordinates for which $ ds^2=dz\,d\bar z+dx^+\,dx^- $. We can then identify $\widetilde\lambda=(p_-,p_{\bar{z}})$ as the only nonzero momenta, and we can imagine that $x_i$ lies on $\bar z=x^-=0$, thanks to translation invariance. This is a totally null zero-area sheet. The closed polygon chain $P=(x_1,\ldots,x_n)$ is the null Wilson-loop
contour associated with the color-ordered amplitude: the $x_i$ are its
cusps. 

The answer is, as we now see, an integer number $A(P)$ for each polygon $P$.
\textit{What is the question?} 

The rules of the game are clearly stated by introducing momentum twistor coordinates \cite{Hodges}, which manifest the dual conformal property. Recall that via the embedding formalism we can represent each $x_i$ by its line in twistor space \cite{WittenTwistor}, $X_i^{AB}=Y_i^{[A}Z_i^{B]}$. Indeed, it is easy to see that $\lambda_i=\lambda=(0,1)$ leads to the choice of representative\footnote{The standard assignment $Y_i=Z_{i+1}$ is degenerate in the half-collinear regime, as $Z_{i+1}\to Z_i$.}
\begin{equation}
 \begin{gathered}
 Y:=(0,1, 0 ),\qquad Z_i:=(1,0,\mu_i),\qquad
 X_i^{AB}=Y^{[A}Z_i^{B]},\\
 \mu_{i+1}-\mu_i=\widetilde\lambda_i .
 \end{gathered}
 \label{eq:twistorcurve}
\end{equation}
Thus $Z_i$ can be identified with the
vertices of a closed plane polygon. Note that by keeping $Y$ fixed, the polygon lies in the quotient of the real
twistor space $\mathbb P\!\left(\mathbb T_{\mathbb R}/\langle Y\rangle\right)\simeq\mathbb{RP}^{2}$. Keeping $Y$ fixed induces the $PGL^+(3,\mathbb R)$ action on the quotient. This is the dual conformal property: it can change the angles between $\{\widetilde\lambda_i\}$ but not their orientation.  Two affine
representatives of this half-collinear null Wilson loop are shown in
Fig.~\ref{fig:wilsonloop}, the polygons have the same amplitude $A(P)=A(P')$.

\input{figures_v9_wilson_loop}

The triple brackets of the vertices are
\begin{equation}
 \langle ijk\rangle:=\langle Z_iZ_jZ_kY\rangle\,.
 \label{eq:tripbracket}
\end{equation}
We can think of these brackets as coordinates of the polygon. They are slightly overdetermined, in two senses. First, they satisfy Schouten relations as one expects. Second, configurations with different values of some of brackets actually correspond to the same physical kinematics, as we now explain.

\section{Chirotopes and the gluon amplitude}
\label{sec:chirotopeamplitude}

We now discuss the amplitude value $A(P)$ obtained in \cite{GLSSW}. Here we formulate it in a slightly more compressed way. We find it convenient to discuss the polygon $P$ by introducing its rank-three chirotope coordinates, defined as
\begin{equation} \label{eq:threebrackets}
\chi_{ijk} :=\operatorname{sg}\langle ijk \rangle = \operatorname{sg}\left([\mu_j-\mu_i,\mu_j-\mu_k]\right)\,,
\end{equation}
which is invariant under dual conformal transformations. The second equality is our frame choice. Here  $\operatorname{sg}x\in\{-1,0,1\}$. In particular $\chi_{i,i+1,i+2}$  records the polygon turning left or right as we walk the edge $\widetilde\lambda_i$ following its orientation. 

As we mentioned, the amplitude formula introduced in \cite{GLSSW} is akin to a sum over Feynman diagrams. In particular it requires the ``gauge choice" of picking a privileged label even though the end result is cyclic invariant. 

We now introduce the vertex formula. Consider an ordered set of momentum edges
$\widetilde\lambda_u,\ldots,\widetilde\lambda_v$. We partition it into $m$ nonempty sub-blocks that meet the polygon at the boundary vertices
$a_0,\ldots,a_m$: $a_0=u$ and $a_m=v{+}1$, with $a_k$ the vertex
where the $k$-th sub-block ends.  A tree vertex for this partition carries the weight
\begin{equation}
 \begin{aligned}
 V^{\pm}(a_0,\ldots,a_m)
 &=\prod_{k=1}^{m-1}\Bigl[
   \chi_{a_{k+1},a_k,a_{k-1}}\\[-2pt]
 &\qquad\times\Theta\!\left(
   \mp\chi_{a_m,a_k,a_0}\chi_{a_{k+1},a_k,a_{k-1}}\right)\Bigr].
 \end{aligned}
 \label{eq:amplitudevertex}
\end{equation}
For
$m=1$ both products are empty and $V^{\pm}=1$. Consider now the set
$\mathfrak T(P)$ of the planar trees rooted in
$\widetilde\lambda_n$. The root vertex $v_*$ carries weight $V^{-}$; every
other internal vertex carries $V^{+}$ and one minus sign. The half-collinear
formula of \cite{GLSSW} is the finite sum
\begin{equation}
 \begin{aligned}
 A(P)&=\sum_{T\in\mathfrak T(P)}(-1)^{N_+(T)}\,\mathcal W(T),\\[-2pt]
 \mathcal W(T)&=
 \raisebox{-1.28em}{%
 \begin{tikzpicture}[baseline=(vb.base),x=.64cm,y=.45cm,line width=.4pt,
   every node/.style={font=\scriptsize,inner sep=.5pt}]
   \node (mleg) at (-1.05,0) {$\widetilde\lambda_n$};
   \node[draw=pieceA,fill=pieceA!8,circle,minimum size=12pt] (vb) at (0,0) {$V^{-}$};
   \node[draw,circle,minimum size=11pt] (vu) at (1.05,.78) {$V^{+}$};
   \node[draw,circle,minimum size=11pt] (vd) at (1.05,-.78) {$V^{+}$};
   \node (f1) at (2.30,1.35) {$\widetilde\lambda_1$};
   \node (f2) at (2.30,.78) {$\widetilde\lambda_2$};
   \node (fu) at (2.16,.25) {$\cdots$};
   \node (mid) at (1.25,0) {$\cdots$};
   \node (fd) at (2.16,-.25) {$\cdots$};
   \node (fn2) at (2.30,-.78) {$\widetilde\lambda_{n-2}$};
   \node (fn1) at (2.30,-1.35) {$\widetilde\lambda_{n-1}$};
   \draw (mleg)--(vb) (vb)--(vu) (vb)--(mid) (vb)--(vd)
         (vu)--(f1) (vu)--(f2) (vu)--(fu)
         (vd)--(fd) (vd)--(fn2) (vd)--(fn1);
 \end{tikzpicture}}
 \quad=V^{-}_{v_\star}(T)\!\prod_{v\ne v_\star}V^{+}_v(T)
 \end{aligned}
 \label{eq:amplitude}
\end{equation}
Here $N_+(T)$ is the number of
$V^{+}$-vertices (we discard $V^+$ propagators, which would make the sum infinite). The finite sum is integral and depends only on $\chi$.
Recall that singling out the last particle is immaterial: rooting the
trees at any other edge gives the same value, a consequence of the
supersymmetry.

It will be useful sometimes to consider $\overline{P}$, the polygon with reverse orientation. With this transformation each factor $\chi\to -\chi$. We can see that 
\begin{equation}\label{eq:reflx}
A(\overline P)=(-1)^nA(P)\,.
\end{equation}

\section{Polygons and their chambers}
\label{sec:chambers}

In two dimensions, our Wilson loops self-intersect generically. In this section we characterize them. We will say that the
polygon $P$ is \textit{generic} when every edge is of nonzero length, every
turn is strictly left or right ($\chi_{i,i+1,i+2}\neq0)$, and each self-intersection is transverse and
not a vertex itself, which we denote as
$\widetilde\lambda_i\pitchfork\widetilde\lambda_j$ and is determined by the condition
\begin{equation}
\chi_{i,i+1,j}\chi_{i,i+1,j+1}=\chi_{j,j+1,i}\chi_{j,j+1,i+1}=-1\,.
\end{equation}
for non-adjacent edges $\widetilde\lambda_i,\,\widetilde\lambda_j$. We will further assume each intersection only involves two edges.\footnote{The definition of $A(P)$
\eqref{eq:amplitude} allows many concurrent edges. This genericity conditions aids the construction of
the state sum of knot invariants.}

We imagine moving the vertices $\{\mu_i\}$ smoothly. The chamber of $P$ are the generic polygons connected to $P$ in this sense. There are finitely many open chambers in the vertex configuration space
$(\mathbb R^2)^n$. The boundaries of chambers are degenerate configurations we call walls, where the amplitude may jump. 

It will be useful to understand coarse connected components of many chambers, obtained by introducing an extra label: 
The rotation number $r(P)$ is the total signed turning as we travel through the polygon. If $\theta_i\in(-\pi,\pi)$ is the turning angle at cusp $\mu_i$, then
\begin{equation}
 r(P)=\frac{1}{2\pi}\sum_i\theta_i .
 \label{eq:rotdef}
\end{equation}
The individual angles
depend on the affine chart, yet the integer \eqref{eq:rotdef} is topological. In particular it is conformal invariant, as we can see by using Whitney's formula to write it in terms of the chirotope. If $\mu_1$ is any vertex of (the convex hull of) the polygon \cite{AHTT}, then
\begin{equation}
 r(P)=\chi_{n,1,2}
 +\sum_{\substack{1\le i<j\le n\\
      \widetilde\lambda_i\pitchfork\widetilde\lambda_j}}
      \chi_{i,i+1,j} .
  \label{eq:rotation}
\end{equation}
We will study the cells of fixed $(n,r)$ labels. The zeroth order question is how many chambers does the
space of half-collinear loops have for fixed $(n,r)$.\footnote{This problem touches upon the
chamber structure of realizable rank-three oriented matroids, namely the sign patterns of the $3\times3$ minors $\langle ijk\rangle$ of
the matrix $( Z_1 \,\cdots\,Z_n)$ (quotiented by $Y$), a crucial
problem in combinatorial geometry \cite{RichterSturmfels,AAK}.} For instance, the three representatives below lie in the distinct chambers of $(n,r)=(5,2)$:
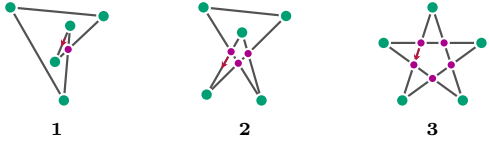
\begin{center}
\begin{tikzpicture}[x=1.08cm,y=1.08cm]
  \path[use as bounding box] (-.64,-.82) rectangle (5.24,.64);
  \VSixRtwoRow
\end{tikzpicture}
\end{center}
Table~\ref{tab:census} shows the number of chambers up to $n=7$.  For $|r|>0$, each
column combines the $r$ and $-r$ chambers, obtained by reversing orientation.  Reference~\cite{companion} gives the enumeration details and an exact
atlas of the 24 five-vertex chambers.

\begin{table}[ht]
\caption{Census of visible chambers, mod. cyclic relabeling. In parenthesis: reflexion-invariant chambers appear at $r=0$.}
\label{tab:census}
\centering
\setlength{\tabcolsep}{4.2pt}
\begin{tabular}{c@{\quad}rrrr@{\quad}r}
\toprule
$n\backslash |r|$ & $0$ & $1$ & $2$ & $3$ & total\\
\midrule
3 & ---          & 2    & --- & --- & 2\\
4 & $1\,(1)$     & 4    & --- & --- & 5\\
5 & $8\,(0)$     & 10   & 6   & --- & 24\\
6 & $57\,(9)$    & 84   & 68  & --- & 209\\
7 & $772\,(0)$   & 1510 & 840 & 302 & 3424\\
\bottomrule
\end{tabular}
\end{table}

The rapid growth of the chambers with $n$ makes the problem numerically intractable.  The point of
wall-crossing is to replace this proliferating list of chambers by a small
calculus. To set some intuition, recall the Whitney--Graustein theorem: Two oriented immersed circles are
regularly homotopic if and only if they have the same rotation number
\cite{Whitney,Wesenberg}. Regular here means that at no instant in the homotopy we fold a tangent direction backwards ($\theta=\pi$). In exploring wall-crossing at fixed $r$, we would like to think of the regular homotopy as smoothly varying the kinematics $\{\tilde{\lambda}_i\}$. For this to work we need the theorem to be restricted to the space of polygons.  Luckily, a stronger version of that theorem holds up to cyclic relabeling \cite{companion}, allowing the regular homotopy to take place entirely in $(n,r)$! For instance, in the above example for $(5,2)$ we have the following moves connecting the 3 chambers:
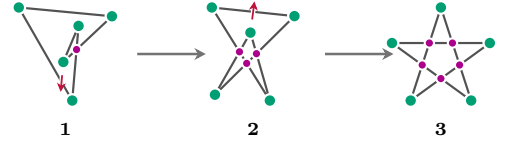
\begin{center}
\begin{tikzpicture}[x=1.08cm,y=1.08cm]
  \path[use as bounding box] (-.64,-.82) rectangle (5.24,.70);
  \VSixRtwoMoveRow
\end{tikzpicture}
\end{center}

This is a polygonal version of Arnold's wall
calculus \cite{Arnold}. The reader versed in combinatorics may wonder if our classification is contained in the division $\langle ijk\rangle=0$ which is typical of matroids. It is. Yet the ``visible'' requirement makes some of those boundaries completely trivial. For instance, a
vertex meeting the projected line of a remote edge beyond its finite segment is \emph{silent}, i.e. an spurious singularity! \cite{companion}. Ignoring the silent walls, one can easily read off the remaining chirotope coordinates $\chi_{ijk}$ from a chamber, and so the value of $A(P)$ is constant in each chamber. 

\section{Wall-crossing laws}
\label{sec:walllaws}
\input{figures_v9_walls}

We can think of the wall-crossing calculus as playing the role of an Equation of Motion (EOM) satisfied by the amplitude. Integrating this EOM, we will find in next sections a novel expression that is equivalent to $A(P)$.

Let us make this more precise. The definition \eqref{eq:amplitude} boils down to the EOM of Yang-Mills theory restricted to the single-minus sector. It factors at singularities, which in this case are walls of configuration space. This can be read off from the vertex weights $V^\pm$ in \eqref{eq:amplitudevertex}. Consider a label of the polygon, say $\mu_k$, crossing the edge
$\widetilde\lambda_n$, without loss of generality. For a fixed diagram in \eqref{eq:amplitude} there is at most one affected vertex, for which the jump of its step function as $\chi_{a_m a_k a_0}$ changes sign gives
\begin{equation}
 \Delta V^{\pm}=\pm V^{\pm}_L V^{\pm}_R,
 \label{eq:vertexwallsplit}
\end{equation}
where $L,R$ are the two pieces separated at $a_k$. What remains is the independent
sum over rooted diagrams on each of the two smaller polygons, hence the
product of their amplitudes. Carrying out the bookkeeping leads to our first wall-crossing law:

\begin{enumerate}

\item\label{law:s7}\textbf{W1 (vertex--edge).}
A vertex $\mu_j$ crosses a nonadjacent oriented edge
$(\mu_i,\mu_{i+1})$, taking $P_-$ to $P_+$. Reconnecting the strands at
the contact compatibly with their orientations gives two smaller polygons
$Q_L,Q_R$. The jump is given by
\begin{equation}
 A(P_+)-A(P_-)=\chi A(Q_L)A(Q_R).
 \label{eq:incidence}
\end{equation}
Here $\chi:=\chi_{i,i+1,j}$ is evaluated at $P_-$. There are two subcases we call `bigon' and `sliding', shown in
Fig.~\ref{fig:walls}(a) and (b), respectively.
\end{enumerate}
Note $\chi=1$ when the vertex starts to the left of
the oriented edge. This is precisely the rule \eqref{eq:spuni} in the introduction. 

When the three collinear vertices are consecutive, one side of the split
is a single branch of weight $1$. A similar argument therefore gives the
amplitude of the deletion $P\setminus j$.   We obtain, for $n\geq4$, 

\begin{enumerate}
\setcounter{enumi}{1}
\item\label{law:s3}\textbf{W2 (flat).}
The corner at $\mu_j$ straightens, with $\mu_j$ between its neighbors at
the wall, as shown in Fig.~\ref{fig:walls}(c). Denoting the two sides by
their left or right turn at $\mu_j$,
\begin{equation}
 A(P_{j:\mathrm{right}})-A(P_{j:\mathrm{left}})
   =A(P\setminus j).
 \label{eq:flat}
\end{equation}
\end{enumerate}
The final degeneration of the polygon space involves having three edges concurrent. It can be easily seen that the chirotope coordinates $\chi_{ijk}$ are insensitive to this,\footnote{One would need data about the order in which crossings occur, i.e. the Gauss word of the polygon, or alternatively (the signs of)
$\langle \ell_i\ell_j\ell_k\rangle$, where
$\ell_i=*Z_i\wedge Z_{i+1}$.} and so we obtain 
\begin{enumerate}
\setcounter{enumi}{2}
\item\label{law:r3}\textbf{W3 (triple).}
At a generic triple wall, three pairwise nonadjacent edges become
concurrent and the three crossings exchange their order. No vertex triple
becomes collinear, so the chirotope and every term of
Eq.~\eqref{eq:amplitude} remain unchanged, as illustrated in
Fig.~\ref{fig:r3}:
\begin{equation}
 A(P_+)=A(P_-).
 \label{eq:rthree}
\end{equation}

\end{enumerate}

\VSixRThreeFigure

It turns out, these three rules are integrable in the sense that small cocycles, closed paths around walls in the configuration space, vanish. In fact there is a unique function that satisfies them, which can be fixed to be precisely $A(P)$ as we will see.

\FloatBarrier
\section{Soft theorem across rotation classes}
\label{sec:softrotation}

We introduce one last tool, which allows moving between different $(n,r)$ classes: The
soft theorem discussed in Ref.~\cite{GLSSW}.  This is actually
Weinberg's soft-gluon theorem in the frame \eqref{eq:superfactor}.  We claim
that together with the wall laws above they fully fix the amplitude in every chamber.  Let us
see how this works.

The soft theorem is easy to state in this language. Given a polygon $P$ we
insert a soft edge $\varepsilon\widetilde\lambda_s$ between
$\widetilde\lambda_i$ and $\widetilde\lambda_{i+1}$:
\begin{equation}
 \begin{gathered}
 \lim_{\varepsilon\to0^+}
 A(\ldots,\widetilde\lambda_i,\varepsilon\widetilde\lambda_s,
   \widetilde\lambda_{i+1}-\varepsilon\widetilde\lambda_s,\ldots)\\[-2pt]
 =\frac{\chi_-+\chi_+}{2}\,A(P),
 \qquad\text{with }\chi_\pm=\chi_{s,i+1,i+1\pm1}.
 \end{gathered}
 \label{eq:softinsertionlaw}
\end{equation}
This limit fixes the value of $A$ for every
sufficiently short new edge.  The identity follows from a
simple diagrammatic argument using the vertices $V^\pm$ in
Eq.~\eqref{eq:amplitudevertex}, by using
\begin{equation}
\lim_{\varepsilon\to0^+}
V^\pm(1,2,a_2,\ldots,n)
=
\frac{\chi_{21a_2}\mp \chi_{21n}}{2}\,
V^\pm(1,a_2,\ldots,n)
\end{equation}

We now describe how the soft theorem moves between rotation numbers.  Put
$\chi:=\chi_{i,i+1,i+2}$, the turn of the corner before the soft insertion,
and rotate the soft edge as shown in Fig.~\ref{fig:soft-flow}.

\input{figures_v9_soft_properties}
The last step generates a tiny loop and crosses a wall when
$\widetilde\lambda_i$ and $\widetilde\lambda_s$ are back-to-back.  Write $P_\ell,P_0$ for the
loop and no-loop sides.  The soft theorem gives the local wall-crossing rule
\begin{equation}
 \begin{aligned}
 A(P_\ell)-A(P_0)&=\chi A(P),\\
 r(P_\ell)-r(P_0)&=-\chi.
 \end{aligned}
 \label{eq:softlocaljump}
\end{equation}
The preceding observations give a sharp uniqueness statement. Thanks to the cell space $(n,r)$ being connected, we can indeed move over the full configuration space if we start from a low $n$ anchor. A simple anchor is just the $n=3$ triangle. This shows

\begin{theorem}
Let $C$ be a chamber-constant function satisfying the
wall-crossing laws \eqref{eq:incidence}, \eqref{eq:flat}, and
\eqref{eq:rthree}, together with the Weinberg soft theorem
\eqref{eq:softinsertionlaw} and the reflexion identity \eqref{eq:reflx}.  If $C$ agrees with $A$ for a triangle, then
\begin{equation}
 \boxed{A(P)=C(P).}
 \label{eq:softuniqueness}
\end{equation}
for any generic polygon $P$.
\end{theorem}
Physically, we think of this as solving the EOMs by starting from a lower-point source insertion. The wall-crossing laws are the unitarity rules. In the next section we construct a `canonical' quantity $C(P)$ using the knot lift of $P$. It is engineered to satisfy the wall-crossing laws. Thanks to the theorem, this gives a novel representation of the single-minus amplitude.

To grasp the power of the soft theorem in this context, let us close this section by arguing that \eqref{eq:softlocaljump} is indeed a new wall-crossing law across $r\to r+1$, that is, independent of the rotating edge being small/soft. Compare stages 2 and 3 of Fig.~\ref{fig:soft-flow}, just above and below threshold, and imagine extending the soft edge in any direction. The only singularity encountered can be a W1 wall, at which the polygon factors into two. That product is the same above and below the threshold so it does not affect the law. Thus other strands may cross the cusp region without changing the law, as
illustrated in Fig.~\ref{fig:s4-cusp}. We then have
\begin{enumerate}
\setcounter{enumi}{3}
\item\label{law:s4}\textbf{W4 (cusp law).}
For $n\ge4$, suppose the two edges adjacent to $\mu_j$ double back. Write $P_{\mathrm{loop}}$
for the side with the newborn loop and $P_{\mathrm{no\,loop}}$ for the
other side. Then \eqref{eq:softlocaljump} holds. 
\end{enumerate}
The law suggests the soft theorem extends to the full $S$-algebra \cite{GuevaraHimwichPateStrominger,StromingerW}, analogous to the $w_{1+\infty}$ algebra in the gravitational case \cite{GLSSW}.
\VSixCuspFigure

\section{Knots and the state sum}
\label{sec:statesum}

\input{figures_v9_state_sum}

We now present a new description of the amplitude as a state sum that counts certain admissible states. The states are represented by positive 3d links of knots. %
The sum is over resolutions of the same immersed Wilson
loop, in the sense that they are decompositions that all project down to a given polygon $P$ in 2d.  We use them to
construct a second chamber-constant function $C(P)$ that agrees with the amplitude.

It is convenient to first represent all crossings in $P$ as positive. Both strands entering a crossing carry the orientation of $P$. In knot theory, a crossing in a diagram is
\emph{positive} when the under-strand is obtained from the over-strand by a
counterclockwise turn of less than a half-turn.  A plane curve becomes a link in three dimensions once every
crossing is told which strand passes over. More formally we pick a smooth height function $h(t)$ for the curve, using that the crossings are disjoint due to its genericity. Choosing the positive option at
each crossing gives the \emph{positive lift}.  The fourteen-gon $P$ below is
the example we use throughout this section:
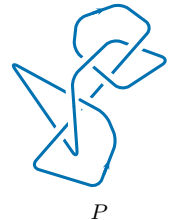
\vstatepolygonfigure

\emph{Smoothing} a crossing
reconnects its two strands compatibly with their orientations, exactly the
reconnection of the W1 law (Sec.~\ref{sec:walllaws}), 
\vsmoothingcartoon

Smoothing crossings in a link may break it or reconnect it. A \emph{decomposition} is a set $S$ of crossings whose
smoothings split $P$ into $|S|+1$ positive knots $Q_i$, associated to oriented subpolygons under 2d projection. Figure~\ref{fig:decomp} shows decompositions of the above polygon; colors distinguish the resulting subpolygons. Panel (d) is not a decomposition. An equivalent way to state it is to demand that every smoothing separates and none rejoins. More formally, we demand the crossings $S$ to be \emph{pairwise unlinked}: two crossings of $P$ are
linked when their four visits alternate as we tour the polygon, and unlinked otherwise.

 \vthreestatesumfigure
Smoothing one of two linked
crossings turns the other into a crossing between different knots $Q_i$. In the following we may ignore such crossings. They do not contribute to the limit we will be interested in.

Call the decomposition \emph{uniform} when every one of its knot diagrams
turns uniformly, all left or all right. In Fig.~\ref{fig:decomp}, cases (a) and (b) are uniform. Let $m_Q$ be
its number of retained self-crossings,
$r_Q$ its rotation number, and let $H^+_Q(a,z)$ be the HOMFLY--PT polynomial
of that lift \cite{HOMFLY,PT}. It is normalized by
\begin{equation}
 \begin{gathered}
 H(\bigcirc)=1,\qquad
 H(L\sqcup\bigcirc)=\frac{a-a^{-1}}{z}H(L).
 \end{gathered}
 \label{eq:skein}
\end{equation}
where $\bigcirc$ denotes the unknot and $L\sqcup\bigcirc$ its formal union with $L$, meaning they are unlinked. For any oriented link, with $L_+$, $L_-$,
and $L_0$ containing a positive crossing, a negative (mirror) crossing, and its oriented
smoothing, it satisfies the Skein relations
\begin{equation}
aH(L_+)-a^{-1}H(L_-)=zH(L_0)\,.
\label{eq:skein2}
\end{equation}
These laws fix the HOMFLY polynomial just like the wall-crossing fixes the amplitude. They are indeed related as we will see in the next section. Here we will point out the key result which allows us to define the state sum: 
\begin{theorem}\label{thm:corner}
For every subpolygon $Q$ whose turn signs agree with at most one exception,
its positive lift satisfies
\begin{equation}\label{eq:mdnx}
 \mindeg_a H^+_Q
   \ge 1-m_Q-|r_Q|=:d_Q \,,
\end{equation}
\begin{equation}
 \mindeg_z H^+_Q
   \ge 0\,.
 \label{eq:zfloor}
\end{equation}
\end{theorem}
The second condition holds for knots. We sketch the proof of the first condition below. Together, these properties motivate the \emph{corner coefficient} of $Q$, defined here as
\begin{equation}
c(Q):=\bigl[a^{\,d_Q}z^0\bigr]H^+_Q(a,z).
 \label{eq:factor}
\end{equation}
For instance, the three subpolygons in Fig.~\ref{fig:decomp}(a) give
\begin{equation}
 c\!\left(\vstatebluecartoon\right)=-1,\qquad
 c\!\left(\vstateorangecartoon\right)
 =c\!\left(\vstategreencartoon\right)=1.
 \label{eq:decompcorners}
\end{equation}
With this in hand, we can finally define the state sum over uniform decompositions:
\begin{equation}
 \boxed{\displaystyle
 C(P)=(-1)^{\ell}
 \sum_{\substack{P_S\text{ uniform}}}
 (-1)^{|S|}\prod_{Q\in P_S}c(Q)}.
 \label{eq:C}
\end{equation}
Here $\ell$ is the number of left turns of the original polygon. As we will outline in the next section, the skein relation \eqref{eq:skein2} makes this sum satisfy the wall-crossing and soft identities required by the uniqueness theorem, hence it provides our new representation of the single-minus amplitude.

We can elaborate on the meaning of the \textit{corner} coefficient. The Newton polygon of $H^+_Q$ is the convex hull of the exponent pairs $(i,j)$ for which $[a^iz^j]H^+_Q\neq0$. Thus $(d_Q,0)$ is the southwest slot in which the corner coefficient is read. Figure~\ref{fig:slot} realizes the four possible patterns.
\vthreecornerfigure

We now give a sketch of the proof of Theorem~\ref{thm:corner}, and rely on Ref.~\cite{companion} for the details. It follows from the well known MFW bound applied to a transverse knot \cite{Bennequin,Etnyre,Ng,Morton,FranksWilliams}, $\operatorname{mindeg}_aH_Q\geq 1+\operatorname{sl}(Q)$, the self-linking number. The transverse knot is a lift of the polygon obtained as follows. One picks a contact 1-form
\begin{equation}\label{eq:1frm}
    \alpha = dx^++h\,dz\,,
\end{equation}
The knot is transverse if we find a height function $h$, with over-strands at smaller $h$, such that at each point we cross $\alpha$ planes upwards i.e. $\alpha>0$. This is not possible at the downward tangents, which have $dx^+<0, dz=0$. The solution is to replace each such tangent by a curl

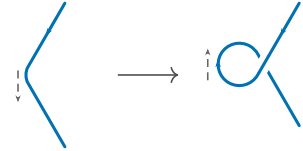
\input{figures_v9_transverse_curl}

Crucially, for uniformly rotating polygons there are $|r|$ such tangents! Further accounting for the orientation of the curl and the $|m|$ intersections, the self-linking number is
\begin{equation}
    \operatorname{sl}(Q)=-m-|r|
\end{equation}
and \eqref{eq:mdnx} follows.
\section{The knot origin of wall-crossing}

We now argue that the state sum \eqref{eq:C} satisfies the ingredients of
our uniqueness theorem by virtue of its lift to knots. We will not formally prove here the wall-crossing laws but rather give a sketch of how they are related to the laws of knots. In particular we highlight how they rely on the three Reidemeister moves, which do not alter the HOMFLY of a knot \cite{HOMFLY,PT}. The formal proof is given in \cite{companion}, with a Lean library available in \cite{CornerLawsLean}. 

\textbf{W1 walls from skein.} We focus on the bigon law, Fig \ref{fig:walls}(a). Fix a state in the sum in which both newborn crossings remain on one subpolygon.  Let $Q_+$ and $Q_-$ be that
subpolygon on the two-crossing and zero-crossing sides, respectively. The positive member of the skein triple \eqref{eq:skein2} is
$L_+=Q_+$. To relate the negative member, call it $L_-$, to the knot lift of the subpolygon $Q_-$ we use the \textit{Reidemeister-II} move as in the figure below. Finally, for $L_0$ we obtain a smoothing which leads precisely to $Q_L$ and $Q_R$ knots, belonging to the two factor amplitudes $A_L$ and $A_R$ of the wall-crossing formula.

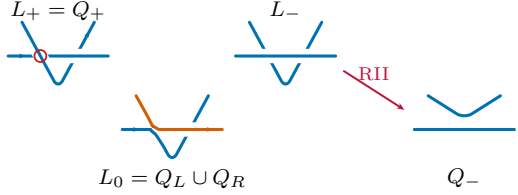
\input{figures_v9_s7_skein}
Let us show the factorization precisely. Write $H_\pm=H_{Q_\pm}$. For the intersection numbers $m_+=m_-+2$, while for the rotation numbers $r_+=r_-$, hence \mbox{$d_+=d_--2$}.  The skein
relation and the Lickorish--Millett factorization 
\cite{LickorishMillett} give
\begin{equation}
 \begin{aligned}
 H_+&=a^{-2}H_-+a^{-1}zH_0,\\
H_0
 &=a^{-2\operatorname{lk}(L_0)}\frac{a-a^{-1}}{z}
H^+_{Q_L}H^+_{Q_R}+\mathcal{O}(z^0).
 \end{aligned}
 \label{eq:cornerskein}
\end{equation}
where $\operatorname{lk}$ is the linking number, namely half the signed count of crossings. All crossings in the two
components of $L_0$ are positive. We deduce
\begin{align}
2\operatorname{lk}(L_0)&=1+m_- -m_L-m_R \nonumber \\
&=d_L+d_R-d_-\,.
\end{align}
In the second line we used that the oriented smoothing gives $r_-=r_L+r_R$; uniformity makes these rotations
have the same sign, so $|r_L|+|r_R|=|r_-|$. Taking $a,z\to0$ in the second
line of Eq.~\eqref{eq:cornerskein}, the $+a$ term in the last line would
require degree $d_L+d_R-2$ and vanishes by Theorem~2, whereas the
$-a^{-1}$ term selects the two corner coefficients.  Thus
\begin{equation}
 c_+-c_-=[a^{d_--1}z^{-1}]H_0=-c_Lc_R .
\end{equation}
which translates to the factorization law \eqref{eq:incidence}. The other polygons in such state are spectators. The bookkeeping of the sum over states leading to W1 is detailed in Ref.~\cite{companion}.

\textbf{Soft theorem.} We again derive the result \eqref{eq:softinsertionlaw} for a fixed subpolygon, while the others are simply spectators, for sufficiently small $\varepsilon$. Reading Fig.~\ref{fig:soft-flow} from left to
right, the first sector carries the states of $P$ unchanged, while the middle
sector our subpolygon contains mixed turns and thus vanishes. We thus focus on the loop sector, step 3 of the figure.  Fix a state in which the newborn crossing remains
on one uniformly turning subpolygon $Q$.  Its positive lift contains an empty
\textit{Reidemeister-I} curl.
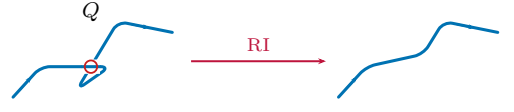
\input{figures_v9_soft_state_sum}
Removing the curl preserves $H_Q^+$, lowers both $m_Q$
and $|r_Q|$ by one, and leaves one dissenting turn.  Theorem~2 therefore gives
\begin{equation}
 \mindeg_aH_Q^+\ge d_Q+2
 \qquad\Longrightarrow\qquad c(Q)=0 .
 \label{eq:nocurl}
\end{equation}
What we have just proven is that, in the state-sum, subpolygons with curls are indeed not allowed. Thus to get a non-zero result we must smooth the circled crossing. This splits off an unknot with $c(\bigcirc)=1$.  The turn parity gives
$-\chi=\chi_\pm$ in the first sector; the additional smoothing changes it to
$\chi=\chi_\pm$ in the loop sector.  Hence, accounting for smoothing and spectator subpolygons,
\begin{equation}
 C(P_\varepsilon)=\frac{\chi_-+\chi_+}{2}\,C(P) .
 \label{eq:Csoft}
\end{equation}
\textbf{W3.} The W3 law has two subcases depending orientation. We will focus on the `transitive' one depicted below.
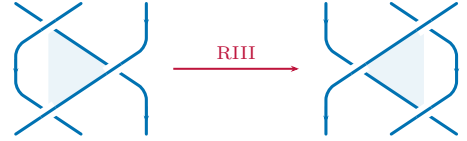
\input{figures_v9_r3_transitive}
It is easy to see that if in the state sum the strands do not belong to the same polygon, the above move is trivial: subpolygons do not link with each other. On the other hand, when a single subpolygon $Q$ contains the three strands, the \textit{Reidemeister-III} move shows that the $c(Q)$ factor is invariant under W3. The states that smooth one or two of the crossings require further cancellations shown in Ref.~\cite{companion}. 

On the other hand, when the three strands have coherent orientation, exact scans of hundreds
of local configurations up to $n=9$, including nontrivial cyclic walls, have found
complete cancellations leading to W3, but we do not prove this case here. W3 suggests an interesting connection to braiding/Yang-Baxter relations.

We close this section by pointing out the reflexion property \eqref{eq:reflx} for $C$. Reversal bijects the decompositions and preserves
linking, uniformity, and $|S|$.  Both tangents reverse at every crossing,
so positivity, $H^+_Q$, $m_Q$, $|r_Q|$, and hence $c(Q)$ are unchanged.
Since $\ell(\overline P)=n-\ell(P)$, the prefactor in \eqref{eq:C} gives
$C(\overline P)=(-1)^nC(P)$.

\section{Examples: infinite families}

We have so far emphasized a conceptual reframing of the gluon amplitude in connection to knot theory. In this section we instead focus on explicit numerical results we can derive at arbitrary $n$ in special kinematics. We obtain an explicit perfect match between the two formulations, i.e. \eqref{eq:amplitude} and \eqref{eq:C}. 

\subsection{Catalan Stars}

We consider star polygons that lie at the boundary of the $(n,r)$ lattice, as shown in Fig.~\ref{fig:v3lattice}. More precisely, for a fixed rotation $r$ the minimal vertex count $n$ that can achieve it is
\begin{equation}
 n_{\min}(r)=
 \begin{cases}
 2|r|+1,&r\ne0,\\
 4,&r=0.
 \end{cases}
 \label{eq:nmin}
\end{equation}
Even though this defines an open region, here it is very useful to construct the regular polygon. For $r\ge1$, place marked points in a circle at angles $\theta_j=\frac{2\pi j}{2r+1}$ and let the polygon visit the
vertices in the order
\begin{equation}
 K_r=(u_0,u_r,u_{2r},\ldots,u_{(2r)r}),
 \qquad\text{indices modulo }2r+1.
 \label{eq:star}
\end{equation}
We depict the first cases of $K_r$ in Fig.~\ref{fig:anchors}.  

Very nicely, these
``minimal stars" lead to a closed form of the amplitude \eqref{eq:amplitude}. We can formulate the Feynman diagram sum as the Berends-Giele (BG) recursion of \cite{GLSSW}:
\begin{equation}\label{eq:BGG}
\bar{A}_{S}=-\!\!\sum_{\substack{S=(S_1|\cdots|S_m)\\m\ge3}}
V^+(a_0,\ldots,a_m)\prod_{\alpha=1}^{m}\bar A_{S_\alpha}.
\end{equation}
where $\bar{A}_S$  is the \textit{preamplitude} just made of $V^+$ vertices, that is, without the $V^-$ root. We note that for the special cyclic kinematics \eqref{eq:star}, the preamplitude $\bar{A}_s$ depends only on the size $s=|S|$.

The key point of the star kinematics \cite{companion} is that $V^+=0$ for $s$ even and $V^+=1$ for $s$ odd, assuming the $s$ parts are odd themselves. We then group the odd values into the generating function
\begin{equation}
    G(t):=\sum_{a=0}^\infty t^a \bar{A}_{2a+1}\,.
\end{equation}
Then the BG recursion \eqref{eq:BGG} takes the closed form
\begin{equation}
G=1-\sum_{a\ge1}t^aG^{\,2a+1} =1-\frac{tG^3}{1-tG^2}\,,
\end{equation}
and is solved by the Catalan numbers $\operatorname{Cat}_a$
\begin{equation}
G(t)=\frac{\sqrt{1+4t}-1}{2t}
=\sum_{a\ge0}(-1)^a\operatorname{Cat}_a t^a\,.
\end{equation}
We now reattach the final $V^-$ root to the preamplitudes. We find that only the cubic vertex survives and that $V^-=-1$, thus
\begin{align}\label{eq:cats}
A(K_r) &=-\sum_{a=0}^{r-1}
(-1)^a\operatorname{Cat}_a\,
(-1)^{r-1-a}\operatorname{Cat}_{r-1-a}\nonumber\\
&=(-1)^r\operatorname{Cat}_r.
\end{align}
As a first corollary of this result, note that $\operatorname{Cat}_r\sim4^r$ shows that the amplitude can grow quickly with $n=2r{+}1$. Secondly, this largely explains the complexity of the original explicit formulas of \cite{GLSSW}: It comes from resolving wall crossing. Indeed, as we cross a large number of walls to `reach the stars' (starting from e.g. convex polygons), it is natural to expect that for fixed odd $n$ we are actually saturating a certain bound, namely
\begin{equation}
    |A(P_n)|\leq \operatorname{Cat}_r\,,\quad\,n=2r+1\,.
\end{equation}
We have verified this bound exhaustively up to $n=9$.

Catalan numbers are well known to emerge in scattering amplitudes through the counting of \textit{cubic} diagrams, and for a scalar theory, in special kinematics, they \textit{are} the amplitude. It is interesting to encounter them for gluons! Strikingly, from the perspective of the knot formula \eqref{eq:C} we have a completely orthogonal origin: For $C(P)$, only the
empty smoothing $|S|=0$ survives! This means that for $r\ge2$ its positive lift \textit{is} the
$(r,2r+1)$ torus knot, whose corner coefficient follows from the analysis of Brini--Eynard--Mari\~no \cite{BEM}. We find perfect agreement with \eqref{eq:cats}.

\input{figures_v9_anchors}

\FloatBarrier

\subsection{Decay loops}

Along the vertical columns of the $(n,r)$ lattice of Fig.~\ref{fig:v3lattice} we encounter \textit{decay loops}. More precisely, 
the one-incoming decay region of
Ref.~\cite{GLSSW} is defined as
\begin{equation}
   \mathcal R_1: \qquad \tilde\lambda_i^+>0\,,\quad\,i\neq 1\,.
\end{equation}
It was observed that in this region $|A(P)|\leq1$. This has a simple polygon interpretation, where we can see that $r=0,\pm 1$. Choose an affine frame with the edge $\widetilde\lambda_1$
horizontal and directed right-to-left.  The return chain, the remaining
edges $\widetilde\lambda_2,\ldots,\widetilde\lambda_n$, is strictly
left-to-right monotone and embedded. Let $k$ be the number of crossings along $\widetilde\lambda_1$. They are pairwise unlinked and their full set $S_{\max}$, with
$k=|S_{\max}|$, is the maximal decomposition. We cannot smooth a smaller set, this will necessarily have mixed turns and vanish. Thus we have  $k+1$
embedded subpolygons as shown in Fig.~\ref{fig:decay}, each with $c(Q)=1$ unless they contain mixed turns. The result is
\begin{equation}
 \left.C(P)\right|_{\mathcal R_1}=
 \begin{cases}
  (-1)^{\ell(P)+k},
    & \text{uniform},\\
  0, & \text{otherwise}.
 \end{cases}
 \label{eq:decayregion}
\end{equation}
Thus $C(P)\in\{-1,0,1\}$ in the decay region, in agreement with \cite{GLSSW}.

\input{figures_v9_decay}

\input{figures_v9_proof_outline}
\vthreelatticefigure
\FloatBarrier

\section{Topological String Dictionary}

We have reformulated the single-minus gluon amplitude as a state sum of knots. To wrap up our discussion we will make a few remarks allowing us to interpret this as an observable of topological field theory \cite{WittenJones} and, via a large $N$ duality, in the topological string \cite{WittenCSString,GopakumarVafa,OoguriVafa,LabastidaMarinoVafa,MarinoCSReview}. The identifications of this section are supported but conjectural.

In \cite{GLMS} a two-parameter deformation of the gravitational amplitude $B_2(y,z)$ was identified as a twisted half-BPS index. It is natural to conceive a similar deformation in this case. We will denote it by $\hat{\mathcal{C}}_P(a,z)$ and recover the amplitude as $a,z\to0$.

We can interpret the contribution from each decomposition in terms of expectation values of Wilson loops. More precisely, note that each decomposition $S$ is indeed a link of many knots.
We write, schematically,
\begin{equation}\label{eq:topstringex}
\frac{\hat{\mathcal{C}}_P(a,z)}{z}=(-1)^{\ell(P)+1}
\sum_{S\ {\rm uniform}}
z^{|S|}\langle W_{L_S}\rangle\,,
\end{equation}
$W_{L_S}$ is a Wilson loop embedded in 3d space, its value given by $U(N)$ Chern-Simons theory \cite{WittenJones}. In fact, the parameters here have a natural interpretation in such case
\begin{align}
z=q^{1/2}-q^{-1/2}\,,\quad a=q^{N/2}\,,
\end{align}
with $q:=e^{\frac{2\pi i}{\kappa+N}}$, and $\kappa$ the level. Through the large $N$ duality \cite{GopakumarVafa,OoguriVafa}, $\langle W_S\rangle$ is recovered from a topological string theory, where $q=e^{g_s}$. To reach our limit $a,z\to 0$, this identification requires a particular analytic continuation of the theory, e.g. $g_s\to 0^-$ with $Ng_s^2$ fixed, which requires a complex level parameter. Analytic  continuation was studied in \cite{WittenAnalyticContinuation}.

The crucial connection is that for a knot $Q$ or a link $L$, the fundamental $U(N)$ Wilson loop is the HOMFLY polynomial times the unknot expectation,
\begin{equation}
 \langle W(L)\rangle_0=a^{2\Lambda_S}\frac{a-a^{-1}}{z}\,H_L(a,z)\,.
 \label{eq:wilsonknot}
\end{equation}
Here $\Lambda_S$ is the sum of the mutual linking numbers, and $\langle \ldots\rangle_0$ represents zero framing as explained below. The factor of $z^{|S|}\sim g_s^{|S|}$ in \eqref{eq:topstringex} is now easy to understand from a disconnected prescription \cite{LabastidaMarinoVafa,WittenTwistor,CSW,GukovMotlNeitzke}. Furthermore, we can actually argue for the rule `every smoothing separates, none rejoins' which we used in Section~\ref{sec:statesum} to define our decompositions $S$. 

Our smoothing is the inverse process to a \textit{band surgery} between knots, e.g. inserting a gluon propagator \cite{GuadagniniMartelliniMintchev},
\input{figures_v9_band_surgery}
\vbandsurgerycartoon

Now suppose that after applying $|S|$ smoothings to the original link we obtain $k$ pieces.
By virtue of the LM formula \eqref{eq:cornerskein}, the lowest power of $z$ is indeed $z^{-k}$ and
\begin{equation}
 [z^{-k}]\,\langle W(L_S)\rangle_0=\prod_{i=1}^{k}\,\,[z^{-1}]\langle W(Q_i)\rangle_0,
 \label{eq:wilsonlink}
\end{equation}
so at this order the mutual linking between knots drops out. To glue all the pieces together we insert propagators, each yielding a power of $z\sim g_s$, and a total power counting
\begin{equation}
    g_s^{|S|-k} =g_s^{-\chi}\,,
\end{equation}
where we have invoked the genus expansion: $\chi$ is the Euler characteristic of the surface obtained by joining the knots via the bands.\footnote{More precisely, in the putative large-$N$ setup each knot is the boundary of a disk. The disk can live in the embedding higher-dimensional space so as to not self-intersect. } Focusing on the tree-level contribution immediately gives
\begin{equation}
    k=|S|+1\,.
\end{equation}
We see that allowing further band surgeries (i.e. increasing $|S|$ at fixed $k$) would imply a loop in the surface.

It remains to argue the factor of $a^{1-d_Q}$ in Theorem~\ref{thm:corner}. This comes from the fact that for each knot, the expectation value of its Wilson loop requires a point-splitting procedure given by specifying a framing \cite{WittenJones}. For us, this amounts to displace the knot $Q\to Q'$ in a small direction orthogonal to $\eqref{eq:1frm}$ at each point. This provides an explicit calculation of $\operatorname{lk}(Q,Q')=-\operatorname{sl}(Q)$. For our positive knot diagram on the plane, we simply draw $Q'$ as accompanying $Q$ to its right. The result along a curl looks like this
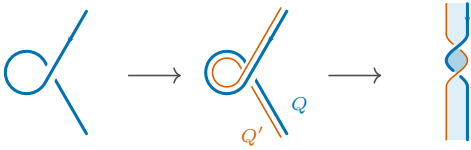
\input{figures_v9_ribbon_curl}
Each curl is one full twist of the ribbon, one unit of framing. A unit of framing multiplies the fundamental Wilson loop by $a$ \cite{MarinoVafa}, and a uniformly turning component carries $|r_Q|$ curls on top of its $m_Q$ crossings: our ribbon framing gives $\operatorname{lk}(Q,Q')=-\operatorname{sl}(Q)=m_Q+|r_Q|=1-d_Q$. Thus the expectation value becomes
\begin{equation}
     \langle W(L)\rangle{=}a^{-\operatorname{sl}(L)}\frac{a{-}a^{{-}1}}{z}H_L(a,z),\quad \operatorname{sl}(L){=}\sum_i \operatorname{sl}(Q_i){-} 2\Lambda
 \label{eq:wilsonframed}
\end{equation}
This then justifies the definition \eqref{eq:topstringex} and shows that $\hat{\mathcal{C}}_P(0,0)=A(P)$. 

The construction further suggests that $\hat{\mathcal{C}}$ is a `motivic amplitude' and may be obtained by promoting the single-minus gluon Feynman rules to a non-commutative background \cite{GLMS}. It also bears some resemblance to the spectral network construction of GMN for BPS states counting, which is particularly illuminating near the Argyres--Douglas point \cite{GaiottoMooreNeitzke}. We leave these fascinating directions, as well as the precise construction of the string dual, to future work. Next we discuss some further speculations.

In AdS/CFT, null Wilson polygons bound minimal string worldsheets
\cite{AldayMaldacena}.  Here the boundary data already span
$\bar z=x^-=0$, whose induced metric and area vanish. This differs from the two-dimensional kinematics of Refs.~\cite{AldayMaldacena,CaronHuotHe}, where the polygon lies in $\mathbb R^{1,1}$ and the minimal area is nontrivial.  In this degenerate
sense the surface is already minimal, with no area to minimize.\footnote{We thank J. Maldacena for providing this perspective.}  The
surviving knot invariant therefore suggests a protected nonperturbative
correction on the null sheet.

We expect supersymmetry to protect the single-minus correspondence, making
it nonperturbative.  On the other hand, it can be shown that the wall calculus supplies both $U(1)$ decoupling and
the Kleiss--Kuijf relations between single trace terms at large $N$ \cite{companion}, and the correspondence
turns them into identities among the knot state sums.

The twistor description contains a natural candidate for the extra direction
of the lift function in \eqref{eq:1frm}. Parametrize the contour by $t$, with $Z(t)$ the twistor of the
running point, linear along each side.  The remaining freedom
\begin{equation}
 \widehat Z(t)=Z(t)+h(t)Y,
 \qquad
 Y\wedge\widehat Z(t)=Y\wedge Z(t)
 \label{eq:Ylift}
\end{equation}
leaves the spacetime null contour unchanged.  Once the quotient plane and the real
$Y$-fiber are oriented, the positive-lift knot type is canonical.

This suggests, at least conjecturally, that a sharper description could be holomorphic Chern--Simons theory in
complex twistor space \cite{WittenTwistor,CostelloTwistor}.  Holomorphic Wilson loops there compute scattering
data through holomorphic linking, and their loop equations are holomorphic
skein relations \cite{BullimoreSkinner}.  On the half-collinear split-real
slice, this may reduce to ordinary over--under data in the $Y$-fiber: W1 would
be loop splitting, W3 topological invariance (likely related to braiding/Yang-Baxter condition), and the positive lift a chiral
resolution of the singular null Wilson loop.  In this picture the lift would be part of the quantum definition of the
operator.

\begin{acknowledgments}
We thank N.~Arkani-Hamed, F.~Coronado,  A. ~Lupsasca, J.~Maldacena, G. ~Salvatori, D. ~Skinner and A.~Strominger for many
enlightening discussions. %

The structural form of equation \eqref{eq:C} was originally devised by Black-Box-Arena, an automated AI research harness used to obtain new formulas in physics and mathematics. Although the formula was incorrect, the author managed to patch several details and provide the interpretation presented here. The open-source package as well as the development of the formula are being documented in \href{https://github.com/aguevara22/black-box-arena}{https://github.com/aguevara22/black-box-arena}.

We thank Anthropic, OpenAI and Google Deepmind for generously providing AI assistance during various stages of this work. The author is supported in part by the Simons foundation.

\end{acknowledgments}

\bibliography{refs_v9}

\end{document}

%% file: figures_v9_palette.tex
\definecolor{pieceA}{RGB}{0,114,178}
\definecolor{pieceB}{RGB}{213,94,0}
\definecolor{pieceC}{RGB}{0,158,115}
\definecolor{pieceD}{RGB}{204,121,167}
\definecolor{curlmark}{RGB}{200,30,30}
\definecolor{curlfill}{RGB}{250,214,214}
\definecolor{cornermark}{RGB}{170,0,140}
\definecolor{rowhi}{RGB}{255,190,60}
\definecolor{movearrow}{RGB}{190,20,60}
\definecolor{beforecol}{RGB}{150,190,225}
\definecolor{discardring}{RGB}{145,145,145}

\newcommand{\knotgapwidth}{5.4pt}
\newcommand{\knotround}{3.4pt}
\tikzset{
  knotstrand/.style={line width=1.15pt,line cap=round,line join=round,rounded corners=\knotround},
  knotgap/.style={white,line width=\knotgapwidth,line cap=butt},
}

%% file: figures_v9_chamber_components.tex
\newcommand{\VSixPanelCore}[5]{%
  \begin{scope}[shift={#2}]
    \foreach \x/\y/\col [count=\i from 0] in {#3}{%
      \coordinate (atlasv\i) at (\x,\y);}
    \draw[black!68,line width=.82pt,line join=round]
      (atlasv0)--(atlasv1)--(atlasv2)--(atlasv3)--(atlasv4)--cycle;
    #5
    \foreach \x/\y/\op in {#4}{%
      \fill[white,opacity=\op] (\x,\y) circle (2.15pt);
      \fill[cornermark,opacity=\op] (\x,\y) circle (1.35pt);}
    \foreach \x/\y/\col in {#3}{%
      \fill[white] (\x,\y) circle (2.8pt);
      \fill[\col] (\x,\y) circle (1.95pt);}
    \node[font=\scriptsize\bfseries,anchor=north] at (0,-.735) {#1};
  \end{scope}}
\newcommand{\VSixAtlasPanel}[6]{%
  \VSixPanelCore{#1}{#2}{#3}{#4}{%
    \draw[movearrow,line width=.62pt,-{Stealth[length=2.6pt,width=2.5pt]}] #5--#6;}}
\newcommand{\VSixPlainPanel}[4]{%
  \VSixPanelCore{#1}{#2}{#3}{#4}{}}
\newcommand{\VSixMovePanel}[6]{%
  \VSixPanelCore{#1}{#2}{#3}{#4}{%
    \draw[white,line width=1.7pt,shorten <=5pt] #5--#6;
    \draw[movearrow,line width=.68pt,shorten <=5pt,-{Stealth[length=3.4pt,width=3.4pt]}] #5--#6;}}

\newcommand{\VSixRtwoRow}{%
  \VSixAtlasPanel{1}{(0.000,0.000)}{0.1615/0.3446/pieceC,-0.0251/-0.0976/pieceC,0.5707/0.4623/pieceC,-0.5707/0.5700/pieceC,0.0840/-0.5700/pieceC}{0.1369/0.0547/1}{(0.0981,0.1942)}{(0.0533,0.0881)}
  \VSixAtlasPanel{2}{(2.300,0.000)}{-0.0330/0.2622/pieceC,-0.4676/-0.4987/pieceC,0.5059/0.4643/pieceC,-0.5059/0.5700/pieceC,0.2048/-0.5700/pieceC}{-0.1674/0.0270/1,-0.0792/-0.1145/1,0.0407/0.0041/1}{(-0.1808,0.0035)}{(-0.2851,-0.1791)}
  \VSixAtlasPanel{3}{(4.600,0.000)}{0.0000/0.5700/pieceC,-0.3705/-0.5700/pieceC,0.5993/0.1345/pieceC,-0.5993/0.1345/pieceC,0.3705/-0.5700/pieceC}{-0.1415/0.1345/1,-0.2290/-0.1345/1,0.0000/-0.3008/1,0.2290/-0.1345/1,0.1415/0.1345/1}{(-0.1260,0.1824)}{(-0.2149,-0.0912)}}

\newcommand{\VSixRtwoMoveRow}{%
  \VSixMovePanel{1}{(0.000,0.000)}{0.1615/0.3446/pieceC,-0.0251/-0.0976/pieceC,0.5707/0.4623/pieceC,-0.5707/0.5700/pieceC,0.0840/-0.5700/pieceC}{0.1369/0.0547/1}{(-0.0251,-0.0976)}{(-0.0579,-0.4642)}
  \VSixMovePanel{2}{(2.300,0.000)}{-0.0330/0.2622/pieceC,-0.4676/-0.4987/pieceC,0.5059/0.4643/pieceC,-0.5059/0.5700/pieceC,0.2048/-0.5700/pieceC}{-0.1674/0.0270/1,-0.0792/-0.1145/1,0.0407/0.0041/1}{(-0.0330,0.2622)}{(0.0165,0.6446)}
  \VSixPlainPanel{3}{(4.600,0.000)}{0.0000/0.5700/pieceC,-0.3705/-0.5700/pieceC,0.5993/0.1345/pieceC,-0.5993/0.1345/pieceC,0.3705/-0.5700/pieceC}{-0.1415/0.1345/1,-0.2290/-0.1345/1,0.0000/-0.3008/1,0.2290/-0.1345/1,0.1415/0.1345/1}
  \draw[black!55,line width=.95pt,-{Stealth[length=4.3pt,width=4.5pt]}]
    (.88,0)--(1.72,0);
  \draw[black!55,line width=.95pt,-{Stealth[length=4.3pt,width=4.5pt]}]
    (3.18,0)--(4.02,0);}

%% file: figures_v9_intro.tex
\par\onecolumngrid
\noindent\begin{minipage}[b]{\textwidth}\centering
\scalebox{0.85}{\begin{tikzpicture}[x=1cm,y=1cm]
  \path[use as bounding box] (-2.3,-2.05) rectangle (15.75,1.95);
  \begin{scope}[shift={(-0.9,0)}]
    \fill[pieceA!14] (0,0) circle (.46); \draw[pieceA!70!black,line width=.7pt] (0,0) circle (.46);
    \draw[black!80,line width=.6pt,-{Stealth[length=2.4pt]}] (207.0:1.18)--(207.0:.50); \node[font=\tiny,inner sep=1pt] at (207.0:1.31) {$1$};
    \draw[black!80,line width=.6pt,-{Stealth[length=2.4pt]}] (198.0:1.18)--(198.0:.50); \node[font=\tiny,inner sep=1pt] at (198.0:1.31) {};
    \draw[black!80,line width=.6pt,-{Stealth[length=2.4pt]}] (189.0:1.18)--(189.0:.50); \node[font=\tiny,inner sep=1pt] at (189.0:1.31) {};
    \draw[black!80,line width=.6pt,-{Stealth[length=2.4pt]}] (180.0:1.18)--(180.0:.50); \node[font=\tiny,inner sep=1pt] at (180.0:1.31) {};
    \draw[black!80,line width=.6pt,-{Stealth[length=2.4pt]}] (171.0:1.18)--(171.0:.50); \node[font=\tiny,inner sep=1pt] at (171.0:1.31) {};
    \draw[black!80,line width=.6pt,-{Stealth[length=2.4pt]}] (162.0:1.18)--(162.0:.50); \node[font=\tiny,inner sep=1pt] at (162.0:1.31) {};
    \draw[black!80,line width=.6pt,-{Stealth[length=2.4pt]}] (153.0:1.18)--(153.0:.50); \node[font=\tiny,inner sep=1pt] at (153.0:1.31) {};
    \draw[black!80,line width=.6pt,-{Stealth[length=2.4pt]}] (-27.0:.48)--(-27.0:1.18); \node[font=\tiny,inner sep=1pt] at (-27.0:1.31) {};
    \draw[black!80,line width=.6pt,-{Stealth[length=2.4pt]}] (-18.0:.48)--(-18.0:1.18); \node[font=\tiny,inner sep=1pt] at (-18.0:1.31) {};
    \draw[black!80,line width=.6pt,-{Stealth[length=2.4pt]}] (-9.0:.48)--(-9.0:1.18); \node[font=\tiny,inner sep=1pt] at (-9.0:1.31) {};
    \draw[black!80,line width=.6pt,-{Stealth[length=2.4pt]}] (0.0:.48)--(0.0:1.18); \node[font=\tiny,inner sep=1pt] at (0.0:1.31) {};
    \draw[black!80,line width=.6pt,-{Stealth[length=2.4pt]}] (9.0:.48)--(9.0:1.18); \node[font=\tiny,inner sep=1pt] at (9.0:1.31) {};
    \draw[black!80,line width=.6pt,-{Stealth[length=2.4pt]}] (18.0:.48)--(18.0:1.18); \node[font=\tiny,inner sep=1pt] at (18.0:1.31) {};
    \draw[black!80,line width=.6pt,-{Stealth[length=2.4pt]}] (27.0:.48)--(27.0:1.18); \node[font=\tiny,inner sep=1pt] at (27.0:1.31) {$14$};
  \end{scope}
  \node[font=\scriptsize,anchor=north] at (-0.9,-1.72) {(a)};
  \begin{scope}[shift={(3.05,0.05)},scale=1.45]
    \draw[pieceA,line width=1.10pt,line join=round] (-0.4825,-0.4507)--(-0.4445,-1.0021)--(-0.0322,-0.9950)--(0.7279,0.4219)--(0.4383,0.8633)--(-0.0196,0.4302)--(0.8563,-0.2347)--(0.6289,-0.9066)--(-0.8563,0.5245)--(-0.2051,0.7554)--(-0.1579,0.3824)--(-0.5271,0.3113)--(-0.5100,1.0021)--(0.4462,0.5449)--cycle;
    \foreach \p in {(-0.4825,-0.4507),(-0.4445,-1.0021),(-0.0322,-0.9950),(0.7279,0.4219),(0.4383,0.8633),(-0.0196,0.4302),(0.8563,-0.2347),(0.6289,-0.9066),(-0.8563,0.5245),(-0.2051,0.7554),(-0.1579,0.3824),(-0.5271,0.3113),(-0.5100,1.0021),(0.4462,0.5449)} \fill[pieceB!92!black] \p circle (1.6pt);
    \draw[pieceA,line width=.88pt,-{Stealth[length=3pt]}] (-0.4654,-0.6988)--(-0.4589,-0.7925);
    \node[font=\scriptsize\bfseries,text=black!72] at (-0.806,1.122) {$P$};
  \end{scope}
  \node[font=\scriptsize,anchor=north] at (3.05,-1.72) {(b)};
  \begin{scope}[shift={(7.95,0.05)},scale=1.45]
    \draw[pieceA,knotstrand]
      (-0.8563,0.5245) --
      (-0.2051,0.7554) --
      (-0.1579,0.3824) --
      (-0.5271,0.3113) --
      (-0.5100,1.0021) --
      (0.4462,0.5449) --
      (-0.1833,-0.0162) [rounded corners=3.4pt] -- cycle;
    \draw[pieceB,knotstrand]
      (-0.4825,-0.4507) --
      (-0.4445,-1.0021) --
      (-0.0322,-0.9950) --
      (0.1353,-0.5307) [rounded corners=3.4pt] --
      (-0.1774,-0.2374) [rounded corners=3.4pt] -- cycle;
    \draw[pieceC,knotstrand]
      (0.5412,-0.0968) [rounded corners=3.4pt] --
      (0.8563,-0.2347) [rounded corners=\knotround] --
      (0.6289,-0.9066) --
      (0.3134,-0.5028) [rounded corners=3.4pt] -- cycle;
    \draw[pieceD,knotstrand]
      (0.7279,0.4219) --
      (0.4383,0.8633) --
      (-0.0196,0.4302) --
      (0.4883,0.1460) [rounded corners=3.4pt] -- cycle;
    \draw[knotgap] (0.2721,0.7062)--(0.1625,0.6025);
    \draw[pieceD,line width=1.15pt,line cap=round] (0.2721,0.7062)--(0.1625,0.6025);
    \draw[knotgap] (0.2420,0.3628)--(0.1331,0.2658);
    \draw[pieceA,line width=1.15pt,line cap=round] (0.2420,0.3628)--(0.1331,0.2658);
    \draw[knotgap] (-0.5899,0.6190)--(-0.4479,0.6693);
    \draw[pieceA,line width=1.15pt,line cap=round] (-0.5899,0.6190)--(-0.4479,0.6693);
    \draw[pieceA,line width=1.15pt,-{Stealth[length=3.2pt]}] (-0.0797,0.7963)--(0.0637,0.7277);
    \draw[pieceB,line width=1.15pt,-{Stealth[length=3.2pt]}] (-0.4654,-0.6988)--(-0.4597,-0.7815);
    \draw[pieceC,line width=1.15pt,-{Stealth[length=3.2pt]}] (0.7540,-0.5370)--(0.7199,-0.6378);
    \draw[pieceD,line width=1.15pt,-{Stealth[length=3.2pt]}] (0.2322,0.6684)--(0.1635,0.6035);
    \draw[black!55,densely dashed,line width=.5pt,rounded corners=2pt] (-0.956,-0.116) rectangle (0.546,1.102);
  \end{scope}
  \node[font=\scriptsize,anchor=north] at (7.95,-1.72) {(c)};
  \begin{scope}[shift={(13.45,0.05)}]
    \draw[pieceA,double=pieceA!14,double distance=5.2pt,line width=0.8pt,line join=round,rounded corners=6.0pt] (-1.6933,0.0821)--(-0.0000,0.6824)--(0.1227,-0.2874)--(-0.8372,-0.4722)--(-0.7930,1.3238)--(1.6933,0.1351)--(0.0566,-1.3238)--cycle;
    \draw[white,line width=11.6pt,line cap=butt] (-1.0683,0.3037)--(-0.5636,0.4826);
    \draw[pieceA,double=pieceA!14,double distance=5.2pt,line width=0.8pt,line join=round,line cap=butt] (-1.0683,0.3037)--(-0.5636,0.4826);
    \begin{scope}[shift={(0.4501,0.7294)},rotate=-25.55]
      \fill[white] (-0.260,-0.135) rectangle (0.260,0.135);
      \fill[pieceA!14] plot[domain=-0.260:0.260,samples=41,variable=\s] (\s,{0.1054*cos(360*(\s+0.260)/0.520)}) -- plot[domain=0.260:-0.260,samples=41,variable=\s] (\s,{-0.1054*cos(360*(\s+0.260)/0.520)}) -- cycle;
      \fill[pieceA!42] plot[domain=-0.130:0.130,samples=25,variable=\s] (\s,{0.1054*cos(360*(\s+0.260)/0.520)}) -- plot[domain=0.130:-0.130,samples=25,variable=\s] (\s,{-0.1054*cos(360*(\s+0.260)/0.520)}) -- cycle;
      \draw[pieceA,line width=0.8pt] plot[domain=-0.260:0.260,samples=41,variable=\s] (\s,{0.1054*cos(360*(\s+0.260)/0.520)});
      \draw[pieceA,line width=0.8pt] plot[domain=-0.260:0.260,samples=41,variable=\s] (\s,{-0.1054*cos(360*(\s+0.260)/0.520)});
    \end{scope}
    \begin{scope}[shift={(-0.8184,-0.6208)},rotate=141.22]
      \fill[white] (-0.260,-0.135) rectangle (0.260,0.135);
      \fill[pieceA!14] plot[domain=-0.260:0.260,samples=41,variable=\s] (\s,{0.1054*cos(360*(\s+0.260)/0.520)}) -- plot[domain=0.260:-0.260,samples=41,variable=\s] (\s,{-0.1054*cos(360*(\s+0.260)/0.520)}) -- cycle;
      \fill[pieceA!42] plot[domain=-0.130:0.130,samples=25,variable=\s] (\s,{0.1054*cos(360*(\s+0.260)/0.520)}) -- plot[domain=0.130:-0.130,samples=25,variable=\s] (\s,{-0.1054*cos(360*(\s+0.260)/0.520)}) -- cycle;
      \draw[pieceA,line width=0.8pt] plot[domain=-0.260:0.260,samples=41,variable=\s] (\s,{0.1054*cos(360*(\s+0.260)/0.520)});
      \draw[pieceA,line width=0.8pt] plot[domain=-0.260:0.260,samples=41,variable=\s] (\s,{-0.1054*cos(360*(\s+0.260)/0.520)});
    \end{scope}
    \draw[pieceA,line width=1.0pt,-{Stealth[length=3.2pt]}] (-0.5444,1.2049)--(-0.2957,1.0860);
    \draw[black!55,densely dashed,line width=.5pt,rounded corners=2pt] (-1.913,-1.544) rectangle (1.913,1.544);
    \node[font=\scriptsize,pieceA,anchor=north west] at (-1.85,1.50) {$Q$};
  \end{scope}
  \draw[black!45,line width=.4pt] (8.742,1.648)--(11.537,1.594);
  \draw[black!45,line width=.4pt] (8.742,-0.119)--(11.537,-1.494);
  \node[font=\scriptsize,anchor=north] at (13.45,-1.72) {(d)};
\end{tikzpicture}}
\par\vspace{-4pt}

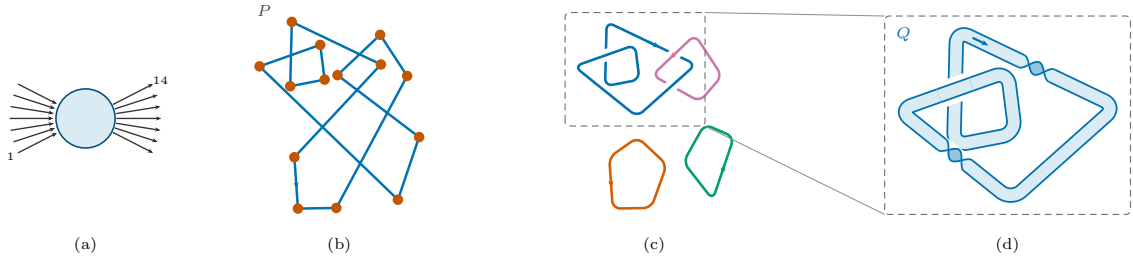
\captionof{figure}{\label{fig:intro}(a) Gluon scattering; (b) its null Wilson polygon $P$; (c) a decomposition of the positive lift of $P$ into four linked knots; (d) the framed knot $Q$ embedded in a topological theory, including twists.}\end{minipage}\par
\clearpage
\twocolumngrid

%% file: figures_v9_wilson_loop.tex
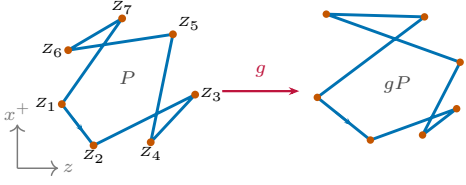
\begin{figure}[t]
\centering
\begin{tikzpicture}[x=1cm,y=1cm]
  \path (-3.48,-1.26) rectangle (3.48,1.16);

  \newcommand{\WilsonImagePoint}[3]{%
    \pgfmathsetmacro{\WilsonDen}{1+.38*(#2)-.18*(#3)}%
    \coordinate (#1) at
      ({((#2)+.18*(#3)+.12)/\WilsonDen},
       {(-.24*(#2)+.92*(#3)-.06)/\WilsonDen});%
  }

  \coordinate (axisorigin) at (-3.22,-1.08);
  \draw[->,black!52,line width=.55pt]
    (axisorigin)--++(.54,0)
    node[right,font=\scriptsize,inner sep=1.5pt] {$z$};
  \draw[->,black!52,line width=.55pt]
    (axisorigin)--++(0,.58)
    node[above,font=\scriptsize,inner sep=1.5pt] {$x^+$};

  \begin{scope}[shift={(-1.75,.02)},scale=.84]
    \coordinate (Z0) at (-1.05,-.30);
    \coordinate (Z1) at (-.55,-.95);
    \coordinate (Z2) at (1.05,-.15);
    \coordinate (Z3) at (.35,-.90);
    \coordinate (Z4) at (.70,.80);
    \coordinate (Z5) at (-.95,.55);
    \coordinate (Z6) at (-.10,1.05);

    \draw[pieceA,line width=1.10pt,line join=round]
      (Z0)--(Z1)--(Z2)--(Z3)--(Z4)--(Z5)--(Z6)--cycle;
    \draw[pieceA,line width=.88pt,-{Stealth[length=3.0pt]}]
      (-.88,-.52)--(-.71,-.74);
    \foreach \p in {Z0,Z1,Z2,Z3,Z4,Z5,Z6}
      \fill[pieceB!92!black] (\p) circle (1.75pt);

    \node[font=\tiny,anchor=east,inner sep=1.5pt] at (Z0) {$Z_1$};
    \node[font=\tiny,anchor=north,inner sep=1.5pt] at (Z1) {$Z_2$};
    \node[font=\tiny,anchor=west,inner sep=1.5pt] at (Z2) {$Z_3$};
    \node[font=\tiny,anchor=north,inner sep=1.5pt] at (Z3) {$Z_4$};
    \node[font=\tiny,anchor=south west,inner sep=1.5pt] at (Z4) {$Z_5$};
    \node[font=\tiny,anchor=east,inner sep=1.5pt] at (Z5) {$Z_6$};
    \node[font=\tiny,anchor=south,inner sep=1.5pt] at (Z6) {$Z_7$};
    \node[font=\scriptsize\bfseries,text=black!72] at (-.02,.12) {$P$};
  \end{scope}

  \draw[-{Stealth[length=4pt]},movearrow,line width=.85pt]
    (-.51,-.06)--(.51,-.06);
  \node[font=\scriptsize,text=movearrow,anchor=south]
    at (0,.02) {$g$};

  \begin{scope}[shift={(1.95,-.04)},scale=.80]
    \WilsonImagePoint{W0}{-1.05}{-.30}
    \WilsonImagePoint{W1}{-.55}{-.95}
    \WilsonImagePoint{W2}{1.05}{-.15}
    \WilsonImagePoint{W3}{.35}{-.90}
    \WilsonImagePoint{W4}{.70}{.80}
    \WilsonImagePoint{W5}{-.95}{.55}
    \WilsonImagePoint{W6}{-.10}{1.05}

    \draw[pieceA,line width=1.10pt,line join=round]
      (W0)--(W1)--(W2)--(W3)--(W4)--(W5)--(W6)--cycle;
    \draw[pieceA,line width=.88pt,-{Stealth[length=3.0pt]}]
      (-1.20,-.37)--(-.93,-.59);
    \foreach \p in {W0,W1,W2,W3,W4,W5,W6}
      \fill[pieceB!92!black] (\p) circle (1.75pt);

    \node[font=\scriptsize\bfseries,text=black!72] at (-.18,.08) {$gP$};
  \end{scope}
\end{tikzpicture}
\caption{\label{fig:wilsonloop}Two affine representatives of
the same half-collinear null Wilson loop, related by a dual conformal
transformation $g$; the axes span the totally null sheet.}
\end{figure}

%% file: figures_v9_walls.tex
\newcommand{\VSixCuspFigure}{%
\begin{center}
\begin{tikzpicture}[x=.72cm,y=.72cm]
  \path[use as bounding box] (-1.82,-1.85) rectangle (5.72,1.95);
  \node[font=\footnotesize] at (0,1.65)
    {\textcolor{beforecol}{before} $\to$ after};
  \node[font=\footnotesize] at (3.90,1.65) {deletion};
  \begin{scope}[shift={(0,0)}]
    \draw[beforecol,line width=1.05pt,line join=round]
      (.5004,-.19425)--(-.4332,-.744)--(-1.4736,1.27275)--
      (1.074,-.3855)--(1.2936,-1.27275)--(1.4736,.6975)--
      (-.6192,1.248)--cycle;
    \draw[black!88,line width=.68pt,line join=round]
      (.5004,-.19425)--(-.4332,-.744)--(-1.4736,1.27275)--
      (1.074,-.3855)--(-.1104,-1.12275)--(1.4736,.6975)--
      (-.6192,1.248)--cycle;
    \draw[white,line width=2.65pt,line cap=butt]
      (.1704,.231)--(.2148,.174)--(.2592,.117);
    \draw[black!88,line width=.68pt,line cap=butt]
      (.1704,.231)--(.2148,.174)--(.2592,.117);
    \draw[white,line width=2.65pt,line cap=butt]
      (.617455,-.088332)--(.837455,-.231531);
    \draw[black!88,line width=.68pt,line cap=butt]
      (.617455,-.088332)--(.837455,-.231531);
    \draw[black!88,line width=.68pt,-{Stealth[length=2.5pt]}]
      (.1896,-.37725)--(-.0672,-.528);
    \node[circle,fill=beforecol,inner sep=0pt,minimum size=3pt]
      at (1.2936,-1.27275) {};
    \node[circle,fill=movearrow,inner sep=0pt,minimum size=3pt]
      at (-.1104,-1.12275) {};
  \end{scope}
  \begin{scope}[shift={(3.90,0)}]
    \draw[black!12,line width=.50pt,line join=round]
      (.5004,-.19425)--(-.4332,-.744)--(-1.4736,1.27275)--
      (1.074,-.3855)--(1.2936,-1.27275)--(1.4736,.6975)--
      (-.6192,1.248)--cycle;
    \draw[pieceC,line width=.90pt,line join=round]
      (.5004,-.19425)--(-.4332,-.744)--(-1.4736,1.27275)--
      (1.074,-.3855)--(1.4736,.6975)--(-.6192,1.248)--cycle;
    \draw[white,line width=2.8pt,line cap=butt]
      (.1704,.231)--(.2148,.174)--(.2592,.117);
    \draw[pieceC,line width=.90pt,line cap=butt]
      (.1704,.231)--(.2148,.174)--(.2592,.117);
    \draw[pieceC,line width=.90pt,-{Stealth[length=2.5pt]}]
      (.1896,-.37725)--(-.0672,-.528);
  \end{scope}
  \node[font=\footnotesize] at (0,-1.65) {$A:0\to1$};
  \node[font=\footnotesize] at (3.90,-1.65) {$A(P\setminus j)=1$};
\end{tikzpicture}
\end{center}
}

\begin{figure}[!b]
\centering
\begin{tikzpicture}[x=.72cm,y=.72cm]
  \path[use as bounding box] (-1.82,-5.35) rectangle (9.62,2.05);
  \node[font=\footnotesize] at (0,1.82) {(a) W1 (bigon)};
  \node[font=\footnotesize] at (3.90,1.82) {(b) W1 (sliding)};
  \node[font=\footnotesize] at (7.80,1.82) {(c) W2 (flat)};

  \begin{scope}[shift={(7.528,-.145)}]
\draw[beforecol,line width=1.05pt,line join=round]
    (.689,-.467)--(-.089,-1.200)--(-.956,1.489)--(1.167,-.722)--
    (1.489,-.033)--(1.500,.722)--(-.244,1.456)--cycle;
  \draw[black!88,line width=.68pt,line join=round]
    (.689,-.467)--(-.089,-1.200)--(-.956,1.489)--(1.167,-.722)--
    (1.178,.033)--(1.500,.722)--(-.244,1.456)--cycle;
  \draw[white,line width=2.65pt,line cap=butt]
    (.414,.100)--(.451,.024)--(.488,-.052);
  \draw[black!88,line width=.68pt,line cap=butt]
    (.414,.100)--(.451,.024)--(.488,-.052);
  \draw[black!88,line width=.68pt,-{Stealth[length=2.5pt]}]
    (.430,-.711)--(.216,-.912);
  \node[circle,fill=beforecol,inner sep=0pt,minimum size=3pt] at (1.489,-.033) {};
  \node[circle,fill=movearrow,inner sep=0pt,minimum size=3pt] at (1.178,.033) {};
  \end{scope}
  \begin{scope}[shift={(0,-.773)},yscale=1.6]
\draw[beforecol,line width=1.05pt,line join=round]
    (-1.556,0)--(1.556,0)--(1.011,.922)--(1.344,1.200)--
    (1.344,.722)--(0,.333)--(.133,.589)--(-.378,1.300)--
    (-.878,1.144)--cycle;
  \draw[black!88,line width=.68pt,line join=round]
    (-1.556,0)--(1.556,0)--(1.011,.922)--(1.344,1.200)--
    (1.344,.722)--(0,-.333)--(.133,.589)--(-.378,1.300)--
    (-.878,1.144)--cycle;
  \draw[white,line width=2.65pt,line cap=butt] (-.030,0)--(.126,0);
  \draw[black!88,line width=.68pt,line cap=butt] (-.030,0)--(.126,0);
  \draw[white,line width=2.65pt,line cap=butt] (.347,0)--(.503,0);
  \draw[black!88,line width=.68pt,line cap=butt] (.347,0)--(.503,0);
  \draw[white,line width=2.65pt,line cap=butt]
    (1.242,.531)--(1.197,.607)--(1.152,.683);
  \draw[black!88,line width=.68pt,line cap=butt]
    (1.242,.531)--(1.197,.607)--(1.152,.683);
  \draw[black!88,line width=.68pt,-{Stealth[length=2.5pt]}]
    (-.95,0)--(-.65,0);
  \node[circle,fill=beforecol,inner sep=0pt,minimum size=3pt] at (0,.333) {};
  \node[circle,fill=movearrow,inner sep=0pt,minimum size=3pt] at (0,-.333) {};
  \node[font=\scriptsize,text=beforecol,anchor=south east] at (-.03,.35) {$P_-$};
  \node[font=\scriptsize,text=black!88,anchor=north west] at (.03,-.35) {$P_+$};
  \end{scope}
  \begin{scope}[shift={(3.90,0)}]
    \draw[beforecol,line width=1.05pt,line join=round]
      (-1.5,0)--(1.5,0)--(1,1.3)--(-1,.8)--(0,.3)--
      (.3,-1.3)--(-1.3,-.8)--cycle;
    \draw[black!88,line width=.68pt,line join=round]
      (-1.5,0)--(1.5,0)--(1,1.3)--(-1,.8)--(0,-.3)--
      (.3,-1.3)--(-1.3,-.8)--cycle;
    \draw[white,line width=2.65pt,line cap=butt]
      (-.365,0)--(-.180,0) (-.038,0)--(.150,0);
    \draw[black!88,line width=.68pt,line cap=butt]
      (-.365,0)--(-.180,0) (-.038,0)--(.150,0);
    \draw[black!88,line width=.68pt,-{Stealth[length=2.5pt]}]
      (-1.05,0)--(-.77,0);
    \node[circle,fill=beforecol,inner sep=0pt,minimum size=3pt] at (0,.3) {};
    \node[circle,fill=movearrow,inner sep=0pt,minimum size=3pt] at (0,-.3) {};
    \node[font=\scriptsize,text=beforecol,anchor=south east] at (-.05,.32) {$P_-$};
    \node[font=\scriptsize,text=black!88,anchor=north west] at (.05,-.32) {$P_+$};
  \end{scope}

  \node[font=\scriptsize,text=beforecol,anchor=west] at (9.06,.20) {$P_-$};
  \node[font=\scriptsize,text=black!88,anchor=west] at (9.05,-.55) {$P_+$};

  \begin{scope}[shift={(7.528,-3.495)}]
\draw[black!12,line width=.50pt,line join=round]
    (.689,-.467)--(-.089,-1.200)--(-.956,1.489)--(1.167,-.722)--
    (1.489,-.033)--(1.500,.722)--(-.244,1.456)--cycle;
  \draw[pieceC,line width=.90pt,line join=round]
    (.689,-.467)--(-.089,-1.200)--(-.956,1.489)--(1.167,-.722)--
    (1.500,.722)--(-.244,1.456)--cycle;
  \draw[white,line width=2.8pt,line cap=butt]
    (.414,.100)--(.451,.024)--(.488,-.052);
  \draw[pieceC,line width=.90pt,line cap=butt]
    (.414,.100)--(.451,.024)--(.488,-.052);
  \draw[pieceC,line width=.90pt,-{Stealth[length=2.5pt]}]
    (.430,-.711)--(.216,-.912);
  \end{scope}
  \begin{scope}[shift={(0,-4.123)},yscale=1.6]
\draw[black!12,line width=.50pt,line join=round]
    (-1.556,0)--(1.556,0)--(1.011,.922)--(1.344,1.200)--
    (1.344,.722)--(0,0)--(.133,.589)--(-.378,1.300)--
    (-.878,1.144)--cycle;
  \draw[pieceA,line width=.90pt,line join=round]
    (0,0)--(1.556,0)--(1.011,.922)--(1.344,1.200)--
    (1.344,.722)--cycle;
  \draw[white,line width=2.8pt,line cap=butt]
    (1.226,.558)--(1.181,.634)--(1.136,.710);
  \draw[pieceA,line width=.90pt,line cap=butt]
    (1.226,.558)--(1.181,.634)--(1.136,.710);
  \draw[pieceA,line width=.90pt,-{Stealth[length=2.5pt]}]
    (.45,0)--(.75,0);
  \draw[pieceB,line width=.90pt,line join=round]
    (0,0)--(.133,.589)--(-.378,1.300)--(-.878,1.144)--
    (-1.556,0)--cycle;
  \draw[pieceB,line width=.90pt,-{Stealth[length=2.5pt]}]
    (-.80,0)--(-.53,0);
  \end{scope}
  \begin{scope}[shift={(3.90,-3.35)}]
    \draw[black!12,line width=.50pt,line join=round]
      (-1.5,0)--(1.5,0)--(1,1.3)--(-1,.8)--(0,0)--
      (.3,-1.3)--(-1.3,-.8)--cycle;
    \draw[pieceA,line width=.90pt,line join=round]
      (0,0)--(1.5,0)--(1,1.3)--(-1,.8)--cycle;
    \draw[pieceA,line width=.90pt,-{Stealth[length=2.5pt]}]
      (.45,0)--(.75,0);
    \draw[pieceB,line width=.90pt,line join=round]
      (0,0)--(.3,-1.3)--(-1.3,-.8)--(-1.5,0)--cycle;
    \draw[pieceB,line width=.90pt,-{Stealth[length=2.5pt]}]
      (-.85,0)--(-.57,0);
  \end{scope}
  \node[font=\footnotesize] at (0,-5.08) {$A:0\to1$};
  \node[font=\footnotesize] at (3.90,-5.08) {$A:0\to1$};
  \node[font=\footnotesize] at (7.80,-5.08) {$A:-1\to0$};
\end{tikzpicture}
\caption{\label{fig:walls}Top: before (pale blue) and after (black).
Bottom: deletion or oriented smoothing.}
\end{figure}
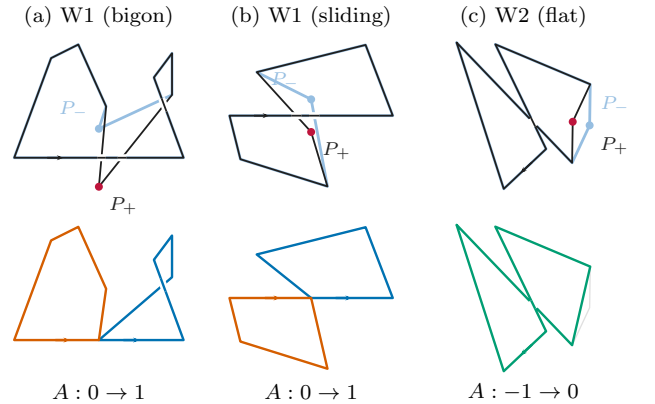

\newcommand{\VSixRThreeFigure}{%
\begin{figure}[!ht]
\centering
\begin{tikzpicture}
  \path (-1.92,-1.62) rectangle (1.92,1.62);
  \fill[curlfill,opacity=.85] (-0.138,0.057) -- (-0.605,0.291) -- (-0.364,-0.176) -- cycle;
  \draw[black!88,line width=1.05pt,line join=round] (1.297,-0.661) -- (-0.138,0.057) -- (-0.605,0.291) -- (-0.732,0.354) -- (-1.226,-1.065) -- (-0.364,-0.176) -- (-0.138,0.057) -- (1.118,1.350) -- (0.227,1.120) -- (-0.861,0.785) -- (-0.605,0.291) -- (-0.364,-0.176) -- (0.282,-1.420) -- cycle;
  \draw[black!88,line width=1.05pt,-{Stealth[length=3.4pt]}] (0.325,0.635) -- (0.506,0.822);
  \draw[white,line width=3.85pt,line cap=butt] (-0.511,0.244) -- (-0.605,0.291) -- (-0.699,0.338);
  \draw[black!88,line width=1.05pt,line join=round] (-0.511,0.244) -- (-0.605,0.291) -- (-0.699,0.338);
  \draw[white,line width=3.85pt,line cap=butt] (-0.189,0.004) -- (-0.138,0.057) -- (-0.064,0.132);
  \draw[black!88,line width=1.05pt,line join=round] (-0.189,0.004) -- (-0.138,0.057) -- (-0.064,0.132);
  \draw[white,line width=3.85pt,line cap=butt] (-0.412,-0.083) -- (-0.364,-0.176) -- (-0.315,-0.269);
  \draw[black!88,line width=1.05pt,line join=round] (-0.412,-0.083) -- (-0.364,-0.176) -- (-0.315,-0.269);
  \draw[curlmark,line width=.7pt] (-0.138,0.057) circle (2.2pt);
  \draw[curlmark,line width=.7pt] (-0.605,0.291) circle (2.2pt);
  \draw[curlmark,line width=.7pt] (-0.364,-0.176) circle (2.2pt);
  \node[font=\footnotesize\bfseries,anchor=north] at (0,-1.62) {$P_-$};
\end{tikzpicture}
\hfill
\begin{tikzpicture}
  \path (-1.92,-1.62) rectangle (1.92,1.62);
  \fill[curlfill,opacity=.85] (-0.336,0.057) -- (0.131,-0.177) -- (-0.110,0.290) -- cycle;
  \draw[black!88,line width=1.05pt,line join=round] (1.257,-0.739) -- (0.131,-0.177) -- (-0.336,0.057) -- (-0.772,0.276) -- (-1.328,-0.965) -- (-0.336,0.057) -- (-0.110,0.290) -- (1.016,1.450) -- (0.227,1.120) -- (-0.387,0.825) -- (-0.110,0.290) -- (0.131,-0.177) -- (0.446,-0.786) -- cycle;
  \draw[black!88,line width=1.05pt,-{Stealth[length=3.4pt]}] (0.326,0.636) -- (0.507,0.823);
  \draw[white,line width=3.85pt,line cap=butt] (0.238,-0.231) -- (0.131,-0.177) -- (0.024,-0.123);
  \draw[black!88,line width=1.05pt,line join=round] (0.238,-0.231) -- (0.131,-0.177) -- (0.024,-0.123);
  \draw[white,line width=3.85pt,line cap=butt] (-0.420,-0.029) -- (-0.336,0.057) -- (-0.252,0.143);
  \draw[black!88,line width=1.05pt,line join=round] (-0.420,-0.029) -- (-0.336,0.057) -- (-0.252,0.143);
  \draw[white,line width=3.85pt,line cap=butt] (-0.165,0.397) -- (-0.110,0.290) -- (-0.055,0.183);
  \draw[black!88,line width=1.05pt,line join=round] (-0.165,0.397) -- (-0.110,0.290) -- (-0.055,0.183);
  \draw[curlmark,line width=.7pt] (0.131,-0.177) circle (2.2pt);
  \draw[curlmark,line width=.7pt] (-0.336,0.057) circle (2.2pt);
  \draw[curlmark,line width=.7pt] (-0.110,0.290) circle (2.2pt);
  \node[font=\footnotesize\bfseries,anchor=north] at (0,-1.62) {$P_+$};
\end{tikzpicture}
\caption{\label{fig:r3}The two chambers adjacent to a W3 wall; the circled
triangle collapses at the wall.}
\end{figure}
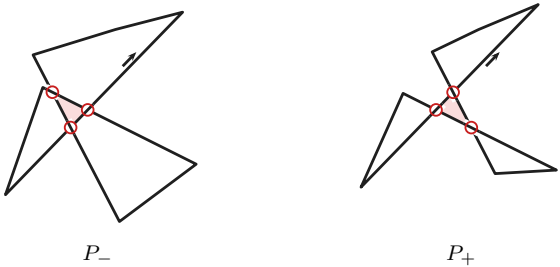
}

%% file: figures_v9_soft_properties.tex
\begin{figure}[!t]
\centering
\begin{tikzpicture}[x=.78cm,y=.78cm]
  \tikzset{
    sfedge/.style={black!88,line width=.88pt,line cap=round,line join=round},
    sfflow/.style={movearrow,line width=.76pt,-{Stealth[length=2.8pt]}},
    sfdot/.style={circle,fill=pieceA!82!black,inner sep=0pt,minimum size=3pt},
    sfnew/.style={circle,fill=movearrow,inner sep=0pt,minimum size=3.2pt}
  }
  \path[use as bounding box] (-1.03,-.62) rectangle (7.03,1.35);

  \begin{scope}[shift={(0,0)}]
    \node[font=\footnotesize\bfseries] at (0,1.16) {1};
    \coordinate (a) at (-.88,0);
    \coordinate (o) at (0,0);
    \coordinate (s) at (.25,.22);
    \coordinate (b) at (.14,.92);
    \draw[sfedge] (a)--(o)--(s)--(b);
    \draw[sfedge,-{Stealth[length=2.4pt]}] (-.72,0)--(-.45,0);
    \node[sfdot] at (o) {};
    \node[sfnew] at (s) {};
    \node[font=\scriptsize] at (0,-.42) {$(-\chi,-\chi)$};
  \end{scope}

  \draw[sfflow] (1.00,.22)--(1.88,.22);

  \begin{scope}[shift={(2.88,0)}]
    \node[font=\footnotesize\bfseries] at (0,1.16) {2};
    \coordinate (a) at (-.88,0);
    \coordinate (o) at (0,0);
    \coordinate (s) at (-.31,.16);
    \coordinate (b) at (.14,.92);
    \draw[sfedge] (a)--(o)--(s)--(b);
    \draw[sfedge,-{Stealth[length=2.4pt]}] (-.72,0)--(-.45,0);
    \node[sfdot] at (o) {};
    \node[sfnew] at (s) {};
    \node[font=\scriptsize] at (0,-.42) {$(-\chi,\chi)$};
  \end{scope}

  \draw[sfflow] (3.88,.22)--(4.76,.22);

  \begin{scope}[shift={(5.76,0)}]
    \node[font=\footnotesize\bfseries] at (0,1.16) {3};
    \coordinate (a) at (-.88,0);
    \coordinate (o) at (0,0);
    \coordinate (s) at (-.31,-.16);
    \coordinate (x) at (-.111,0);
    \coordinate (b) at (.14,.92);
    \fill[curlfill,opacity=.90] (o)--(s)--(x)--cycle;
    \draw[sfedge] (a)--(o)--(s)--(b);
    \draw[white,line width=2.65pt,line cap=butt] (-.165,0)--(-.060,0);
    \draw[sfedge] (-.165,0)--(-.060,0);
    \draw[sfedge,-{Stealth[length=2.4pt]}] (-.72,0)--(-.45,0);
    \node[sfdot] at (o) {};
    \node[sfnew] at (s) {};
    \node[font=\scriptsize] at (0,-.42) {$(\chi,\chi)$};
  \end{scope}
\end{tikzpicture}
\caption{\label{fig:soft-flow}Rotating the soft edge. We show the values of $(\chi_-,\chi_+)$ entering \eqref{eq:softinsertionlaw}.}
\end{figure}
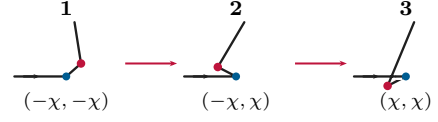

\renewcommand{\VSixCuspFigure}{%
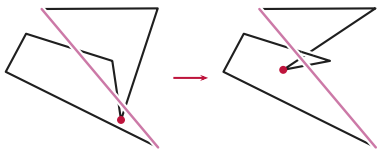
\begin{figure}[!h]
\centering
\begin{tikzpicture}[x=.72cm,y=.72cm]
  \path[use as bounding box] (-1.58,-1.48) rectangle (5.58,1.48);

  \begin{scope}[shift={(0,0)}]
    \draw[black!88,line width=.80pt,line join=round]
      (1.4,-1.277670)--(-1.4,.111976)--
      (-1.025974,.811976)--(.558158,.312727)--
      (.718124,-.772209)--(1.393117,1.277670)--
      (-.739893,1.266196);
    \draw[white,line width=2.45pt,line cap=round]
      (-.739893,1.266196)--(1.4,-1.277670);
    \draw[pieceD,line width=.92pt,line cap=round]
      (-.739893,1.266196)--(1.4,-1.277670);
    \node[circle,fill=movearrow,inner sep=0pt,minimum size=3pt]
      at (.718124,-.772209) {};
  \end{scope}

  \draw[movearrow,line width=.76pt,-{Stealth[length=3pt]}]
    (1.68,0)--(2.32,0);

  \begin{scope}[shift={(4,0)}]
    \draw[black!88,line width=.80pt,line join=round]
      (1.4,-1.277670)--(-1.4,.111976)--
      (-1.025974,.811976)--(.558158,.312727)--
      (-.301325,.148264)--(1.393117,1.277670)--
      (-.739893,1.266196);
    \draw[white,line width=2.55pt,line cap=butt]
      (.037,.477)--(.247,.411);
    \draw[black!88,line width=.80pt,line cap=butt]
      (.037,.477)--(.247,.411);
    \draw[white,line width=2.45pt,line cap=round]
      (-.739893,1.266196)--(1.4,-1.277670);
    \draw[pieceD,line width=.92pt,line cap=round]
      (-.739893,1.266196)--(1.4,-1.277670);
    \node[circle,fill=movearrow,inner sep=0pt,minimum size=3pt]
      at (-.301325,.148264) {};
  \end{scope}
\end{tikzpicture}
\caption{\label{fig:s4-cusp}A spectator strand passing through the W4 cusp region.}
\end{figure}
}

%% file: figures_v9_state_sum.tex
\newcommand{\vstatecoords}{%
  \coordinate (p0) at (-1.18287,0.39316);
  \coordinate (p1) at (-0.28443,-0.90297);
  \coordinate (p2) at (-0.36046,0.18479);
  \coordinate (p3) at (0.17475,0.69731);
  \coordinate (p4) at (0.93565,0.41870);
  \coordinate (p5) at (0.29940,0.00291);
  \coordinate (p6) at (-0.34136,0.54844);
  \coordinate (p7) at (-0.29646,0.95629);
  \coordinate (p8) at (0.28927,1.16567);
  \coordinate (p9) at (0.71225,0.85307);
  \coordinate (p10) at (-0.91069,-1.00219);
  \coordinate (p11) at (-0.00727,-1.27681);
  \coordinate (p12) at (0.23688,-0.64853);
  \coordinate (p13) at (0.08918,-0.29998);
  \coordinate (x0) at (-0.52292,-0.55891);
  \coordinate (o0a) at (-0.45922,-0.48609);
  \coordinate (o0b) at (-0.58662,-0.63173);
  \coordinate (x1) at (-0.32434,-0.33191);
  \coordinate (o1a) at (-0.31268,-0.49874);
  \coordinate (o1b) at (-0.33601,-0.16508);
  \coordinate (x2) at (-0.34304,-0.06446);
  \coordinate (o2a) at (-0.33372,-0.19772);
  \coordinate (o2b) at (-0.35235,0.06879);
  \coordinate (x3) at (-0.15045,0.38590);
  \coordinate (o3a) at (-0.21012,0.32876);
  \coordinate (o3b) at (-0.09078,0.44304);
  \coordinate (x4) at (0.47865,0.58603);
  \coordinate (o4a) at (0.55074,0.66844);
  \coordinate (o4b) at (0.40657,0.50363);
  \coordinate (x5) at (0.10978,0.16435);
  \coordinate (o5a) at (0.16627,0.11625);
  \coordinate (o5b) at (0.05328,0.21245);
  \coordinate (x6) at (-0.17194,-0.15769);
  \coordinate (o6a) at (-0.08929,-0.20273);
  \coordinate (o6b) at (-0.25460,-0.11266);
}
\newcommand{\vstateparent}[1]{%
  \draw[#1,knotstrand]
    (p0)--(p1)--(p2)--(p3)--(p4)--(p5)--(p6)--(p7)--(p8)--(p9)--
    (p10)--(p11)--(p12)--(p13)--cycle;
}
\newcommand{\vstateover}[3]{%
  \draw[knotgap] (#2)--(#3);
  \draw[#1,line width=1.15pt,line cap=round] (#2)--(#3);
}
\newcommand{\vstatebluearrow}{%
  \draw[pieceA,line width=1.15pt,-{Stealth[length=3.4pt]}]
    (-.09145,1.02957)--(.08426,1.09239);}
\newcommand{\vstategreenarrow}{%
  \draw[pieceC,line width=1.15pt,-{Stealth[length=3.4pt]}]
    (.06109,-1.10089)--(.13434,-.91241);}
\newcommand{\vstateorangearrow}{%
  \draw[pieceB,line width=1.15pt,-{Stealth[length=3.4pt]}]
    (-1.02115,.15986)--(-.90435,-.00864);}
\newcommand{\vstategrayarrows}{%
  \draw[black!55,line width=1.15pt,-{Stealth[length=3.4pt]}]
    (-.09145,1.02957)--(.08426,1.09239);
  \draw[black!55,line width=1.15pt,-{Stealth[length=3.4pt]}]
    (.06109,-1.10089)--(.13434,-.91241);}

\newcommand{\vstatebluecartoon}{%
\begin{tikzpicture}[baseline=(current bounding box.center),x=.10cm,y=.10cm]
  \draw[pieceA,line width=.75pt,line cap=round,
    smooth,variable=\t,domain=0:360,samples=181]
    plot ({sin(\t)+2*sin(2*\t)},{cos(\t)-2*cos(2*\t)+.5});
  \foreach \lo/\hi in {8/23,128/143,248/263}{
    \draw[white,line width=2.65pt,line cap=butt,
      smooth,variable=\t,domain=\lo:\hi,samples=9]
      plot ({sin(\t)+2*sin(2*\t)},{cos(\t)-2*cos(2*\t)+.5});
    \draw[pieceA,line width=.75pt,line cap=round,
      smooth,variable=\t,domain=\lo:\hi,samples=9]
      plot ({sin(\t)+2*sin(2*\t)},{cos(\t)-2*cos(2*\t)+.5});}
\end{tikzpicture}}
\newcommand{\vstateorangecartoon}{%
\begin{tikzpicture}[baseline=(current bounding box.center),x=.10cm,y=.10cm]
  \draw[pieceB,line width=.75pt] (0,0) circle (2.45);
\end{tikzpicture}}
\newcommand{\vstategreencartoon}{%
\begin{tikzpicture}[baseline=(current bounding box.center),x=.10cm,y=.10cm]
  \draw[pieceC,line width=.75pt] (0,0) circle (2.45);
\end{tikzpicture}}

\newcommand{\vstatepolygonfigure}{%
\begin{center}
\begin{tikzpicture}
  \path (-1.50,-1.34) rectangle (1.50,1.24);
  \vstatecoords
  \vstateparent{pieceA}
  \vstateover{pieceA}{o0a}{o0b}
  \vstateover{pieceA}{o1a}{o1b}
  \vstateover{pieceA}{o2a}{o2b}
  \vstateover{pieceA}{o3a}{o3b}
  \vstateover{pieceA}{o4a}{o4b}
  \vstateover{pieceA}{o5a}{o5b}
  \vstateover{pieceA}{o6a}{o6b}
  \vstatebluearrow
  \draw[pieceA,line width=1.15pt,-{Stealth[length=3.4pt]}]
    (.06109,-1.10089)--(.13434,-.91241);
  \node[font=\footnotesize,anchor=north] at (0,-1.38) {$P$};
\end{tikzpicture}
\end{center}
}
\newcommand{\vsmoothingcartoon}{%
\begin{center}
\begin{tikzpicture}[x=.55cm,y=.55cm,line width=1.05pt,>={Stealth[length=3.2pt]}]
  \draw[pieceA,->] (1,-1)--(-1,1);
  \draw[knotgap] (-.45,-.45)--(.45,.45);
  \draw[pieceA,->] (-1,-1)--(1,1);
  \node[font=\scriptsize,anchor=north] at (0,-1.25) {positive crossing};
  \draw[black!70,-{Stealth[length=3pt]},line width=.6pt] (1.7,0)--(2.9,0);
  \node[font=\scriptsize,anchor=south] at (2.3,.08) {smooth};
  \begin{scope}[shift={(4.6,0)}]
    \draw[pieceA,->,rounded corners=6pt] (-1,-1)--(-.26,0)--(-1,1);
    \draw[pieceC,->,rounded corners=6pt] (1,-1)--(.26,0)--(1,1);
  \end{scope}
  \node[font=\scriptsize,anchor=north] at (4.6,-1.25) {oriented smoothing};
\end{tikzpicture}
\end{center}
}
\newcommand{\vthreestatesumfigure}{%
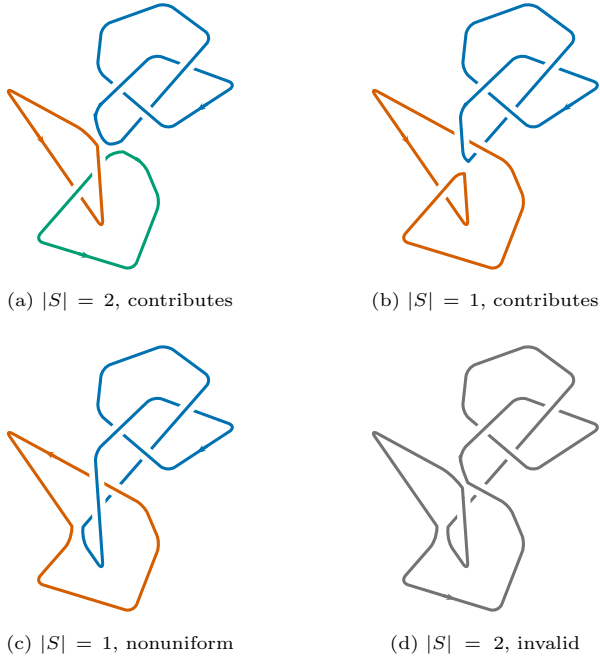
\begin{figure}[!ht]
\centering
\begin{tikzpicture}[scale=1.45]
  \path (-1.36,-1.42) rectangle (1.12,1.30);
  \vstatecoords
  \draw[pieceB,knotstrand]
    (-0.85290,-0.08288)
    -- (p1)
    -- (-0.33177,-0.22566)
    [sharp corners]
    .. controls (-0.33797,-0.13700) and (-0.42789,-0.01823) .. (-0.53159,0.03828)
    [rounded corners=\knotround]
    -- (p0)
    -- cycle;
  \draw[pieceA,knotstrand]
    (-0.25545,0.28535)
    -- (p3)
    -- (p4)
    -- (p5)
    -- (p6)
    -- (p7)
    -- (p8)
    -- (p9)
    -- (-0.06334,-0.03354)
    [sharp corners]
    .. controls (-0.12307,-0.10183) and (-0.20396,-0.14025) .. (-0.24308,-0.11893)
    -- (-0.24947,-0.11545)
    .. controls (-0.30093,-0.08741) and (-0.34967,0.03036) .. (-0.35777,0.14625)
    [rounded corners=\knotround]
    -- (p2)
    -- cycle;
  \draw[pieceC,knotstrand]
    (-0.45898,-1.13950)
    -- (p11)
    -- (p12)
    -- (p13)
    -- (-0.02710,-0.23662)
    [sharp corners]
    .. controls (-0.10676,-0.19321) and (-0.20888,-0.19992) .. (-0.25402,-0.25152)
    [rounded corners=\knotround]
    -- (p10)
    -- cycle;
  \draw[knotgap] (-0.49307,-0.52479)--(-0.55277,-0.59304);
  \draw[pieceC,line width=1.15pt,line cap=round] (-0.49307,-0.52479)--(-0.55277,-0.59304);
  \draw[knotgap] (-0.31743,-0.43090)--(-0.33126,-0.23292);
  \draw[pieceB,line width=1.15pt,line cap=round] (-0.31743,-0.43090)--(-0.33126,-0.23292);
  \draw[knotgap] (-0.17590,0.36153)--(-0.12500,0.41027);
  \draw[pieceA,line width=1.15pt,line cap=round] (-0.17590,0.36153)--(-0.12500,0.41027);
  \draw[knotgap] (0.51466,0.62720)--(0.44264,0.54487);
  \draw[pieceA,line width=1.15pt,line cap=round] (0.51466,0.62720)--(0.44264,0.54487);
  \draw[knotgap] (0.13213,0.14532)--(0.08742,0.18338);
  \draw[pieceA,line width=1.15pt,line cap=round] (0.13213,0.14532)--(0.08742,0.18338);
  \draw[pieceB,line width=1.15pt,-{Stealth[length=3.4pt]}] (-0.88629,-0.03470)--(-0.81950,-0.13105);
  \draw[pieceA,line width=1.15pt,-{Stealth[length=3.4pt]}] (0.66660,0.24287)--(0.56845,0.17874);
  \draw[pieceC,line width=1.15pt,-{Stealth[length=3.4pt]}] (-0.51507,-1.12245)--(-0.40289,-1.15655);
  \node[font=\scriptsize,anchor=north,align=center,text width=3.6cm] at (-0.12,-1.40) {(a) $|S|=2$, contributes};
\end{tikzpicture}
\hfill
\begin{tikzpicture}[scale=1.45]
  \path (-1.36,-1.42) rectangle (1.12,1.30);
  \vstatecoords
  \draw[pieceB,knotstrand]
    (-0.85290,-0.08288)
    -- (p1)
    -- (-0.31420,-0.47699)
    [sharp corners]
    .. controls (-0.31978,-0.39720) and (-0.36743,-0.38117) .. (-0.42010,-0.44137)
    [rounded corners=\knotround]
    -- (p10)
    -- (p11)
    -- (p12)
    -- (p13)
    -- (p0)
    -- cycle;
  \draw[pieceA,knotstrand]
    (-0.25545,0.28535)
    -- (p3)
    -- (p4)
    -- (p5)
    -- (p6)
    -- (p7)
    -- (p8)
    -- (p9)
    -- (-0.22859,-0.22245)
    [sharp corners]
    .. controls (-0.28126,-0.28265) and (-0.32891,-0.26663) .. (-0.33449,-0.18683)
    [rounded corners=\knotround]
    -- (p2)
    -- cycle;
  \draw[knotgap] (-0.49307,-0.52479)--(-0.55277,-0.59304);
  \draw[pieceB,line width=1.15pt,line cap=round] (-0.49307,-0.52479)--(-0.55277,-0.59304);
  \draw[knotgap] (-0.33795,-0.13720)--(-0.34812,0.00827);
  \draw[pieceA,line width=1.15pt,line cap=round] (-0.33795,-0.13720)--(-0.34812,0.00827);
  \draw[knotgap] (-0.17590,0.36153)--(-0.12500,0.41027);
  \draw[pieceA,line width=1.15pt,line cap=round] (-0.17590,0.36153)--(-0.12500,0.41027);
  \draw[knotgap] (0.51466,0.62720)--(0.44264,0.54487);
  \draw[pieceA,line width=1.15pt,line cap=round] (0.51466,0.62720)--(0.44264,0.54487);
  \draw[knotgap] (0.13213,0.14532)--(0.08742,0.18338);
  \draw[pieceA,line width=1.15pt,line cap=round] (0.13213,0.14532)--(0.08742,0.18338);
  \draw[knotgap] (-0.13380,-0.17848)--(-0.21009,-0.13691);
  \draw[pieceB,line width=1.15pt,line cap=round] (-0.13380,-0.17848)--(-0.21009,-0.13691);
  \draw[pieceB,line width=1.15pt,-{Stealth[length=3.4pt]}] (-0.88629,-0.03470)--(-0.81950,-0.13105);
  \draw[pieceA,line width=1.15pt,-{Stealth[length=3.4pt]}] (0.66660,0.24287)--(0.56845,0.17874);
  \node[font=\scriptsize,anchor=north,align=center,text width=3.6cm] at (-0.12,-1.40) {(b) $|S|=1$, contributes};
\end{tikzpicture}
\par\vspace{3pt}
\begin{tikzpicture}[scale=1.45]
  \path (-1.36,-1.42) rectangle (1.12,1.30);
  \vstatecoords
  \draw[pieceB,knotstrand]
    (-0.90067,-0.01396)
    -- (-0.61846,-0.42108)
    [sharp corners]
    .. controls (-0.56591,-0.49689) and (-0.57261,-0.61571) .. (-0.63334,-0.68514)
    [rounded corners=\knotround]
    -- (p10)
    -- (p11)
    -- (p12)
    -- (p13)
    -- (p0)
    -- cycle;
  \draw[pieceA,knotstrand]
    (-0.30439,-0.61744)
    -- (p2)
    -- (p3)
    -- (p4)
    -- (p5)
    -- (p6)
    -- (p7)
    -- (p8)
    -- (p9)
    -- (-0.41250,-0.43269)
    [sharp corners]
    .. controls (-0.47323,-0.50211) and (-0.47993,-0.62094) .. (-0.42738,-0.69674)
    [rounded corners=\knotround]
    -- (p1)
    -- cycle;
  \draw[knotgap] (-0.31743,-0.43090)--(-0.33126,-0.23292);
  \draw[pieceA,line width=1.15pt,line cap=round] (-0.31743,-0.43090)--(-0.33126,-0.23292);
  \draw[knotgap] (-0.33795,-0.13720)--(-0.34812,0.00827);
  \draw[pieceA,line width=1.15pt,line cap=round] (-0.33795,-0.13720)--(-0.34812,0.00827);
  \draw[knotgap] (-0.17590,0.36153)--(-0.12500,0.41027);
  \draw[pieceA,line width=1.15pt,line cap=round] (-0.17590,0.36153)--(-0.12500,0.41027);
  \draw[knotgap] (0.51466,0.62720)--(0.44264,0.54487);
  \draw[pieceA,line width=1.15pt,line cap=round] (0.51466,0.62720)--(0.44264,0.54487);
  \draw[knotgap] (0.13213,0.14532)--(0.08742,0.18338);
  \draw[pieceA,line width=1.15pt,line cap=round] (0.13213,0.14532)--(0.08742,0.18338);
  \draw[knotgap] (-0.13380,-0.17848)--(-0.21009,-0.13691);
  \draw[pieceB,line width=1.15pt,line cap=round] (-0.13380,-0.17848)--(-0.21009,-0.13691);
  \draw[pieceB,line width=1.15pt,-{Stealth[length=3.4pt]}] (-0.71148,0.13630)--(-0.81443,0.19240);
  \draw[pieceA,line width=1.15pt,-{Stealth[length=3.4pt]}] (0.66660,0.24287)--(0.56845,0.17874);
  \node[font=\scriptsize,anchor=north,align=center,text width=3.6cm] at (-0.12,-1.40) {(c) $|S|=1$, nonuniform};
\end{tikzpicture}
\hfill
\begin{tikzpicture}[scale=1.45]
  \path (-1.36,-1.42) rectangle (1.12,1.30);
  \vstatecoords
  \draw[black!55,knotstrand]
    (-0.90067,-0.01396)
    -- (-0.61846,-0.42108)
    [sharp corners]
    .. controls (-0.56591,-0.49689) and (-0.57261,-0.61571) .. (-0.63334,-0.68514)
    [rounded corners=\knotround]
    -- (p10)
    -- (p11)
    -- (p12)
    -- (p13)
    -- (-0.21648,-0.13343)
    [sharp corners]
    .. controls (-0.28609,-0.09550) and (-0.34967,0.03036) .. (-0.35777,0.14625)
    [rounded corners=\knotround]
    -- (p2)
    -- (p3)
    -- (p4)
    -- (p5)
    -- (p6)
    -- (p7)
    -- (p8)
    -- (p9)
    -- (-0.41250,-0.43269)
    [sharp corners]
    .. controls (-0.47323,-0.50211) and (-0.47993,-0.62094) .. (-0.42738,-0.69674)
    [rounded corners=\knotround]
    -- (p1)
    -- (-0.33177,-0.22566)
    [sharp corners]
    .. controls (-0.33797,-0.13700) and (-0.42789,-0.01823) .. (-0.53159,0.03828)
    [rounded corners=\knotround]
    -- (p0)
    -- cycle;
  \draw[knotgap] (-0.31743,-0.43090)--(-0.33126,-0.23292);
  \draw[black!55,line width=1.15pt,line cap=round] (-0.31743,-0.43090)--(-0.33126,-0.23292);
  \draw[knotgap] (-0.17590,0.36153)--(-0.12500,0.41027);
  \draw[black!55,line width=1.15pt,line cap=round] (-0.17590,0.36153)--(-0.12500,0.41027);
  \draw[knotgap] (0.51466,0.62720)--(0.44264,0.54487);
  \draw[black!55,line width=1.15pt,line cap=round] (0.51466,0.62720)--(0.44264,0.54487);
  \draw[knotgap] (0.13213,0.14532)--(0.08742,0.18338);
  \draw[black!55,line width=1.15pt,line cap=round] (0.13213,0.14532)--(0.08742,0.18338);
  \draw[knotgap] (-0.13380,-0.17848)--(-0.21009,-0.13691);
  \draw[black!55,line width=1.15pt,line cap=round] (-0.13380,-0.17848)--(-0.21009,-0.13691);
  \draw[black!55,line width=1.15pt,-{Stealth[length=3.4pt]}] (-0.51507,-1.12245)--(-0.40289,-1.15655);
  \node[font=\scriptsize,anchor=north,align=center,text width=3.6cm] at (-0.12,-1.40) {(d) $|S|=2$, invalid};
\end{tikzpicture}
\caption{\label{fig:decomp}Four smoothing choices for $P$.}
\end{figure}
}

\newcommand{\vthreecornerfigure}{%
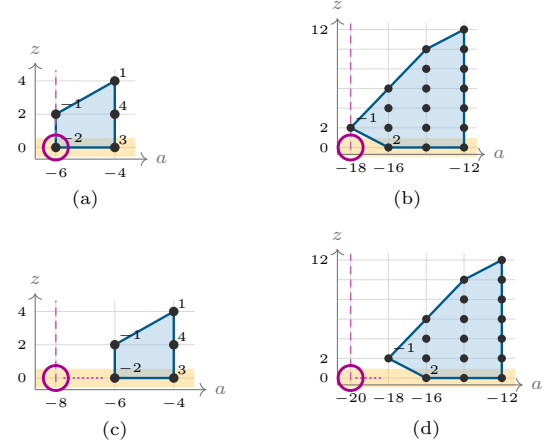
\begin{figure}[!ht]
\centering
\begin{tikzpicture}[x=1cm,y=1cm]
  \begin{scope}[shift={(.78cm,0)},x=.78cm,y=.44cm]
    \draw[help lines,black!15] (-.35,-.35) grid[xstep=1,ystep=1] (1.35,2.35);
    \draw[rowhi,line width=7pt,opacity=.35] (-.35,0) -- (1.35,0);
    \fill[beforecol,opacity=.42] (0,0)--(1,0)--(1,2)--(0,1)--cycle;
    \draw[pieceA!78!black,line width=.9pt,line join=round]
      (0,0)--(1,0)--(1,2)--(0,1)--cycle;
    \draw[cornermark!65,densely dashed,line width=.55pt] (0,-.12)--(0,2.32);
    \draw[->,black!55] (-.35,-.35)--(1.55,-.35) node[right,font=\scriptsize] {$a$};
    \draw[->,black!55] (-.35,-.35)--(-.35,2.55) node[above,font=\scriptsize] {$z$};
    \foreach \x/\y/\c in {0/0/-2,0/1/-1,1/0/3,1/1/4,1/2/1}{
      \fill[black!82] (\x,\y) circle (1.8pt);
      \node[font=\tiny,anchor=south west,inner sep=1.2pt] at (\x,\y) {$\c$};}
    \draw[cornermark,line width=1.1pt] (0,0) circle (4.7pt);
    \node[font=\tiny,anchor=north] at (0,-.35) {$-6$};
    \node[font=\tiny,anchor=north] at (1,-.35) {$-4$};
    \node[font=\tiny,anchor=east] at (-.35,0) {$0$};
    \node[font=\tiny,anchor=east] at (-.35,1) {$2$};
    \node[font=\tiny,anchor=east] at (-.35,2) {$4$};
    \node[font=\scriptsize,anchor=north,align=center] at (.5,-1.02)
      {(a)};
  \end{scope}

  \begin{scope}[shift={(4.68cm,0)},x=.50cm,y=.26cm]
    \draw[help lines,black!15] (-.35,-.35) grid[xstep=1,ystep=1] (3.35,6.35);
    \draw[rowhi,line width=7pt,opacity=.35] (-.35,0)--(3.35,0);
    \fill[beforecol,opacity=.42] (0,1)--(1,0)--(3,0)--(3,6)--(2,5)--cycle;
    \draw[pieceA!78!black,line width=.9pt,line join=round]
      (0,1)--(1,0)--(3,0)--(3,6)--(2,5)--cycle;
    \draw[cornermark!65,densely dashed,line width=.55pt] (0,-.12)--(0,6.30);
    \draw[->,black!55] (-.35,-.35)--(3.55,-.35) node[right,font=\scriptsize] {$a$};
    \draw[->,black!55] (-.35,-.35)--(-.35,6.58) node[above,font=\scriptsize] {$z$};
    \foreach \x/\y in {0/1,1/0,1/1,1/2,1/3,
                         2/0,2/1,2/2,2/3,2/4,2/5,
                         3/0,3/1,3/2,3/3,3/4,3/5,3/6}
      \fill[black!82] (\x,\y) circle (1.55pt);
    \node[font=\tiny,anchor=south west,inner sep=1.2pt] at (0,1) {$-1$};
    \node[font=\tiny,anchor=south west,inner sep=1.2pt] at (1,0) {$2$};
    \draw[cornermark,line width=1.1pt] (0,0) circle (4.7pt);
    \node[font=\tiny,anchor=north] at (0,-.35) {$-18$};
    \node[font=\tiny,anchor=north] at (1,-.35) {$-16$};
    \node[font=\tiny,anchor=north] at (3,-.35) {$-12$};
    \node[font=\tiny,anchor=east] at (-.35,0) {$0$};
    \node[font=\tiny,anchor=east] at (-.35,1) {$2$};
    \node[font=\tiny,anchor=east] at (-.35,6) {$12$};
    \node[font=\scriptsize,anchor=north,align=center] at (1.5,-1.72)
      {(b)};
  \end{scope}

  \begin{scope}[shift={(.78cm,-3.05cm)},x=.78cm,y=.44cm]
    \draw[help lines,black!15] (-.35,-.35) grid[xstep=1,ystep=1] (2.35,2.35);
    \draw[rowhi,line width=7pt,opacity=.35] (-.35,0)--(2.35,0);
    \fill[beforecol,opacity=.42] (1,0)--(2,0)--(2,2)--(1,1)--cycle;
    \draw[pieceA!78!black,line width=.9pt,line join=round]
      (1,0)--(2,0)--(2,2)--(1,1)--cycle;
    \draw[cornermark!65,densely dashed,line width=.55pt] (0,-.12)--(0,2.32);
    \draw[cornermark!70,densely dotted,line width=.75pt] (.13,0)--(.82,0);
    \draw[->,black!55] (-.35,-.35)--(2.55,-.35) node[right,font=\scriptsize] {$a$};
    \draw[->,black!55] (-.35,-.35)--(-.35,2.55) node[above,font=\scriptsize] {$z$};
    \foreach \x/\y/\c in {1/0/-2,1/1/-1,2/0/3,2/1/4,2/2/1}{
      \fill[black!82] (\x,\y) circle (1.8pt);
      \node[font=\tiny,anchor=south west,inner sep=1.2pt] at (\x,\y) {$\c$};}
    \draw[cornermark,line width=1.1pt] (0,0) circle (4.7pt);
    \node[font=\tiny,anchor=north] at (0,-.35) {$-8$};
    \node[font=\tiny,anchor=north] at (1,-.35) {$-6$};
    \node[font=\tiny,anchor=north] at (2,-.35) {$-4$};
    \node[font=\tiny,anchor=east] at (-.35,0) {$0$};
    \node[font=\tiny,anchor=east] at (-.35,1) {$2$};
    \node[font=\tiny,anchor=east] at (-.35,2) {$4$};
    \node[font=\scriptsize,anchor=north,align=center] at (1,-1.02)
      {(c)};
  \end{scope}

  \begin{scope}[shift={(4.68cm,-3.05cm)},x=.50cm,y=.26cm]
    \draw[help lines,black!15] (-.35,-.35) grid[xstep=1,ystep=1] (4.35,6.35);
    \draw[rowhi,line width=7pt,opacity=.35] (-.35,0)--(4.35,0);
    \fill[beforecol,opacity=.42] (1,1)--(2,0)--(4,0)--(4,6)--(3,5)--cycle;
    \draw[pieceA!78!black,line width=.9pt,line join=round]
      (1,1)--(2,0)--(4,0)--(4,6)--(3,5)--cycle;
    \draw[cornermark!65,densely dashed,line width=.55pt] (0,-.12)--(0,6.30);
    \draw[cornermark!70,densely dotted,line width=.75pt] (.13,0)--(.82,0);
    \draw[->,black!55] (-.35,-.35)--(4.55,-.35) node[right,font=\scriptsize] {$a$};
    \draw[->,black!55] (-.35,-.35)--(-.35,6.58) node[above,font=\scriptsize] {$z$};
    \foreach \x/\y in {1/1,2/0,2/1,2/2,2/3,
                         3/0,3/1,3/2,3/3,3/4,3/5,
                         4/0,4/1,4/2,4/3,4/4,4/5,4/6}
      \fill[black!82] (\x,\y) circle (1.55pt);
    \node[font=\tiny,anchor=south west,inner sep=1.2pt] at (1,1) {$-1$};
    \node[font=\tiny,anchor=south west,inner sep=1.2pt] at (2,0) {$2$};
    \draw[cornermark,line width=1.1pt] (0,0) circle (4.7pt);
    \node[font=\tiny,anchor=north] at (0,-.35) {$-20$};
    \node[font=\tiny,anchor=north] at (1,-.35) {$-18$};
    \node[font=\tiny,anchor=north] at (2,-.35) {$-16$};
    \node[font=\tiny,anchor=north] at (4,-.35) {$-12$};
    \node[font=\tiny,anchor=east] at (-.35,0) {$0$};
    \node[font=\tiny,anchor=east] at (-.35,1) {$2$};
    \node[font=\tiny,anchor=east] at (-.35,6) {$12$};
    \node[font=\scriptsize,anchor=north,align=center] at (2,-1.72)
      {(d)};
  \end{scope}
\end{tikzpicture}
\caption{\label{fig:slot}The four sharpness patterns. Dots are
monomials and the magenta circle marks $(d_Q,0)$.}
\end{figure}
}

%% file: figures_v9_transverse_curl.tex
\begin{center}
\begin{tikzpicture}[x=1cm,y=1cm]
  \tikzset{
    curlorient/.style={pieceA,line width=.85pt,-{Stealth[length=2.8pt]}},
    curlmark/.style={black!60,line width=.6pt,-{Stealth[length=2.6pt]}}
  }
  \path[use as bounding box] (-.95,-1.12) rectangle (4.20,1.12);

  \begin{scope}
    \draw[pieceA,knotstrand] (.55,.95)--(0,0)--(.55,-.95);
    \draw[curlorient] (.44,.76)--(.30,.52);
    \draw[curlmark,densely dashed] (-.07,.06)--(-.07,-.36);
  \end{scope}

  \draw[->,black!70,line width=.6pt] (1.25,0)--(2.05,0);

  \begin{scope}[shift={(3.10,0)}]
    \draw[pieceA,line width=1.15pt,line cap=round,line join=round] (.55,.95)--(0,0)
      arc[start angle=330,end angle=30,radius=.28]
      --(.55,-.67);
    \draw[knotgap] (.141,.244)--(.021,.036);
    \draw[pieceA,line width=1.15pt,line cap=round] (.141,.244)--(.021,.036);
    \draw[curlorient] (.44,.76)--(.30,.52);
    \draw[curlorient] (-.5225,.08)--(-.5225,.20);
    \draw[curlmark,densely dashed] (-.66,-.07)--(-.66,.35);
  \end{scope}
\end{tikzpicture}
\end{center}

%% file: figures_v9_s7_skein.tex
\begin{center}
\begin{tikzpicture}[x=.92cm,y=.72cm]
  \tikzset{
    ssevenknot/.style={pieceA,knotstrand},
    ssevenhalf/.style={knotstrand},
    ssevenrii/.style={movearrow,line width=.75pt,-{Stealth[length=3pt]}},
    ssevenorient/.style={line width=.75pt,-{Stealth[length=2.5pt]}}
  }
  \path[use as bounding box] (-.12,-2.38) rectangle (7.42,1.02);

  \begin{scope}[shift={(.78,.03)}]
    \draw[ssevenknot] (-.72,0)--(.72,0);
    \draw[ssevenknot] (-.52,.62)--(0,-.60)--(.52,.62);
    \draw[knotgap] (-.30,.11)--(-.21,-.11);
    \draw[ssevenknot] (-.30,.11)--(-.21,-.11);
    \draw[knotgap] (.17,0)--(.34,0);
    \draw[ssevenknot] (.17,0)--(.34,0);
    \draw[pieceA,ssevenorient] (-.66,0)--(-.47,0);
    \draw[pieceA,ssevenorient] (.36,.25)--(.45,.46);
    \draw[curlmark,line width=.68pt] (-.256,0) circle (2.25pt);
  \end{scope}
  \node[font=\footnotesize] at (.78,.86) {$L_+=Q_+$};

  \begin{scope}[shift={(4.05,.03)}]
    \draw[ssevenknot] (-.72,0)--(.72,0);
    \draw[ssevenknot] (-.52,.62)--(0,-.60)--(.52,.62);
    \draw[knotgap] (-.34,0)--(-.17,0);
    \draw[ssevenknot] (-.34,0)--(-.17,0);
    \draw[knotgap] (.17,0)--(.34,0);
    \draw[ssevenknot] (.17,0)--(.34,0);
  \end{scope}
  \node[font=\footnotesize] at (4.05,.86) {$L_-$};

  \begin{scope}[shift={(6.62,-1.32)}]
    \draw[ssevenknot] (-.72,0)--(.72,0);
    \draw[ssevenknot] (-.52,.62)--(0,.20)--(.52,.62);
  \end{scope}
  \node[font=\footnotesize] at (6.62,-2.22) {$Q_-$};

  \draw[ssevenrii] (4.88,-.24)--(5.76,-.96)
    node[midway,above,font=\scriptsize,text=movearrow] {RII};

  \begin{scope}[shift={(2.42,-1.32)}]
    \draw[pieceA,ssevenhalf]
      (-.72,0)--(-.44,0)
      .. controls (-.33,0) and (-.24,-.09) .. (-.17,-.21)
      --(0,-.60)--(.52,.62);
    \draw[pieceB,ssevenhalf]
      (-.52,.62)--(-.34,.21)
      .. controls (-.29,.08) and (-.20,0) .. (-.06,0)
      --(.72,0);
    \draw[knotgap] (.17,0)--(.34,0);
    \draw[pieceB,ssevenhalf] (.17,0)--(.34,0);
    \draw[pieceA,ssevenorient] (-.66,0)--(-.47,0);
    \draw[pieceB,ssevenorient] (.40,0)--(.59,0);
  \end{scope}
  \node[font=\footnotesize] at (2.42,-2.22)
    {$L_0=Q_L\cup Q_R$};
\end{tikzpicture}
\end{center}

%% file: figures_v9_soft_state_sum.tex
\begin{center}
\begin{tikzpicture}[x=.92cm,y=.72cm]
  \tikzset{
    softknot/.style={pieceA,knotstrand},
    softmove/.style={movearrow,line width=.75pt,-{Stealth[length=3pt]}},
    softorient/.style={line width=.75pt,-{Stealth[length=2.5pt]}}
  }
  \path[use as bounding box] (-.08,-.78) rectangle (7.42,1.12);

  \begin{scope}[shift={(1.28,0)}]
    \coordinate (p0) at (-1.12,-.56);
    \coordinate (p1) at (-.72,0);
    \coordinate (p2) at (.27,0);
    \coordinate (p3) at (-.20,-.42);
    \coordinate (p4) at (.39,.83);
    \coordinate (p5) at (1.17,.63);
    \draw[softknot] (p0)--(p1)--(p2)--(p3)--(p4)--(p5);
    \draw[knotgap] (-.13,0)--(.13,0);
    \draw[softknot] (-.13,0)--(.13,0);
    \draw[pieceA,softorient] (-1.02,-.42)--(-.85,-.18);
    \draw[pieceA,softorient] (.58,.78)--(.84,.71);
    \draw[curlmark,line width=.68pt] (0,0) circle (2.35pt);
  \end{scope}
  \node[font=\footnotesize] at (1.28,1.00) {$Q$};

  \draw[softmove] (2.72,.10)--(4.66,.10)
    node[midway,above,font=\scriptsize,text=movearrow] {RI};

  \begin{scope}[shift={(5.96,0)}]
    \coordinate (p0) at (-1.12,-.56);
    \coordinate (p1) at (-.72,0);
    \coordinate (p2) at (.10,.23);
    \coordinate (p3) at (.39,.83);
    \coordinate (p4) at (1.17,.63);
    \draw[softknot] (p0)--(p1)--(p2)--(p3)--(p4);
    \draw[pieceA,softorient] (-1.02,-.42)--(-.85,-.18);
    \draw[pieceA,softorient] (.58,.78)--(.84,.71);
  \end{scope}
\end{tikzpicture}
\end{center}

%% file: figures_v9_r3_transitive.tex
\begin{center}
\begin{tikzpicture}[x=1.08cm,y=.72cm]
  \tikzset{
    rstrand/.style={pieceA,knotstrand},
    rover/.style={line width=1.15pt,line cap=butt},
    rorient/.style={line width=.78pt,-{Stealth[length=2.7pt]}},
    rflow/.style={movearrow,line width=.76pt,-{Stealth[length=3.1pt]}}
  }
  \path[use as bounding box] (-1.02,-1.28) rectangle (4.82,1.28);

  \begin{scope}
    \fill[pieceA!8] (-.40,.80)--(.40,0)--(-.40,-.80)--cycle;
    \draw[rstrand]
      (-.80,1.20)--(0,.40)--(.80,-.40)--(.80,-1.20);
    \draw[rstrand]
      (0,1.20)--(-.80,.40)--(-.80,-.40)--(0,-1.20);
    \draw[rstrand]
      (.80,1.20)--(.80,.40)--(0,-.40)--(-.80,-1.20);

    \draw[knotgap]
      (-.31,.89)--(-.49,.71);
    \draw[rover,pieceA] (-.31,.89)--(-.49,.71);
    \draw[knotgap]
      (.49,.09)--(.31,-.09);
    \draw[rover,pieceA] (.49,.09)--(.31,-.09);
    \draw[knotgap]
      (-.31,-.71)--(-.49,-.89);
    \draw[rover,pieceA] (-.31,-.71)--(-.49,-.89);

    \draw[rorient,pieceA] (.80,-.69)--(.80,-.99);
    \draw[rorient,pieceA] (-.80,.18)--(-.80,-.12);
    \draw[rorient,pieceA] (.80,1.08)--(.80,.78);
  \end{scope}

  \draw[rflow] (1.13,0)--(2.67,0)
    node[midway,above,font=\scriptsize,text=movearrow] {RIII};

  \begin{scope}[shift={(3.80,0)}]
    \fill[pieceA!8] (.40,.80)--(-.40,0)--(.40,-.80)--cycle;
    \draw[rstrand]
      (-.80,1.20)--(-.80,.40)--(0,-.40)--(.80,-1.20);
    \draw[rstrand]
      (0,1.20)--(.80,.40)--(.80,-.40)--(0,-1.20);
    \draw[rstrand]
      (.80,1.20)--(0,.40)--(-.80,-.40)--(-.80,-1.20);

    \draw[knotgap]
      (.49,.89)--(.31,.71);
    \draw[rover,pieceA] (.49,.89)--(.31,.71);
    \draw[knotgap]
      (-.31,.09)--(-.49,-.09);
    \draw[rover,pieceA] (-.31,.09)--(-.49,-.09);
    \draw[knotgap]
      (.49,-.71)--(.31,-.89);
    \draw[rover,pieceA] (.49,-.71)--(.31,-.89);

    \draw[rorient,pieceA] (-.80,1.08)--(-.80,.78);
    \draw[rorient,pieceA] (.80,.18)--(.80,-.12);
    \draw[rorient,pieceA] (-.80,-.69)--(-.80,-.99);
  \end{scope}
\end{tikzpicture}
\end{center}

%% file: figures_v9_anchors.tex
\begin{figure}[!htbp]
\centering
\begin{tikzpicture}
  \path (-0.98,-1.20) rectangle (0.98,1.20);
  \draw[pieceA,knotstrand] (0.745,0.000) -- (-0.745,0.860) -- (-0.745,-0.860) -- cycle;
  \draw[pieceA,line width=0.9pt,-{Stealth[length=3.4pt]}] (0.113,0.365) -- (-0.113,0.495);
  \node[font=\footnotesize,anchor=north] at (0,-1.20) {$K_1$: $A=-1$};
\end{tikzpicture}
\hfill
\begin{tikzpicture}
  \path (-0.98,-1.20) rectangle (0.98,1.20);
  \draw[pieceA,knotstrand] (-0.860,-0.860) -- (0.000,0.000) -- (0.860,0.860) -- (0.860,-0.860) -- (0.000,0.000) -- (-0.860,0.860) -- cycle;
  \draw[pieceA,line width=0.9pt,-{Stealth[length=3.4pt]}] (0.860,0.130) -- (0.860,-0.130);
  \draw[knotgap] (-0.074,-0.074) -- (0.000,0.000) -- (0.074,0.074);
  \draw[pieceA,line width=1.15pt,line cap=round] (-0.074,-0.074) -- (0.000,0.000) -- (0.074,0.074);
  \node[font=\footnotesize,anchor=north] at (0,-1.20) {$K_0$: $A=-1$};
\end{tikzpicture}
\hfill
\begin{tikzpicture}
  \path (-0.98,-1.20) rectangle (0.98,1.20);
  \draw[pieceA,knotstrand] (0.818,0.000) -- (0.193,0.203) -- (-0.193,0.329) -- (-0.818,0.532) -- (-0.432,0.000) -- (-0.193,-0.329) -- (0.193,-0.860) -- (0.193,-0.203) -- (0.193,0.203) -- (0.193,0.860) -- (-0.193,0.329) -- (-0.432,0.000) -- (-0.818,-0.532) -- (-0.193,-0.329) -- (0.193,-0.203) -- cycle;
  \draw[pieceA,line width=0.9pt,-{Stealth[length=3.4pt]}] (-0.701,0.371) -- (-0.548,0.161);
  \draw[knotgap] (-0.093,0.296) -- (-0.193,0.329) -- (-0.293,0.361);
  \draw[pieceA,line width=1.15pt,line cap=round] (-0.093,0.296) -- (-0.193,0.329) -- (-0.293,0.361);
  \draw[knotgap] (-0.255,-0.244) -- (-0.193,-0.329) -- (-0.131,-0.414);
  \draw[pieceA,line width=1.15pt,line cap=round] (-0.255,-0.244) -- (-0.193,-0.329) -- (-0.131,-0.414);
  \draw[knotgap] (0.193,0.098) -- (0.193,0.203) -- (0.193,0.308);
  \draw[pieceA,line width=1.15pt,line cap=round] (0.193,0.098) -- (0.193,0.203) -- (0.193,0.308);
  \draw[knotgap] (-0.370,0.085) -- (-0.432,0.000) -- (-0.493,-0.085);
  \draw[pieceA,line width=1.15pt,line cap=round] (-0.370,0.085) -- (-0.432,0.000) -- (-0.493,-0.085);
  \draw[knotgap] (0.093,-0.236) -- (0.193,-0.203) -- (0.293,-0.171);
  \draw[pieceA,line width=1.15pt,line cap=round] (0.093,-0.236) -- (0.193,-0.203) -- (0.293,-0.171);
  \node[font=\footnotesize,anchor=north] at (0,-1.20) {$K_2$: $A=+2$};
\end{tikzpicture}
\hfill
\begin{tikzpicture}
  \path (-1.12,-1.20) rectangle (1.12,1.20);
  \begin{scope}[scale=1.15]
  \draw[pieceA,knotstrand] (0.838,0.000) -- (0.240,0.137) -- (0.092,0.170) -- (-0.092,0.212) -- (-0.240,0.246) -- (-0.838,0.383) -- (-0.359,0.000) -- (-0.240,-0.094) -- (-0.092,-0.212) -- (0.026,-0.307) -- (0.506,-0.690) -- (0.240,-0.137) -- (0.174,0.000) -- (0.092,0.170) -- (0.026,0.307) -- (-0.240,0.860) -- (-0.240,0.246) -- (-0.240,0.094) -- (-0.240,-0.094) -- (-0.240,-0.246) -- (-0.240,-0.860) -- (0.026,-0.307) -- (0.092,-0.170) -- (0.174,0.000) -- (0.240,0.137) -- (0.506,0.690) -- (0.026,0.307) -- (-0.092,0.212) -- (-0.240,0.094) -- (-0.359,0.000) -- (-0.838,-0.383) -- (-0.240,-0.246) -- (-0.092,-0.212) -- (0.092,-0.170) -- (0.240,-0.137) -- cycle;
  \draw[pieceA,line width=0.9pt,-{Stealth[length=3.4pt]}] (0.666,0.039) -- (0.412,0.097);
  \draw[knotgap] (-0.019,0.196) -- (-0.092,0.212) -- (-0.152,0.226);
  \draw[pieceA,line width=1.15pt,line cap=round] (-0.019,0.196) -- (-0.092,0.212) -- (-0.152,0.226);
  \draw[knotgap] (-0.181,0.233) -- (-0.240,0.246) -- (-0.343,0.270);
  \draw[pieceA,line width=1.15pt,line cap=round] (-0.181,0.233) -- (-0.240,0.246) -- (-0.343,0.270);
  \draw[knotgap] (-0.152,-0.165) -- (-0.092,-0.212) -- (-0.045,-0.250);
  \draw[pieceA,line width=1.15pt,line cap=round] (-0.152,-0.165) -- (-0.092,-0.212) -- (-0.045,-0.250);
  \draw[knotgap] (-0.021,-0.269) -- (0.026,-0.307) -- (0.108,-0.372);
  \draw[pieceA,line width=1.15pt,line cap=round] (-0.021,-0.269) -- (0.026,-0.307) -- (0.108,-0.372);
  \draw[knotgap] (0.124,0.102) -- (0.092,0.170) -- (0.065,0.225);
  \draw[pieceA,line width=1.15pt,line cap=round] (0.124,0.102) -- (0.092,0.170) -- (0.065,0.225);
  \draw[knotgap] (0.052,0.252) -- (0.026,0.307) -- (-0.020,0.401);
  \draw[pieceA,line width=1.15pt,line cap=round] (0.052,0.252) -- (0.026,0.307) -- (-0.020,0.401);
  \draw[knotgap] (-0.240,-0.019) -- (-0.240,-0.094) -- (-0.240,-0.155);
  \draw[pieceA,line width=1.15pt,line cap=round] (-0.240,-0.019) -- (-0.240,-0.094) -- (-0.240,-0.155);
  \draw[knotgap] (-0.240,-0.186) -- (-0.240,-0.246) -- (-0.240,-0.351);
  \draw[pieceA,line width=1.15pt,line cap=round] (-0.240,-0.186) -- (-0.240,-0.246) -- (-0.240,-0.351);
  \draw[knotgap] (0.141,-0.068) -- (0.174,0.000) -- (0.200,0.055);
  \draw[pieceA,line width=1.15pt,line cap=round] (0.141,-0.068) -- (0.174,0.000) -- (0.200,0.055);
  \draw[knotgap] (0.213,0.082) -- (0.240,0.137) -- (0.285,0.231);
  \draw[pieceA,line width=1.15pt,line cap=round] (0.213,0.082) -- (0.240,0.137) -- (0.285,0.231);
  \draw[knotgap] (-0.181,0.142) -- (-0.240,0.094) -- (-0.288,0.057);
  \draw[pieceA,line width=1.15pt,line cap=round] (-0.181,0.142) -- (-0.240,0.094) -- (-0.288,0.057);
  \draw[knotgap] (-0.311,0.038) -- (-0.359,0.000) -- (-0.441,-0.065);
  \draw[pieceA,line width=1.15pt,line cap=round] (-0.311,0.038) -- (-0.359,0.000) -- (-0.441,-0.065);
  \draw[knotgap] (0.018,-0.187) -- (0.092,-0.170) -- (0.151,-0.157);
  \draw[pieceA,line width=1.15pt,line cap=round] (0.018,-0.187) -- (0.092,-0.170) -- (0.151,-0.157);
  \draw[knotgap] (0.180,-0.150) -- (0.240,-0.137) -- (0.342,-0.113);
  \draw[pieceA,line width=1.15pt,line cap=round] (0.180,-0.150) -- (0.240,-0.137) -- (0.342,-0.113);
  \end{scope}
  \node[font=\footnotesize,anchor=north] at (0,-1.20) {$K_3$: $A=-5$};
\end{tikzpicture}
\caption{\label{fig:anchors}The minimal anchors $K_1,K_0,K_2,K_3$, with
every crossing drawn using the positive lift. These are torus knots whose corner is the amplitude.}
\end{figure}
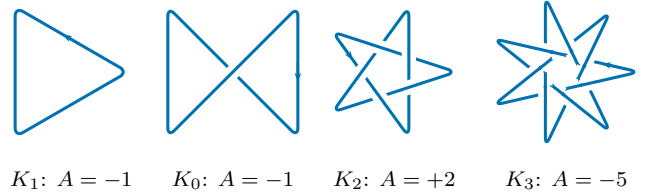

%% file: figures_v9_decay.tex
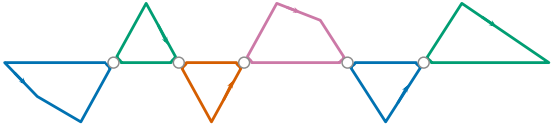
\begin{figure}[!ht]
\centering
\begin{tikzpicture}[x=.72cm,y=.56cm]
  \path[use as bounding box] (-5.18,-1.62) rectangle (5.18,1.62);

  \draw[pieceA,line width=1.05pt,line join=round]
    (-3.14,0)--(-5,0)--(-4.4,-.8)--(-3.6,-1.4)--(-3.06,-.14)
    .. controls (-3.09,-.075) and (-3.14,-.045) .. (-3.14,0);
  \draw[pieceC,line width=1.05pt,line join=round]
    (-2.86,0)
    .. controls (-2.86,.045) and (-2.91,.075) .. (-2.94,.14)
    --(-2.4,1.4)--(-1.86,.14)
    .. controls (-1.89,.075) and (-1.94,.045) .. (-1.94,0)
    --cycle;
  \draw[pieceB,line width=1.05pt,line join=round]
    (-1.66,0)
    .. controls (-1.66,-.045) and (-1.71,-.075) .. (-1.74,-.14)
    --(-1.2,-1.4)--(-.66,-.14)
    .. controls (-.69,-.075) and (-.74,-.045) .. (-.74,0)
    --cycle;
  \draw[pieceD,line width=1.05pt,line join=round]
    (-.46,0)
    .. controls (-.46,.045) and (-.51,.075) .. (-.54,.14)
    --(0,1.4)--(.8,1.0)--(1.24,.12)
    .. controls (1.21,.065) and (1.16,.040) .. (1.16,0)
    --cycle;
  \draw[pieceA,line width=1.05pt,line join=round]
    (1.44,0)
    .. controls (1.44,-.040) and (1.39,-.065) .. (1.36,-.12)
    --(2,-1.4)--(2.64,-.12)
    .. controls (2.61,-.065) and (2.56,-.040) .. (2.56,0)
    --cycle;
  \draw[pieceC,line width=1.05pt,line join=round]
    (5,0)--(2.84,0)
    .. controls (2.84,.040) and (2.79,.065) .. (2.76,.12)
    --(3.4,1.4)--cycle;

  \draw[pieceA,line width=1.05pt,-{Stealth[length=2.9pt]}]
    (-4.82,-.24)--(-4.60,-.53);
  \draw[pieceC,line width=1.05pt,-{Stealth[length=2.9pt]}]
    (-2.18,.89)--(-2.00,.42);
  \draw[pieceB,line width=1.05pt,-{Stealth[length=2.9pt]}]
    (-.98,-.89)--(-.80,-.42);
  \draw[pieceD,line width=1.05pt,-{Stealth[length=2.9pt]}]
    (.18,1.31)--(.44,1.18);
  \draw[pieceA,line width=1.05pt,-{Stealth[length=2.9pt]}]
    (2.22,-.96)--(2.42,-.52);
  \draw[pieceC,line width=1.05pt,-{Stealth[length=2.9pt]}]
    (3.77,1.08)--(4.04,.84);

  \foreach \x in {-3,-1.8,-.6,1.3,2.7}
    \draw[draw=black!45,fill=white,line width=.55pt] (\x,0) circle (2.05pt);
\end{tikzpicture}
\caption{\label{fig:decay}The maximal decomposition $S_{\max}$ in the
decay region $\mathcal R_1$: $k+1$ embedded subpolygons.}
\end{figure}

%% file: figures_v9_proof_outline.tex
\definecolor{vthreeblue}{RGB}{0,114,178}
\definecolor{vthreeorange}{RGB}{213,94,0}
\definecolor{vthreegreen}{RGB}{0,140,105}
\definecolor{vthreepurple}{RGB}{176,60,145}
\definecolor{vthreeink}{RGB}{38,45,52}
\definecolor{vthreelightblue}{RGB}{232,243,250}
\definecolor{vthreelightgreen}{RGB}{232,247,241}
\definecolor{vthreelightorange}{RGB}{252,239,230}

\def\vthreePcoords{(0.9045,-0.6245) (-0.1378,-0.7714) (-0.9045,-0.0365)
  (-0.2716,-1.1961) (0.2135,-1.5755) (0.5135,1.5755)
  (0.2135,0.3265) (-0.9045,-1.2125)}
\def\vthreePverts{0.9045/-0.6245,-0.1378/-0.7714,-0.9045/-0.0365,
  -0.2716/-1.1961,0.2135/-1.5755,0.5135/1.5755,
  0.2135/0.3265,-0.9045/-1.2125}
\def\vthreeFlatcoords{(0.9045,-0.6245) (0.0083,-0.3332) (-0.9045,-0.0365)
  (-0.2716,-1.1961) (0.2135,-1.5755) (0.5135,1.5755)
  (0.2135,0.3265) (-0.9045,-1.2125)}
\def\vthreeFlatverts{0.9045/-0.6245,0.0083/-0.3332,-0.9045/-0.0365,
  -0.2716/-1.1961,0.2135/-1.5755,0.5135/1.5755,
  0.2135/0.3265,-0.9045/-1.2125}
\def\vthreeMidcoords{(0.9045,-0.6245) (0.0245,-0.2845) (-0.9045,-0.0365)
  (-0.2716,-1.1961) (0.2135,-1.5755) (0.5135,1.5755)
  (0.2135,0.3265) (-0.9045,-1.2125)}
\def\vthreeMidverts{0.9045/-0.6245,0.0245/-0.2845,-0.9045/-0.0365,
  -0.2716/-1.1961,0.2135/-1.5755,0.5135/1.5755,
  0.2135/0.3265,-0.9045/-1.2125}
\def\vthreeSsevencoords{(0.9045,-0.6245) (0.0245,-0.2845) (-0.9045,-0.0365)
  (-0.3286,-1.0253) (0.2135,-1.5755) (0.5135,1.5755)
  (0.2135,0.3265) (-0.9045,-1.2125)}
\def\vthreeSsevenverts{0.9045/-0.6245,0.0245/-0.2845,-0.9045/-0.0365,
  -0.3286/-1.0253,0.2135/-1.5755,0.5135/1.5755,
  0.2135/0.3265,-0.9045/-1.2125}
\def\vthreeZcoords{(0.9045,-0.6245) (0.0245,-0.2845) (-0.9045,-0.0365)
  (-0.3855,-0.8545) (0.2135,-1.5755) (0.5135,1.5755)
  (0.2135,0.3265) (-0.9045,-1.2125)}
\def\vthreeZverts{0.9045/-0.6245,0.0245/-0.2845,-0.9045/-0.0365,
  -0.3855/-0.8545,0.2135/-1.5755,0.5135/1.5755,
  0.2135/0.3265,-0.9045/-1.2125}
\def\vthreeQcoords{(0.9045,-0.6245) (-0.9045,-0.0365) (-0.2716,-1.1961)
  (0.2135,-1.5755) (0.5135,1.5755) (0.2135,0.3265)
  (-0.9045,-1.2125)}
\def\vthreeQverts{0.9045/-0.6245,-0.9045/-0.0365,-0.2716/-1.1961,
  0.2135/-1.5755,0.5135/1.5755,0.2135/0.3265,-0.9045/-1.2125}
\def\vthreeLonecoords{(-0.3286,-1.0253) (0.2135,-1.5755)
  (0.5135,1.5755) (0.2135,0.3265) (-0.9045,-1.2125)}
\def\vthreeLoneverts{-0.3286/-1.0253,0.2135/-1.5755,
  0.5135/1.5755,0.2135/0.3265,-0.9045/-1.2125}
\def\vthreeLtwocoords{(-0.3286,-0.6503) (0.9045,-0.2495)
  (0.0245,0.0905) (-0.9045,0.3385)}
\def\vthreeLtwoverts{-0.3286/-0.6503,0.9045/-0.2495,
  0.0245/0.0905,-0.9045/0.3385}

\newcommand{\vthreedrawpolygon}[3][]{%
  \draw[vthreeink!86,line width=.78pt,line join=round,line cap=round,#1]
    plot coordinates {#2} -- cycle;
  \draw[vthreeink!82,#1]
    plot[only marks,mark=*,mark size=.55pt] coordinates {#2};
}

\newcommand{\vthreedrawlargepolygon}[3][]{%
  \draw[vthreeink!88,line width=1.05pt,line join=round,line cap=round,#1]
    plot coordinates {#2} -- cycle;
  \draw[vthreeink!88,#1]
    plot[only marks,mark=*,mark size=1.18pt] coordinates {#2};
}

\newcommand{\vthreeorientationarrow}{%
  \draw[-{Stealth[length=2.8pt]},vthreeink!78,line width=.65pt]
    (.63,-.18) -- (.37,-.11);%
}

\newcommand{\vthreedirectionarrow}[4]{%
  \draw[-{Stealth[length=2.8pt]},vthreeink!78,line width=.65pt]
    (#1,#2)--(#3,#4);%
}

\newcommand{\vthreeKmini}[1]{%
  \begingroup
  \ifnum#1=0
    \coordinate (vthreekzero) at (-.44,-.29);
    \coordinate (vthreekone)  at ( .44, .29);
    \coordinate (vthreektwo)  at (-.44, .29);
    \coordinate (vthreekthree)at ( .44,-.29);
    \draw[vthreeorange!82!black,line width=.88pt,line join=round]
      (vthreekzero)--(vthreekone)--(vthreektwo)--(vthreekthree)--cycle;
    \foreach \p in {vthreekzero,vthreekone,vthreektwo,vthreekthree}
      {\fill[vthreeorange!82!black] (\p) circle[radius=.78pt];}
    \draw[-{Stealth[length=2.9pt,width=2.5pt]},vthreeink!88,line width=.68pt,
          shorten <=2.4pt,shorten >=2.4pt]
      (vthreekzero)--(vthreekone);
  \else
    \pgfmathtruncatemacro{\vthreerank}{abs(#1)}
    \pgfmathtruncatemacro{\vthreecount}{2*\vthreerank+1}
    \pgfmathtruncatemacro{\vthreelast}{\vthreecount-1}
    \foreach \j in {0,...,\vthreelast}{
      \pgfmathsetmacro{\vthreeangle}{90+360*\j*(#1)/\vthreecount}
      \coordinate (vthreestar\j) at
        ({.49*cos(\vthreeangle)},{.49*sin(\vthreeangle)});
    }
    \draw[vthreeorange!82!black,line width=.88pt,line join=round]
      (vthreestar0)
      \foreach \j in {1,...,\vthreelast}{--(vthreestar\j)} -- cycle;
    \foreach \j in {0,...,\vthreelast}
      {\fill[vthreeorange!82!black] (vthreestar\j) circle[radius=.72pt];}
    \draw[-{Stealth[length=2.9pt,width=2.5pt]},vthreeink!88,line width=.68pt,
          shorten <=2.4pt,shorten >=2.4pt]
      (vthreestar0)--(vthreestar1);
  \fi
  \endgroup
}

\newcommand{\vthreeZmini}{%
  \begin{scope}[x=.26cm,y=.26cm]
    \fill[vthreeorange!12]
      (0.2135,-1.5755)--(0.5135,1.5755)--(0.2135,0.3265)--cycle;
    \vthreedrawpolygon[vthreeblue]{\vthreeZcoords}{\vthreeZverts}
    \draw[vthreeblue!48,densely dashed,line width=.65pt]
      (0.2135,-1.5755)--(0.2135,0.3265);
    \draw[vthreeorange,line width=1.0pt,line join=round]
      (0.2135,-1.5755)--(0.5135,1.5755)--(0.2135,0.3265);
    \foreach \x/\y in \vthreeZverts
      {\fill[vthreeblue] (\x,\y) circle[radius=.86pt];}
    \fill[vthreeorange] (0.5135,1.5755) circle[radius=1.05pt];
    \draw[-{Stealth[length=2.5pt]},vthreeink!82,line width=.62pt]
      (.7725,-.5735)--(.3325,-.4035);
  \end{scope}
}

\newcommand{\vthreeDecayMini}[1]{%
  \begingroup
  \ifnum#1=4
    \draw[vthreeblue!75!black,line width=.7pt,line join=round] (0.260,0.000) -- (-0.260,0.000) -- (-0.087,-0.240) -- (0.087,0.240) -- cycle;
    \foreach \p in {(0.260,0.000),(-0.260,0.000),(-0.087,-0.240),(0.087,0.240)}{\fill[vthreeblue!75!black] \p circle[radius=.55pt];}
    \draw[-{Stealth[length=2.2pt,width=2pt]},vthreeink!88,line width=.55pt,shorten <=1.2pt,shorten >=1.6pt] (0.087,0.240)--(0.260,0.000);
  \fi
  \ifnum#1=5
    \draw[vthreeblue!75!black,line width=.7pt,line join=round] (0.260,0.000) -- (-0.260,0.000) -- (-0.130,-0.240) -- (0.000,-0.240) -- (0.130,0.240) -- cycle;
    \foreach \p in {(0.260,0.000),(-0.260,0.000),(-0.130,-0.240),(0.000,-0.240),(0.130,0.240)}{\fill[vthreeblue!75!black] \p circle[radius=.55pt];}
    \draw[-{Stealth[length=2.2pt,width=2pt]},vthreeink!88,line width=.55pt,shorten <=1.2pt,shorten >=1.6pt] (0.130,0.240)--(0.260,0.000);
  \fi
  \ifnum#1=6
    \draw[vthreeblue!75!black,line width=.7pt,line join=round] (0.260,0.000) -- (-0.260,0.000) -- (-0.156,-0.240) -- (-0.052,0.240) -- (0.052,-0.240) -- (0.156,0.240) -- cycle;
    \foreach \p in {(0.260,0.000),(-0.260,0.000),(-0.156,-0.240),(-0.052,0.240),(0.052,-0.240),(0.156,0.240)}{\fill[vthreeblue!75!black] \p circle[radius=.55pt];}
    \draw[-{Stealth[length=2.2pt,width=2pt]},vthreeink!88,line width=.55pt,shorten <=1.2pt,shorten >=1.6pt] (0.156,0.240)--(0.260,0.000);
  \fi
  \ifnum#1=7
    \draw[vthreeblue!75!black,line width=.7pt,line join=round] (0.260,0.000) -- (-0.260,0.000) -- (-0.173,-0.240) -- (-0.087,-0.240) -- (0.000,0.240) -- (0.087,-0.240) -- (0.173,0.240) -- cycle;
    \foreach \p in {(0.260,0.000),(-0.260,0.000),(-0.173,-0.240),(-0.087,-0.240),(0.000,0.240),(0.087,-0.240),(0.173,0.240)}{\fill[vthreeblue!75!black] \p circle[radius=.55pt];}
    \draw[-{Stealth[length=2.2pt,width=2pt]},vthreeink!88,line width=.55pt,shorten <=1.2pt,shorten >=1.6pt] (0.173,0.240)--(0.260,0.000);
  \fi
  \ifnum#1=8
    \draw[vthreeblue!75!black,line width=.7pt,line join=round] (0.260,0.000) -- (-0.260,0.000) -- (-0.186,-0.240) -- (-0.111,0.240) -- (-0.037,-0.240) -- (0.037,0.240) -- (0.111,-0.240) -- (0.186,0.240) -- cycle;
    \foreach \p in {(0.260,0.000),(-0.260,0.000),(-0.186,-0.240),(-0.111,0.240),(-0.037,-0.240),(0.037,0.240),(0.111,-0.240),(0.186,0.240)}{\fill[vthreeblue!75!black] \p circle[radius=.55pt];}
    \draw[-{Stealth[length=2.2pt,width=2pt]},vthreeink!88,line width=.55pt,shorten <=1.2pt,shorten >=1.6pt] (0.186,0.240)--(0.260,0.000);
  \fi
  \endgroup
}

\newcommand{\vthreelatticefigure}{%
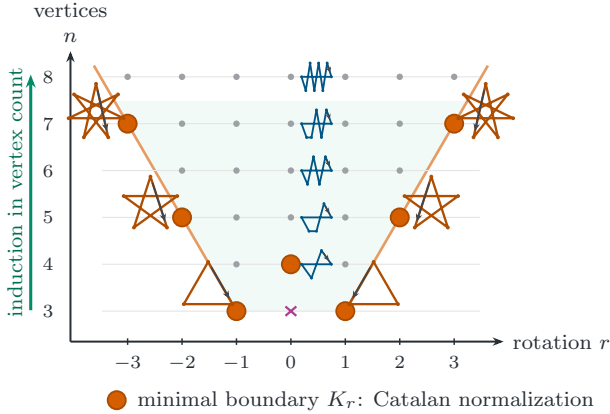
\begin{figure}[!ht]
\centering
\begin{tikzpicture}[
  x=.72cm,
  y=.62cm,
  >=Stealth,
  every node/.style={font=\scriptsize},
  latticepoint/.style={circle,fill=vthreeink!45,minimum size=2.4pt,inner sep=0pt},
  anchorpoint/.style={circle,draw=vthreeorange!80!black,fill=vthreeorange,
                      line width=.7pt,minimum size=6.8pt,inner sep=0pt}
]
  \begin{scope}
    \clip (-3.62,8.24) -- (3.62,8.24) -- (1,3) -- (-1,3) -- cycle;
    \fill[vthreelightgreen,opacity=.58]
      (-4,2.70) rectangle (4,7.48);
  \end{scope}
  \draw[vthreeorange!62,line width=1.0pt]
    (-3.62,8.24) -- (-3,7) -- (-2,5) -- (-1,3);
  \draw[vthreeorange!62,line width=1.0pt]
    (1,3) -- (2,5) -- (3,7) -- (3.62,8.24);

  \foreach \n in {3,...,8}{
    \draw[black!10,line width=.45pt] (-3.47,\n) -- (3.47,\n);
    \node[anchor=east,text=vthreeink] at (-4.20,\n) {$\n$};
  }

  \draw[-{Stealth[length=4.2pt]},vthreeink,line width=.75pt]
    (-4.05,2.35) -- (3.91,2.35)
    node[right,font=\footnotesize] {rotation $r$};
  \draw[-{Stealth[length=4.2pt]},vthreeink,line width=.75pt]
    (-4.05,2.35) -- (-4.05,8.55)
    node[above,font=\footnotesize,align=center] {vertices\\$n$};
  \foreach \r in {-3,...,3}{
    \draw[vthreeink,line width=.55pt] (\r,2.28) -- (\r,2.42);
    \node[anchor=north,text=vthreeink] at (\r,2.23) {$\r$};
  }

  \foreach \r/\n in {
    -3/7,-2/5,-1/3,
     1/3, 2/5, 3/7}{
    \node[anchorpoint] at (\r,\n) {};
  }
  \node[anchorpoint] at (0,4) {};

  \foreach \r in {-1,1}{\node[latticepoint] at (\r,4) {};}
  \foreach \r in {-1,0,1}{\node[latticepoint] at (\r,5) {};}
  \foreach \r in {-2,-1,0,1,2}{\node[latticepoint] at (\r,6) {};}
  \foreach \r in {-2,-1,0,1,2}{\node[latticepoint] at (\r,7) {};}
  \foreach \r in {-3,-2,-1,0,1,2,3}{\node[latticepoint] at (\r,8) {};}

  \draw[vthreepurple,line width=.85pt]
    (-.10,2.90) -- (.10,3.10)
    (-.10,3.10) -- (.10,2.90);

  \draw[-{Stealth[length=5pt]},vthreegreen,line width=1.15pt]
    (-4.78,3.02) -- (-4.78,8.05);
  \node[rotate=90,anchor=south,text=vthreegreen!80!black,
        font=\footnotesize,align=center]
    at (-4.78,5.54) {induction in vertex count};

  \foreach \r/\n/\px/\py in {
    -3/7/-3.58/7.25,-2/5/-2.58/5.27,-1/3/-1.52/3.45,
     1/3/ 1.52/3.45, 2/5/ 2.58/5.27, 3/7/ 3.58/7.25}{
    \begin{scope}[shift={(\px,\py)},x=.76cm,y=.76cm]
      \vthreeKmini{\r}
    \end{scope}
  }
  \foreach \n in {4,...,8}{
    \begin{scope}[shift={(.47,\n)},x=.76cm,y=.76cm]
      \vthreeDecayMini{\n}
    \end{scope}
  }

  \begin{scope}[shift={(0,-.38)}]
    \node[anchorpoint] at (-3.20,1.48) {};
    \node[anchor=west,font=\footnotesize,text=vthreeink]
      at (-2.98,1.48) {minimal boundary $K_r$: Catalan normalization};
  \end{scope}
\end{tikzpicture}
\caption{\label{fig:v3lattice}
The admissible $(n,r)$ lattice at the $n=8$ induction step; the shaded rows
are the completed induction data.}
\end{figure}
}

\newcommand{\vthreechamberfigure}{%
\begin{figure}[!ht]
\centering
\begin{tikzpicture}[x=.80cm,y=.82cm,>=Stealth,
  every node/.style={font=\scriptsize}]
  \node[anchor=south,font=\footnotesize,text=vthreeink]
    at (-1.80,3.45) {a transport route in the fixed $(n,r)=(8,2)$ fiber};
  \draw[vthreeorange,line width=1.55pt,line cap=round]
    (-5.75,3.16)--(-5.30,3.16);
  \node[anchor=west,text=vthreeink] at (-5.20,3.16) {(W2) flat};
  \draw[vthreepurple,line width=1.25pt,line cap=round]
    (-2.72,3.16)--(-2.27,3.16);
  \node[anchor=west,text=vthreeink] at (-2.17,3.16) {(W3) triple};
  \draw[vthreegreen,line width=1.55pt,line cap=round]
    (.18,3.16)--(.63,3.16);
  \node[anchor=west,text=vthreeink] at (.73,3.16) {(W1) vertex--edge};

  \begin{scope}
    \clip[rounded corners=7pt] (-6.60,-2.78) rectangle (3.00,2.78);
    \fill[white] (-7.25,-2.78) rectangle (7.25,2.78);

    \fill[vthreelightblue]
      (-.86,1.76)
      .. controls (.35,1.51) and (1.80,1.68) .. (3.00,1.38)
      .. controls (3.05,.58) and (2.52,-.45) .. (2.92,-1.50)
      .. controls (2.10,-1.61) and (1.35,-1.53) .. (.70,-1.65)
      .. controls (.42,-.98) and (.12,-.42) .. (-.18,.12)
      .. controls (-.54,.85) and (-.72,1.35) .. cycle;

    \draw[vthreepurple!74,line width=.95pt,line cap=round]
      (-7.25,2.36)
      .. controls (-6.18,1.52) and (-5.28,1.54) .. (-4.42,1.83)
      .. controls (-3.30,2.16) and (-2.04,1.47) .. (-.86,1.76)
      .. controls (.35,1.51) and (1.80,1.68) .. (3.00,1.38)
      .. controls (4.55,1.00) and (5.92,1.43) .. (7.25,1.04);
    \draw[vthreepurple!74,line width=.95pt,line cap=round]
      (-7.25,-.82)
      .. controls (-6.30,-1.58) and (-5.18,-1.03) .. (-4.12,-1.30)
      .. controls (-2.85,-1.64) and (-1.35,-2.08) .. (-.22,-1.78)
      .. controls (.15,-1.68) and (.43,-1.62) .. (.70,-1.65)
      .. controls (1.35,-1.53) and (2.10,-1.61) .. (2.92,-1.50)
      .. controls (4.25,-2.03) and (5.70,-1.77) .. (7.25,-2.31);
    \draw[vthreepurple!48,line width=.72pt,line cap=round]
      (-7.25,.48)
      .. controls (-6.62,.15) and (-6.22,-.52) .. (-6.33,-1.18)
      .. controls (-6.43,-1.82) and (-6.78,-2.25) .. (-6.52,-2.78);
    \draw[vthreepurple!48,line width=.72pt,line cap=round]
      (2.30,2.78)
      .. controls (3.14,2.18) and (4.18,2.22) .. (5.02,1.94)
      .. controls (5.93,1.60) and (6.61,1.49) .. (7.25,1.72);

    \draw[vthreegreen!58,line width=.82pt,line cap=round]
      (-7.25,-2.17)
      .. controls (-6.95,-1.50) and (-6.68,-.65) .. (-6.55,.30)
      .. controls (-6.43,1.20) and (-6.33,2.05) .. (-6.12,2.78);
    \draw[vthreeorange,line width=1.58pt,line cap=round]
      (-4.72,-2.78)
      .. controls (-4.47,-1.60) and (-3.95,-.55) .. (-3.55,.24)
      .. controls (-3.06,1.16) and (-2.18,2.08) .. (-1.72,2.78);
    \draw[vthreegreen,line width=1.58pt,line cap=round]
      (-1.50,2.78)
      .. controls (-1.35,2.36) and (-1.10,1.95) .. (-.86,1.76)
      .. controls (-.58,1.28) and (-.38,.55) .. (-.18,.12)
      .. controls (.18,-.52) and (.47,-1.24) .. (.70,-1.65)
      .. controls (.85,-2.00) and (.94,-2.40) .. (1.06,-2.78);
    \draw[vthreeorange!58,line width=.82pt,line cap=round]
      (3.65,2.78)
      .. controls (3.20,2.17) and (3.02,1.86) .. (3.00,1.38)
      .. controls (3.05,.58) and (2.52,-.45) .. (2.92,-1.50)
      .. controls (3.21,-2.03) and (3.55,-2.41) .. (3.38,-2.78);
    \draw[vthreegreen!48,line width=.72pt,line cap=round]
      (5.06,2.78)
      .. controls (5.25,1.88) and (5.83,.94) .. (6.45,.50)
      .. controls (6.72,.31) and (6.99,.20) .. (7.25,.16);
    \draw[vthreeorange!48,line width=.72pt,line cap=round]
      (5.30,-2.78)
      .. controls (5.18,-2.08) and (5.85,-1.49) .. (6.20,-.92)
      .. controls (6.55,-.35) and (6.86,.12) .. (7.25,.43);
    \draw[vthreegreen!48,line width=.72pt,line cap=round]
      (-2.14,-2.78)
      .. controls (-2.86,-2.43) and (-3.45,-2.18) .. (-4.18,-2.13)
      .. controls (-5.10,-2.04) and (-6.25,-2.54) .. (-7.25,-2.34);
  \end{scope}

  \draw[vthreeink!28,rounded corners=7pt,line width=.72pt]
    (-6.60,-2.78) rectangle (3.00,2.78);

  \node[anchor=north west,text=vthreeink!48,inner sep=1.2pt]
    at (-5.62,1.43) {$\mathcal C_0$};
  \node[anchor=north west,text=vthreeink!48,inner sep=1.2pt]
    at (-2.33,1.39) {$\mathcal C_1$};
  \node[anchor=north west,text=vthreeblue!62!black,fill=white,
        fill opacity=.84,text opacity=1,inner sep=1.4pt]
    at (.02,1.40) {$\mathcal C_E$};

  \draw[-{Stealth[length=4pt]},vthreeblue,line width=1.15pt]
    (-4.30,.31) .. controls (-4.02,.30) and (-3.78,.27) .. (-3.64,.25);
  \draw[-{Stealth[length=4pt]},vthreeblue,line width=1.15pt]
    (-3.46,.20) .. controls (-3.23,.15) and (-3.03,.10) .. (-2.78,.06);
  \draw[-{Stealth[length=4pt]},vthreeblue,line width=1.15pt]
    (-1.12,-.03) .. controls (-.82,.00) and (-.50,.06) .. (-.27,.10);
  \draw[-{Stealth[length=4pt]},vthreeblue,line width=1.15pt]
    (-.08,.14) .. controls (.08,.16) and (.23,.18) .. (.39,.19);
  \filldraw[fill=white,draw=vthreeorange,line width=1.05pt]
    (-3.55,.24) circle[radius=2.55pt];
  \filldraw[fill=white,draw=vthreegreen,line width=1.05pt]
    (-.18,.12) circle[radius=2.55pt];
  \node[anchor=north east,text=vthreeorange!84!black,fill=white,
        inner sep=1.2pt] at (-3.66,.02) {(W2)};
  \node[anchor=north west,text=vthreegreen!78!black,fill=white,
        inner sep=1.2pt] at (-.06,-.06) {(W1)};

  \begin{scope}[shift={(-5.25,.32)}]
    \begin{scope}[x=.52cm,y=.52cm,rotate=8,
                  xscale=1.08,yscale=.76,xslant=.24]
      \vthreedrawlargepolygon{\vthreePcoords}{\vthreePverts}
      \vthreedirectionarrow{.70}{-.65}{.28}{-.71}
    \end{scope}
    \node[anchor=east,text=vthreeink,font=\footnotesize] at (-.76,-.77)
      {$P_{8,2}$};
  \end{scope}

  \begin{scope}[shift={(-2.00,-.15)}]
    \begin{scope}[x=.52cm,y=.52cm,rotate=8,
                  xscale=1.08,yscale=.76,xslant=.24]
      \vthreedrawlargepolygon{\vthreeMidcoords}{\vthreeMidverts}
      \vthreedirectionarrow{.7725}{-.5735}{.3325}{-.4035}
    \end{scope}
    \node[anchor=west,text=vthreeink,font=\footnotesize] at (.70,-.84)
      {$P^{(1)}_{8,2}$};
  \end{scope}

  \begin{scope}[shift={(1.15,.08)}]
    \begin{scope}[x=.52cm,y=.52cm,rotate=8,
                  xscale=1.08,yscale=.76,xslant=.24]
      \fill[vthreeorange!12]
        (0.2135,-1.5755)--(0.5135,1.5755)--(0.2135,0.3265)--cycle;
      \vthreedrawlargepolygon[vthreeblue]{\vthreeZcoords}{\vthreeZverts}
      \draw[vthreeblue!48,densely dashed,line width=.7pt]
        (0.2135,-1.5755)--(0.2135,0.3265);
      \draw[vthreeorange,line width=1.35pt,line join=round]
        (0.2135,-1.5755)--(0.5135,1.5755)--(0.2135,0.3265);
      \fill[vthreeorange] (0.5135,1.5755) circle[radius=1.45pt];
      \vthreedirectionarrow{.7725}{-.5735}{.3325}{-.4035}
    \end{scope}
    \node[anchor=north,text=vthreeblue!78!black,font=\footnotesize]
      at (.05,-1.08) {$E_{8,2}$};
  \end{scope}
\end{tikzpicture}
\caption{\label{fig:v3chambers}
One transport route in the fixed $(n,r)=(8,2)$ fiber, from a starting
chamber $\mathcal C_0$ through $\mathcal C_1$ to the anchor chamber
$\mathcal C_E$ containing $E_{8,2}$.  The three polygon portraits are exact
chamber representatives; the surrounding wall network is a schematic
two-dimensional slice of the higher-dimensional chamber space.}
\end{figure}
}

\providecommand{\vthreefigurebreak}{}

%% file: figures_v9_band_surgery.tex
\newcommand{\vbandsurgerycartoon}{%
\begin{center}
\begin{tikzpicture}[x=.6cm,y=.6cm,line width=1.15pt,line cap=round,>={Stealth[length=3.2pt]}]
  \colorlet{bandfront}{black!13}
  \colorlet{bandback}{black!30}
  \colorlet{bandedge}{black!72}
  \draw[pieceA,->,rounded corners=4pt] (-1.25,-1.25)--(-.6,-.6)--(-.6,.6)--(-1.25,1.25);
  \draw[pieceC,->,rounded corners=4pt] (1.25,-1.25)--(.6,-.6)--(.6,.6)--(1.25,1.25);
  \node[font=\scriptsize,anchor=north] at (0,-1.45) {oriented smoothing};
  \draw[black!70,-{Stealth[length=3pt]},line width=.6pt] (1.90,.22)--(3.30,.22);
  \node[font=\scriptsize,anchor=south] at (2.60,.28) {band surgery};
  \draw[black!70,-{Stealth[length=3pt]},line width=.6pt] (3.30,-.22)--(1.90,-.22);
  \node[font=\scriptsize,anchor=north] at (2.60,-.28) {smooth};
  \begin{scope}[shift={(5.2,0)}]
    \fill[bandfront] (-.6,-.6)--(-.6,.6)--(0,0)--cycle;
    \fill[bandback] (0,0)--(.6,.6)--(.6,-.6)--cycle;
    \draw[black!70,line width=.55pt,-{Stealth[length=2.4pt,width=2.2pt]}]
      ([shift={(-.37,0)}]300:.14) arc[start angle=300,end angle=600,radius=.14];
    \draw[black!80,line width=.55pt,-{Stealth[length=2.4pt,width=2.2pt]}]
      ([shift={(.37,0)}]240:.14) arc[start angle=240,end angle=-60,radius=.14];
    \draw[bandedge,line cap=butt] (.57,-.57)--(.15,-.15);
    \draw[bandedge,line cap=butt] (-.15,.15)--(-.57,.57);
    \draw[bandedge,line cap=butt] (-.57,-.57)--(.57,.57);
    \draw[pieceA,->,rounded corners=4pt] (-1.25,-1.25)--(-.6,-.6)--(-.6,.6)--(-1.25,1.25);
    \draw[pieceC,->,rounded corners=4pt] (1.25,-1.25)--(.6,-.6)--(.6,.6)--(1.25,1.25);
  \end{scope}
  \node[font=\scriptsize,anchor=north] at (5.2,-1.45) {positive crossing};
\end{tikzpicture}
\end{center}
}

%% file: figures_v9_ribbon_curl.tex
\begin{center}
\begin{tikzpicture}[x=1cm,y=1cm]
  \tikzset{
    ribQ/.style={pieceA,line width=1.15pt,line cap=round,line join=round},
    ribQp/.style={pieceB,line width=.6pt,line cap=round,line join=round},
    ribgapQp/.style={white,line width=2.6pt,line cap=butt},
    riborient/.style={pieceA,line width=.85pt,-{Stealth[length=2.8pt]}}
  }
  \path[use as bounding box] (-1.05,-1.10) rectangle (6.55,1.20);

  \begin{scope}
    \draw[ribQ] (.55,.95)--(0,0)
      arc[start angle=330,end angle=30,radius=.28]
      --(.55,-.67);
    \draw[knotgap] (.141,.244)--(.021,.036);
    \draw[ribQ] (.141,.244)--(.021,.036);
    \draw[riborient] (.44,.76)--(.30,.52);
  \end{scope}

  \begin{scope}[shift={(2.65,0)}]
    \draw[ribQ] (.55,.95)--(0,0)
      arc[start angle=330,end angle=30,radius=.28]
      --(.55,-.67);
    \draw[ribQp] (.472,.995)--(-.078,.045)
      arc[start angle=330,end angle=30,radius=.19]
      --(.472,-.715);
    \draw[knotgap] (.141,.244)--(0,0);
    \draw[ribQ] (.141,.244)--(0,0);
    \draw[ribgapQp] (.089,.333)--(-.078,.045);
    \draw[ribQp] (.089,.333)--(-.078,.045);
    \draw[riborient] (.44,.76)--(.30,.52);
    \node[font=\scriptsize,pieceA] at (.72,-.30) {$Q$};
    \node[font=\scriptsize,pieceB] at (.10,-.70) {$Q'$};
  \end{scope}

  \draw[->,black!70,line width=.6pt] (1.10,.10)--(1.80,.10);
  \draw[->,black!70,line width=.6pt] (3.75,.10)--(4.45,.10);

  \begin{scope}[shift={(5.45,.10)}]
    \fill[pieceA!12] (-.14,.95) rectangle (.14,-.85);
    \fill[white] (-.16,.55) rectangle (.16,-.15);
    \fill[pieceA!12]
      plot[domain=.55:-.15,samples=41,variable=\y] ({.14*cos(360*(.55-\y)/.7)},\y) --
      plot[domain=-.15:.55,samples=41,variable=\y] ({-.14*cos(360*(.55-\y)/.7)},\y) -- cycle;
    \fill[pieceA!32]
      plot[domain=.375:.025,samples=25,variable=\y] ({.14*cos(360*(.55-\y)/.7)},\y) --
      plot[domain=.025:.375,samples=25,variable=\y] ({-.14*cos(360*(.55-\y)/.7)},\y) -- cycle;
    \draw[ribQ] (.14,.95)--(.14,.55)
      plot[domain=.55:.065,samples=29,variable=\y] ({.14*cos(360*(.55-\y)/.7)},\y);
    \draw[ribQ]
      plot[domain=-.015:-.15,samples=11,variable=\y] ({.14*cos(360*(.55-\y)/.7)},\y) -- (.14,-.85);
    \draw[ribQp] (-.14,.95)--(-.14,.55)
      plot[domain=.55:.415,samples=11,variable=\y] ({-.14*cos(360*(.55-\y)/.7)},\y);
    \draw[ribQp]
      plot[domain=.335:-.15,samples=29,variable=\y] ({-.14*cos(360*(.55-\y)/.7)},\y) -- (-.14,-.85);
    \draw[riborient] (.14,.88)--(.14,.68);
  \end{scope}
\end{tikzpicture}
\end{center}

%% file: wall_laws_letterv9.bbl
\begin{thebibliography}{50}%
\makeatletter
\providecommand \@ifxundefined [1]{%
 \@ifx{#1\undefined}
}%
\providecommand \@ifnum [1]{%
 \ifnum #1\expandafter \@firstoftwo
 \else \expandafter \@secondoftwo
 \fi
}%
\providecommand \@ifx [1]{%
 \ifx #1\expandafter \@firstoftwo
 \else \expandafter \@secondoftwo
 \fi
}%
\providecommand \natexlab [1]{#1}%
\providecommand \enquote  [1]{``#1''}%
\providecommand \bibnamefont  [1]{#1}%
\providecommand \bibfnamefont [1]{#1}%
\providecommand \citenamefont [1]{#1}%
\providecommand \href@noop [0]{\@secondoftwo}%
\providecommand \href [0]{\begingroup \@sanitize@url \@href}%
\providecommand \@href[1]{\@@startlink{#1}\@@href}%
\providecommand \@@href[1]{\endgroup#1\@@endlink}%
\providecommand \@sanitize@url [0]{\catcode `\\12\catcode `\$12\catcode
  `\&12\catcode `\#12\catcode `\^12\catcode `\_12\catcode `\%12\relax}%
\providecommand \@@startlink[1]{}%
\providecommand \@@endlink[0]{}%
\providecommand \url  [0]{\begingroup\@sanitize@url \@url }%
\providecommand \@url [1]{\endgroup\@href {#1}{\urlprefix }}%
\providecommand \urlprefix  [0]{URL }%
\providecommand \Eprint [0]{\href }%
\providecommand \doibase [0]{https://doi.org/}%
\providecommand \selectlanguage [0]{\@gobble}%
\providecommand \bibinfo  [0]{\@secondoftwo}%
\providecommand \bibfield  [0]{\@secondoftwo}%
\providecommand \translation [1]{[#1]}%
\providecommand \BibitemOpen [0]{}%
\providecommand \bibitemStop [0]{}%
\providecommand \bibitemNoStop [0]{.\EOS\space}%
\providecommand \EOS [0]{\spacefactor3000\relax}%
\providecommand \BibitemShut  [1]{\csname bibitem#1\endcsname}%
\let\auto@bib@innerbib\@empty
\bibitem [{\citenamefont {Maldacena}(1998)}]{Maldacena}%
  \BibitemOpen
  \bibfield  {author} {\bibinfo {author} {\bibfnamefont {J.}~\bibnamefont
  {Maldacena}},\ }\bibfield  {title} {\bibinfo {title} {{The large $N$ limit of
  superconformal field theories and supergravity}},\ }\href@noop {} {\bibfield
  {journal} {\bibinfo  {journal} {Adv. Theor. Math. Phys.}\ }\textbf {\bibinfo
  {volume} {2}},\ \bibinfo {pages} {231} (\bibinfo {year} {1998})},\ \Eprint
  {https://arxiv.org/abs/hep-th/9711200} {arXiv:hep-th/9711200 [hep-th]}
  \BibitemShut {NoStop}%
\bibitem [{\citenamefont {Pasterski}\ \emph {et~al.}(2021)\citenamefont
  {Pasterski}, \citenamefont {Pate},\ and\ \citenamefont
  {Raclariu}}]{PasterskiPateRaclariu}%
  \BibitemOpen
  \bibfield  {author} {\bibinfo {author} {\bibfnamefont {S.}~\bibnamefont
  {Pasterski}}, \bibinfo {author} {\bibfnamefont {M.}~\bibnamefont {Pate}},\
  and\ \bibinfo {author} {\bibfnamefont {A.-M.}\ \bibnamefont {Raclariu}},\
  }\href@noop {} {\bibinfo {title} {{Celestial holography}}} (\bibinfo {year}
  {2021}),\ \Eprint {https://arxiv.org/abs/2111.11392} {arXiv:2111.11392
  [hep-th]} \BibitemShut {NoStop}%
\bibitem [{\citenamefont {Witten}(2004)}]{WittenTwistor}%
  \BibitemOpen
  \bibfield  {author} {\bibinfo {author} {\bibfnamefont {E.}~\bibnamefont
  {Witten}},\ }\bibfield  {title} {\bibinfo {title} {{Perturbative gauge theory
  as a string theory in twistor space}},\ }\href@noop {} {\bibfield  {journal}
  {\bibinfo  {journal} {Commun. Math. Phys.}\ }\textbf {\bibinfo {volume}
  {252}},\ \bibinfo {pages} {189} (\bibinfo {year} {2004})},\ \Eprint
  {https://arxiv.org/abs/hep-th/0312171} {arXiv:hep-th/0312171 [hep-th]}
  \BibitemShut {NoStop}%
\bibitem [{\citenamefont {Elvang}\ and\ \citenamefont
  {Huang}(2015)}]{ElvangHuang}%
  \BibitemOpen
  \bibfield  {author} {\bibinfo {author} {\bibfnamefont {H.}~\bibnamefont
  {Elvang}}\ and\ \bibinfo {author} {\bibfnamefont {Y.-t.}\ \bibnamefont
  {Huang}},\ }\href@noop {} {\emph {\bibinfo {title} {{Scattering Amplitudes in
  Gauge Theory and Gravity}}}}\ (\bibinfo  {publisher} {Cambridge University
  Press},\ \bibinfo {address} {Cambridge},\ \bibinfo {year} {2015})\ \Eprint
  {https://arxiv.org/abs/1308.1697} {arXiv:1308.1697 [hep-th]} \BibitemShut
  {NoStop}%
\bibitem [{\citenamefont {Beisert}\ \emph {et~al.}(2012)\citenamefont {Beisert}
  \emph {et~al.}}]{BeisertReview}%
  \BibitemOpen
  \bibfield  {author} {\bibinfo {author} {\bibfnamefont {N.}~\bibnamefont
  {Beisert}} \emph {et~al.},\ }\bibfield  {title} {\bibinfo {title} {{Review of
  AdS/CFT integrability: an overview}},\ }\href@noop {} {\bibfield  {journal}
  {\bibinfo  {journal} {Lett. Math. Phys.}\ }\textbf {\bibinfo {volume} {99}},\
  \bibinfo {pages} {3} (\bibinfo {year} {2012})},\ \Eprint
  {https://arxiv.org/abs/1012.3982} {arXiv:1012.3982 [hep-th]} \BibitemShut
  {NoStop}%
\bibitem [{\citenamefont {Arkani-Hamed}\ \emph {et~al.}(2022)\citenamefont
  {Arkani-Hamed}, \citenamefont {Dixon}, \citenamefont {McLeod}, \citenamefont
  {Spradlin}, \citenamefont {Trnka},\ and\ \citenamefont {Volovich}}]{ADMSTV}%
  \BibitemOpen
  \bibfield  {author} {\bibinfo {author} {\bibfnamefont {N.}~\bibnamefont
  {Arkani-Hamed}}, \bibinfo {author} {\bibfnamefont {L.~J.}\ \bibnamefont
  {Dixon}}, \bibinfo {author} {\bibfnamefont {A.~J.}\ \bibnamefont {McLeod}},
  \bibinfo {author} {\bibfnamefont {M.}~\bibnamefont {Spradlin}}, \bibinfo
  {author} {\bibfnamefont {J.}~\bibnamefont {Trnka}},\ and\ \bibinfo {author}
  {\bibfnamefont {A.}~\bibnamefont {Volovich}},\ }\href@noop {} {\bibinfo
  {title} {{Solving scattering in $\mathcal{N}=4$ super-Yang--Mills theory}}}
  (\bibinfo {year} {2022}),\ \Eprint {https://arxiv.org/abs/2207.10636}
  {arXiv:2207.10636 [hep-th]} \BibitemShut {NoStop}%
\bibitem [{\citenamefont {Bern}\ \emph {et~al.}(2005)\citenamefont {Bern},
  \citenamefont {Dixon},\ and\ \citenamefont {Smirnov}}]{Bern:2005iz}%
  \BibitemOpen
  \bibfield  {author} {\bibinfo {author} {\bibfnamefont {Z.}~\bibnamefont
  {Bern}}, \bibinfo {author} {\bibfnamefont {L.~J.}\ \bibnamefont {Dixon}},\
  and\ \bibinfo {author} {\bibfnamefont {V.~A.}\ \bibnamefont {Smirnov}},\
  }\bibfield  {title} {\bibinfo {title} {{Iteration of planar amplitudes in
  maximally supersymmetric Yang-Mills theory at three loops and beyond}},\
  }\href@noop {} {\bibfield  {journal} {\bibinfo  {journal} {Phys. Rev. D}\
  }\textbf {\bibinfo {volume} {72}},\ \bibinfo {pages} {085001} (\bibinfo
  {year} {2005})},\ \Eprint {https://arxiv.org/abs/hep-th/0505205}
  {arXiv:hep-th/0505205 [hep-th]} \BibitemShut {NoStop}%
\bibitem [{\citenamefont {Alday}\ and\ \citenamefont
  {Maldacena}(2007)}]{AldayMaldacena07}%
  \BibitemOpen
  \bibfield  {author} {\bibinfo {author} {\bibfnamefont {L.~F.}\ \bibnamefont
  {Alday}}\ and\ \bibinfo {author} {\bibfnamefont {J.}~\bibnamefont
  {Maldacena}},\ }\bibfield  {title} {\bibinfo {title} {{Gluon scattering
  amplitudes at strong coupling}},\ }\href@noop {} {\bibfield  {journal}
  {\bibinfo  {journal} {JHEP}\ }\textbf {\bibinfo {volume} {06}},\ \bibinfo
  {pages} {064}},\ \Eprint {https://arxiv.org/abs/0705.0303} {arXiv:0705.0303
  [hep-th]} \BibitemShut {NoStop}%
\bibitem [{\citenamefont {Alday}\ and\ \citenamefont
  {Maldacena}(2009)}]{AldayMaldacena}%
  \BibitemOpen
  \bibfield  {author} {\bibinfo {author} {\bibfnamefont {L.~F.}\ \bibnamefont
  {Alday}}\ and\ \bibinfo {author} {\bibfnamefont {J.}~\bibnamefont
  {Maldacena}},\ }\bibfield  {title} {\bibinfo {title} {{Null polygonal Wilson
  loops and minimal surfaces in Anti-de-Sitter space}},\ }\href@noop {}
  {\bibfield  {journal} {\bibinfo  {journal} {JHEP}\ }\textbf {\bibinfo
  {volume} {11}},\ \bibinfo {pages} {082}},\ \Eprint
  {https://arxiv.org/abs/0904.0663} {arXiv:0904.0663 [hep-th]} \BibitemShut
  {NoStop}%
\bibitem [{\citenamefont {Guevara}\ \emph
  {et~al.}(2026{\natexlab{a}})\citenamefont {Guevara}, \citenamefont
  {Lupsasca}, \citenamefont {Skinner}, \citenamefont {Strominger},\ and\
  \citenamefont {Weil}}]{GLSSW}%
  \BibitemOpen
  \bibfield  {author} {\bibinfo {author} {\bibfnamefont {A.}~\bibnamefont
  {Guevara}}, \bibinfo {author} {\bibfnamefont {A.}~\bibnamefont {Lupsasca}},
  \bibinfo {author} {\bibfnamefont {D.}~\bibnamefont {Skinner}}, \bibinfo
  {author} {\bibfnamefont {A.}~\bibnamefont {Strominger}},\ and\ \bibinfo
  {author} {\bibfnamefont {K.}~\bibnamefont {Weil}},\ }\href@noop {} {\bibinfo
  {title} {{Single-minus gluon tree amplitudes are nonzero}}} (\bibinfo {year}
  {2026}{\natexlab{a}}),\ \Eprint {https://arxiv.org/abs/2602.12176}
  {arXiv:2602.12176 [hep-th]} \BibitemShut {NoStop}%
\bibitem [{\citenamefont {Guevara}\ \emph
  {et~al.}(2026{\natexlab{b}})\citenamefont {Guevara}, \citenamefont
  {Lupsasca}, \citenamefont {Skinner}, \citenamefont {Strominger},\ and\
  \citenamefont {Weil}}]{GLSSWgrav}%
  \BibitemOpen
  \bibfield  {author} {\bibinfo {author} {\bibfnamefont {A.}~\bibnamefont
  {Guevara}}, \bibinfo {author} {\bibfnamefont {A.}~\bibnamefont {Lupsasca}},
  \bibinfo {author} {\bibfnamefont {D.}~\bibnamefont {Skinner}}, \bibinfo
  {author} {\bibfnamefont {A.}~\bibnamefont {Strominger}},\ and\ \bibinfo
  {author} {\bibfnamefont {K.}~\bibnamefont {Weil}},\ }\href@noop {} {\bibinfo
  {title} {{Single-minus graviton tree amplitudes are nonzero}}} (\bibinfo
  {year} {2026}{\natexlab{b}}),\ \Eprint {https://arxiv.org/abs/2603.04330}
  {arXiv:2603.04330 [hep-th]} \BibitemShut {NoStop}%
\bibitem [{\citenamefont {Guevara}\ \emph
  {et~al.}(2026{\natexlab{c}})\citenamefont {Guevara}, \citenamefont
  {Lupsasca}, \citenamefont {Maldacena},\ and\ \citenamefont
  {Strominger}}]{GLMS}%
  \BibitemOpen
  \bibfield  {author} {\bibinfo {author} {\bibfnamefont {A.}~\bibnamefont
  {Guevara}}, \bibinfo {author} {\bibfnamefont {A.}~\bibnamefont {Lupsasca}},
  \bibinfo {author} {\bibfnamefont {J.}~\bibnamefont {Maldacena}},\ and\
  \bibinfo {author} {\bibfnamefont {A.}~\bibnamefont {Strominger}},\
  }\href@noop {} {\bibinfo {title} {{Single-Minus Graviton Amplitudes from
  Matrix Theory}}} (\bibinfo {year} {2026}{\natexlab{c}}),\ \Eprint
  {https://arxiv.org/abs/2608.25239} {arXiv:2608.25239 [hep-th]} \BibitemShut
  {NoStop}%
\bibitem [{\citenamefont {Alday}\ \emph {et~al.}(2011)\citenamefont {Alday},
  \citenamefont {Gaiotto}, \citenamefont {Maldacena}, \citenamefont {Sever},\
  and\ \citenamefont {Vieira}}]{AGMSV}%
  \BibitemOpen
  \bibfield  {author} {\bibinfo {author} {\bibfnamefont {L.~F.}\ \bibnamefont
  {Alday}}, \bibinfo {author} {\bibfnamefont {D.}~\bibnamefont {Gaiotto}},
  \bibinfo {author} {\bibfnamefont {J.}~\bibnamefont {Maldacena}}, \bibinfo
  {author} {\bibfnamefont {A.}~\bibnamefont {Sever}},\ and\ \bibinfo {author}
  {\bibfnamefont {P.}~\bibnamefont {Vieira}},\ }\bibfield  {title} {\bibinfo
  {title} {{An operator product expansion for polygonal null Wilson loops}},\
  }\href@noop {} {\bibfield  {journal} {\bibinfo  {journal} {JHEP}\ }\textbf
  {\bibinfo {volume} {04}},\ \bibinfo {pages} {088}},\ \Eprint
  {https://arxiv.org/abs/1006.2788} {arXiv:1006.2788 [hep-th]} \BibitemShut
  {NoStop}%
\bibitem [{\citenamefont {Dixon}\ and\ \citenamefont
  {Esterlis}(2016)}]{DixonEsterlis}%
  \BibitemOpen
  \bibfield  {author} {\bibinfo {author} {\bibfnamefont {L.~J.}\ \bibnamefont
  {Dixon}}\ and\ \bibinfo {author} {\bibfnamefont {I.}~\bibnamefont
  {Esterlis}},\ }\bibfield  {title} {\bibinfo {title} {{All orders results for
  self-crossing Wilson loops mimicking double parton scattering}},\ }\href@noop
  {} {\bibfield  {journal} {\bibinfo  {journal} {JHEP}\ }\textbf {\bibinfo
  {volume} {07}},\ \bibinfo {pages} {116}},\ \bibinfo {note} {[Erratum: JHEP
  08, 131 (2016)]},\ \Eprint {https://arxiv.org/abs/1602.02107}
  {arXiv:1602.02107 [hep-th]} \BibitemShut {NoStop}%
\bibitem [{\citenamefont {Goldstein}\ and\ \citenamefont
  {Guevara}(2026)}]{CornerLawsLean}%
  \BibitemOpen
  \bibfield  {author} {\bibinfo {author} {\bibfnamefont {M.}~\bibnamefont
  {Goldstein}}\ and\ \bibinfo {author} {\bibfnamefont {A.}~\bibnamefont
  {Guevara}},\ }\href@noop {} {\bibinfo {title} {{Corner state sum: wall laws
  and soft theorem. Lean 4 formalization}}},\ \bibinfo {howpublished}
  {\url{https://github.com/aguevara22/gluon-knot-lean}} (\bibinfo {year}
  {2026})\BibitemShut {NoStop}%
\bibitem [{\citenamefont {Witten}(1989)}]{WittenJones}%
  \BibitemOpen
  \bibfield  {author} {\bibinfo {author} {\bibfnamefont {E.}~\bibnamefont
  {Witten}},\ }\bibfield  {title} {\bibinfo {title} {{Quantum field theory and
  the Jones polynomial}},\ }\href@noop {} {\bibfield  {journal} {\bibinfo
  {journal} {Commun. Math. Phys.}\ }\textbf {\bibinfo {volume} {121}},\
  \bibinfo {pages} {351} (\bibinfo {year} {1989})}\BibitemShut {NoStop}%
\bibitem [{\citenamefont {Brandhuber}\ \emph {et~al.}(2026)\citenamefont
  {Brandhuber}, \citenamefont {Pichini}, \citenamefont {Travaglini},\ and\
  \citenamefont {Wen}}]{BPPTW}%
  \BibitemOpen
  \bibfield  {author} {\bibinfo {author} {\bibfnamefont {A.}~\bibnamefont
  {Brandhuber}}, \bibinfo {author} {\bibfnamefont {P.}~\bibnamefont {Pichini}},
  \bibinfo {author} {\bibfnamefont {G.}~\bibnamefont {Travaglini}},\ and\
  \bibinfo {author} {\bibfnamefont {C.}~\bibnamefont {Wen}},\ }\href@noop {}
  {\bibinfo {title} {{$\mathcal{N}=4$ single-minus superamplitudes and dual
  superconformal symmetry}}} (\bibinfo {year} {2026}),\ \Eprint
  {https://arxiv.org/abs/2603.26609} {arXiv:2603.26609 [hep-th]} \BibitemShut
  {NoStop}%
\bibitem [{\citenamefont {Hodges}(2013)}]{Hodges}%
  \BibitemOpen
  \bibfield  {author} {\bibinfo {author} {\bibfnamefont {A.}~\bibnamefont
  {Hodges}},\ }\bibfield  {title} {\bibinfo {title} {{Eliminating spurious
  poles from gauge-theoretic amplitudes}},\ }\href@noop {} {\bibfield
  {journal} {\bibinfo  {journal} {JHEP}\ }\textbf {\bibinfo {volume} {05}},\
  \bibinfo {pages} {135}},\ \Eprint {https://arxiv.org/abs/0905.1473}
  {arXiv:0905.1473 [hep-th]} \BibitemShut {NoStop}%
\bibitem [{\citenamefont {Arkani-Hamed}\ \emph {et~al.}(2018)\citenamefont
  {Arkani-Hamed}, \citenamefont {Thomas},\ and\ \citenamefont {Trnka}}]{AHTT}%
  \BibitemOpen
  \bibfield  {author} {\bibinfo {author} {\bibfnamefont {N.}~\bibnamefont
  {Arkani-Hamed}}, \bibinfo {author} {\bibfnamefont {H.}~\bibnamefont
  {Thomas}},\ and\ \bibinfo {author} {\bibfnamefont {J.}~\bibnamefont
  {Trnka}},\ }\bibfield  {title} {\bibinfo {title} {{Unwinding the
  amplituhedron in binary}},\ }\href@noop {} {\bibfield  {journal} {\bibinfo
  {journal} {JHEP}\ }\textbf {\bibinfo {volume} {01}},\ \bibinfo {pages}
  {016}},\ \Eprint {https://arxiv.org/abs/1704.05069} {arXiv:1704.05069
  [hep-th]} \BibitemShut {NoStop}%
\bibitem [{\citenamefont {Richter}\ and\ \citenamefont
  {Sturmfels}(1991)}]{RichterSturmfels}%
  \BibitemOpen
  \bibfield  {author} {\bibinfo {author} {\bibfnamefont {J.}~\bibnamefont
  {Richter}}\ and\ \bibinfo {author} {\bibfnamefont {B.}~\bibnamefont
  {Sturmfels}},\ }\bibfield  {title} {\bibinfo {title} {{On the topology and
  geometric construction of oriented matroids and convex polytopes}},\
  }\href@noop {} {\bibfield  {journal} {\bibinfo  {journal} {Trans. Am. Math.
  Soc.}\ }\textbf {\bibinfo {volume} {325}},\ \bibinfo {pages} {389} (\bibinfo
  {year} {1991})}\BibitemShut {NoStop}%
\bibitem [{\citenamefont {Aichholzer}\ \emph {et~al.}(2002)\citenamefont
  {Aichholzer}, \citenamefont {Aurenhammer},\ and\ \citenamefont
  {Krasser}}]{AAK}%
  \BibitemOpen
  \bibfield  {author} {\bibinfo {author} {\bibfnamefont {O.}~\bibnamefont
  {Aichholzer}}, \bibinfo {author} {\bibfnamefont {F.}~\bibnamefont
  {Aurenhammer}},\ and\ \bibinfo {author} {\bibfnamefont {H.}~\bibnamefont
  {Krasser}},\ }\bibfield  {title} {\bibinfo {title} {{Enumerating order types
  for small point sets with applications}},\ }\href@noop {} {\bibfield
  {journal} {\bibinfo  {journal} {Order}\ }\textbf {\bibinfo {volume} {19}},\
  \bibinfo {pages} {265} (\bibinfo {year} {2002})}\BibitemShut {NoStop}%
\bibitem [{\citenamefont {Guevara}(2026)}]{companion}%
  \BibitemOpen
  \bibfield  {author} {\bibinfo {author} {\bibfnamefont {A.}~\bibnamefont
  {Guevara}},\ }\href@noop {} {\bibinfo {title} {{Supplemental Material for
  ``Gluons and Knots''}}},\ \bibinfo {howpublished}
  {\url{https://aguevara22.github.io/GKSM.pdf}} (\bibinfo {year}
  {2026})\BibitemShut {NoStop}%
\bibitem [{\citenamefont {Whitney}(1937)}]{Whitney}%
  \BibitemOpen
  \bibfield  {author} {\bibinfo {author} {\bibfnamefont {H.}~\bibnamefont
  {Whitney}},\ }\bibfield  {title} {\bibinfo {title} {{On regular closed curves
  in the plane}},\ }\href@noop {} {\bibfield  {journal} {\bibinfo  {journal}
  {Compos. Math.}\ }\textbf {\bibinfo {volume} {4}},\ \bibinfo {pages} {276}
  (\bibinfo {year} {1937})}\BibitemShut {NoStop}%
\bibitem [{\citenamefont {Wesenberg}(2020)}]{Wesenberg}%
  \BibitemOpen
  \bibfield  {author} {\bibinfo {author} {\bibfnamefont {D.}~\bibnamefont
  {Wesenberg}},\ }\href@noop {} {\bibinfo {title} {{A new formula for rotation
  number}}} (\bibinfo {year} {2020}),\ \Eprint
  {https://arxiv.org/abs/2010.01422} {arXiv:2010.01422 [math.GT]} \BibitemShut
  {NoStop}%
\bibitem [{\citenamefont {Arnold}(1994)}]{Arnold}%
  \BibitemOpen
  \bibfield  {author} {\bibinfo {author} {\bibfnamefont {V.~I.}\ \bibnamefont
  {Arnold}},\ }\bibfield  {title} {\bibinfo {title} {{Plane curves, their
  invariants, perestroikas and classifications}},\ }\href@noop {} {\bibfield
  {journal} {\bibinfo  {journal} {Adv. Sov. Math.}\ }\textbf {\bibinfo {volume}
  {21}},\ \bibinfo {pages} {33} (\bibinfo {year} {1994})}\BibitemShut {NoStop}%
\bibitem [{\citenamefont {Guevara}\ \emph {et~al.}(2021)\citenamefont
  {Guevara}, \citenamefont {Himwich}, \citenamefont {Pate},\ and\ \citenamefont
  {Strominger}}]{GuevaraHimwichPateStrominger}%
  \BibitemOpen
  \bibfield  {author} {\bibinfo {author} {\bibfnamefont {A.}~\bibnamefont
  {Guevara}}, \bibinfo {author} {\bibfnamefont {E.}~\bibnamefont {Himwich}},
  \bibinfo {author} {\bibfnamefont {M.}~\bibnamefont {Pate}},\ and\ \bibinfo
  {author} {\bibfnamefont {A.}~\bibnamefont {Strominger}},\ }\bibfield  {title}
  {\bibinfo {title} {{Holographic symmetry algebras for gauge theory and
  gravity}},\ }\href@noop {} {\bibfield  {journal} {\bibinfo  {journal} {JHEP}\
  }\textbf {\bibinfo {volume} {11}},\ \bibinfo {pages} {152}},\ \Eprint
  {https://arxiv.org/abs/2103.03961} {arXiv:2103.03961 [hep-th]} \BibitemShut
  {NoStop}%
\bibitem [{\citenamefont {Strominger}(2021)}]{StromingerW}%
  \BibitemOpen
  \bibfield  {author} {\bibinfo {author} {\bibfnamefont {A.}~\bibnamefont
  {Strominger}},\ }\bibfield  {title} {\bibinfo {title} {{$w_{1+\infty}$
  algebra and the celestial sphere: infinite towers of soft graviton, photon,
  and gluon symmetries}},\ }\href@noop {} {\bibfield  {journal} {\bibinfo
  {journal} {Phys. Rev. Lett.}\ }\textbf {\bibinfo {volume} {127}},\ \bibinfo
  {pages} {221601} (\bibinfo {year} {2021})},\ \Eprint
  {https://arxiv.org/abs/2105.14346} {arXiv:2105.14346 [hep-th]} \BibitemShut
  {NoStop}%
\bibitem [{\citenamefont {Freyd}\ \emph {et~al.}(1985)\citenamefont {Freyd},
  \citenamefont {Yetter}, \citenamefont {Hoste}, \citenamefont {Lickorish},
  \citenamefont {Millett},\ and\ \citenamefont {Ocneanu}}]{HOMFLY}%
  \BibitemOpen
  \bibfield  {author} {\bibinfo {author} {\bibfnamefont {P.}~\bibnamefont
  {Freyd}}, \bibinfo {author} {\bibfnamefont {D.}~\bibnamefont {Yetter}},
  \bibinfo {author} {\bibfnamefont {J.}~\bibnamefont {Hoste}}, \bibinfo
  {author} {\bibfnamefont {W.~B.~R.}\ \bibnamefont {Lickorish}}, \bibinfo
  {author} {\bibfnamefont {K.}~\bibnamefont {Millett}},\ and\ \bibinfo {author}
  {\bibfnamefont {A.}~\bibnamefont {Ocneanu}},\ }\bibfield  {title} {\bibinfo
  {title} {{A new polynomial invariant of knots and links}},\ }\href@noop {}
  {\bibfield  {journal} {\bibinfo  {journal} {Bull. Am. Math. Soc.}\ }\textbf
  {\bibinfo {volume} {12}},\ \bibinfo {pages} {239} (\bibinfo {year}
  {1985})}\BibitemShut {NoStop}%
\bibitem [{\citenamefont {Przytycki}\ and\ \citenamefont {Traczyk}(1987)}]{PT}%
  \BibitemOpen
  \bibfield  {author} {\bibinfo {author} {\bibfnamefont {J.~H.}\ \bibnamefont
  {Przytycki}}\ and\ \bibinfo {author} {\bibfnamefont {P.}~\bibnamefont
  {Traczyk}},\ }\bibfield  {title} {\bibinfo {title} {{Invariants of links of
  Conway type}},\ }\href@noop {} {\bibfield  {journal} {\bibinfo  {journal}
  {Kobe J. Math.}\ }\textbf {\bibinfo {volume} {4}},\ \bibinfo {pages} {115}
  (\bibinfo {year} {1987})}\BibitemShut {NoStop}%
\bibitem [{\citenamefont {Bennequin}(1983)}]{Bennequin}%
  \BibitemOpen
  \bibfield  {author} {\bibinfo {author} {\bibfnamefont {D.}~\bibnamefont
  {Bennequin}},\ }\bibfield  {title} {\bibinfo {title} {{Entrelacements et
  \'equations de Pfaff}},\ }\href@noop {} {\bibfield  {journal} {\bibinfo
  {journal} {Ast\'erisque}\ }\textbf {\bibinfo {volume} {107--108}},\ \bibinfo
  {pages} {87} (\bibinfo {year} {1983})}\BibitemShut {NoStop}%
\bibitem [{\citenamefont {Etnyre}(2005)}]{Etnyre}%
  \BibitemOpen
  \bibfield  {author} {\bibinfo {author} {\bibfnamefont {J.~B.}\ \bibnamefont
  {Etnyre}},\ }\bibfield  {title} {\bibinfo {title} {{Legendrian and
  transversal knots}},\ }in\ \href@noop {} {\emph {\bibinfo {booktitle}
  {Handbook of Knot Theory}}}\ (\bibinfo  {publisher} {Elsevier},\ \bibinfo
  {year} {2005})\ pp.\ \bibinfo {pages} {105--185}\BibitemShut {NoStop}%
\bibitem [{\citenamefont {Ng}(2008)}]{Ng}%
  \BibitemOpen
  \bibfield  {author} {\bibinfo {author} {\bibfnamefont {L.}~\bibnamefont
  {Ng}},\ }\bibfield  {title} {\bibinfo {title} {A skein approach to
  {B}ennequin type inequalities},\ }\href@noop {} {\bibfield  {journal}
  {\bibinfo  {journal} {Int. Math. Res. Not. IMRN}\ } (\bibinfo {year}
  {2008})},\ \bibinfo {note} {article rnn116; arXiv:0709.2141}\BibitemShut
  {NoStop}%
\bibitem [{\citenamefont {Morton}(1986)}]{Morton}%
  \BibitemOpen
  \bibfield  {author} {\bibinfo {author} {\bibfnamefont {H.~R.}\ \bibnamefont
  {Morton}},\ }\bibfield  {title} {\bibinfo {title} {{Seifert circles and knot
  polynomials}},\ }\href@noop {} {\bibfield  {journal} {\bibinfo  {journal}
  {Math. Proc. Cambridge Philos. Soc.}\ }\textbf {\bibinfo {volume} {99}},\
  \bibinfo {pages} {107} (\bibinfo {year} {1986})}\BibitemShut {NoStop}%
\bibitem [{\citenamefont {Franks}\ and\ \citenamefont
  {Williams}(1987)}]{FranksWilliams}%
  \BibitemOpen
  \bibfield  {author} {\bibinfo {author} {\bibfnamefont {J.}~\bibnamefont
  {Franks}}\ and\ \bibinfo {author} {\bibfnamefont {R.~F.}\ \bibnamefont
  {Williams}},\ }\bibfield  {title} {\bibinfo {title} {{Braids and the Jones
  polynomial}},\ }\href@noop {} {\bibfield  {journal} {\bibinfo  {journal}
  {Trans. Am. Math. Soc.}\ }\textbf {\bibinfo {volume} {303}},\ \bibinfo
  {pages} {97} (\bibinfo {year} {1987})}\BibitemShut {NoStop}%
\bibitem [{\citenamefont {Lickorish}\ and\ \citenamefont
  {Millett}(1987)}]{LickorishMillett}%
  \BibitemOpen
  \bibfield  {author} {\bibinfo {author} {\bibfnamefont {W.~B.~R.}\
  \bibnamefont {Lickorish}}\ and\ \bibinfo {author} {\bibfnamefont {K.~C.}\
  \bibnamefont {Millett}},\ }\bibfield  {title} {\bibinfo {title} {{A
  polynomial invariant of oriented links}},\ }\href@noop {} {\bibfield
  {journal} {\bibinfo  {journal} {Topology}\ }\textbf {\bibinfo {volume}
  {26}},\ \bibinfo {pages} {107} (\bibinfo {year} {1987})}\BibitemShut
  {NoStop}%
\bibitem [{\citenamefont {Brini}\ \emph {et~al.}(2012)\citenamefont {Brini},
  \citenamefont {Eynard},\ and\ \citenamefont {Mari{\~n}o}}]{BEM}%
  \BibitemOpen
  \bibfield  {author} {\bibinfo {author} {\bibfnamefont {A.}~\bibnamefont
  {Brini}}, \bibinfo {author} {\bibfnamefont {B.}~\bibnamefont {Eynard}},\ and\
  \bibinfo {author} {\bibfnamefont {M.}~\bibnamefont {Mari{\~n}o}},\ }\bibfield
   {title} {\bibinfo {title} {{Torus knots and mirror symmetry}},\ }\href@noop
  {} {\bibfield  {journal} {\bibinfo  {journal} {Ann. Henri Poincar\'e}\
  }\textbf {\bibinfo {volume} {13}},\ \bibinfo {pages} {1873} (\bibinfo {year}
  {2012})}\BibitemShut {NoStop}%
\bibitem [{\citenamefont {Witten}(1995)}]{WittenCSString}%
  \BibitemOpen
  \bibfield  {author} {\bibinfo {author} {\bibfnamefont {E.}~\bibnamefont
  {Witten}},\ }\bibfield  {title} {\bibinfo {title} {{Chern-Simons gauge theory
  as a string theory}},\ }in\ \href@noop {} {\emph {\bibinfo {booktitle} {The
  Floer Memorial Volume}}},\ \bibinfo {series} {Progr. Math.}, Vol.\ \bibinfo
  {volume} {133}\ (\bibinfo  {publisher} {Birkh{\"a}user},\ \bibinfo {year}
  {1995})\ pp.\ \bibinfo {pages} {637--678},\ \Eprint
  {https://arxiv.org/abs/hep-th/9207094} {arXiv:hep-th/9207094 [hep-th]}
  \BibitemShut {NoStop}%
\bibitem [{\citenamefont {Gopakumar}\ and\ \citenamefont
  {Vafa}(1999)}]{GopakumarVafa}%
  \BibitemOpen
  \bibfield  {author} {\bibinfo {author} {\bibfnamefont {R.}~\bibnamefont
  {Gopakumar}}\ and\ \bibinfo {author} {\bibfnamefont {C.}~\bibnamefont
  {Vafa}},\ }\bibfield  {title} {\bibinfo {title} {{On the gauge
  theory/geometry correspondence}},\ }\href@noop {} {\bibfield  {journal}
  {\bibinfo  {journal} {Adv. Theor. Math. Phys.}\ }\textbf {\bibinfo {volume}
  {3}},\ \bibinfo {pages} {1415} (\bibinfo {year} {1999})},\ \Eprint
  {https://arxiv.org/abs/hep-th/9811131} {arXiv:hep-th/9811131 [hep-th]}
  \BibitemShut {NoStop}%
\bibitem [{\citenamefont {Ooguri}\ and\ \citenamefont
  {Vafa}(2000)}]{OoguriVafa}%
  \BibitemOpen
  \bibfield  {author} {\bibinfo {author} {\bibfnamefont {H.}~\bibnamefont
  {Ooguri}}\ and\ \bibinfo {author} {\bibfnamefont {C.}~\bibnamefont {Vafa}},\
  }\bibfield  {title} {\bibinfo {title} {{Knot invariants and topological
  strings}},\ }\href@noop {} {\bibfield  {journal} {\bibinfo  {journal} {Nucl.
  Phys. B}\ }\textbf {\bibinfo {volume} {577}},\ \bibinfo {pages} {419}
  (\bibinfo {year} {2000})},\ \Eprint {https://arxiv.org/abs/hep-th/9912123}
  {arXiv:hep-th/9912123 [hep-th]} \BibitemShut {NoStop}%
\bibitem [{\citenamefont {Labastida}\ \emph {et~al.}(2000)\citenamefont
  {Labastida}, \citenamefont {Mari{\~n}o},\ and\ \citenamefont
  {Vafa}}]{LabastidaMarinoVafa}%
  \BibitemOpen
  \bibfield  {author} {\bibinfo {author} {\bibfnamefont {J.~M.~F.}\
  \bibnamefont {Labastida}}, \bibinfo {author} {\bibfnamefont {M.}~\bibnamefont
  {Mari{\~n}o}},\ and\ \bibinfo {author} {\bibfnamefont {C.}~\bibnamefont
  {Vafa}},\ }\bibfield  {title} {\bibinfo {title} {{Knots, links and branes at
  large $N$}},\ }\href@noop {} {\bibfield  {journal} {\bibinfo  {journal}
  {JHEP}\ }\textbf {\bibinfo {volume} {11}},\ \bibinfo {pages} {007}},\ \Eprint
  {https://arxiv.org/abs/hep-th/0010102} {arXiv:hep-th/0010102 [hep-th]}
  \BibitemShut {NoStop}%
\bibitem [{\citenamefont {Mari{\~n}o}(2005)}]{MarinoCSReview}%
  \BibitemOpen
  \bibfield  {author} {\bibinfo {author} {\bibfnamefont {M.}~\bibnamefont
  {Mari{\~n}o}},\ }\bibfield  {title} {\bibinfo {title} {{Chern-Simons theory
  and topological strings}},\ }\href@noop {} {\bibfield  {journal} {\bibinfo
  {journal} {Rev. Mod. Phys.}\ }\textbf {\bibinfo {volume} {77}},\ \bibinfo
  {pages} {675} (\bibinfo {year} {2005})},\ \Eprint
  {https://arxiv.org/abs/hep-th/0406005} {arXiv:hep-th/0406005 [hep-th]}
  \BibitemShut {NoStop}%
\bibitem [{\citenamefont {Witten}(2010)}]{WittenAnalyticContinuation}%
  \BibitemOpen
  \bibfield  {author} {\bibinfo {author} {\bibfnamefont {E.}~\bibnamefont
  {Witten}},\ }\href@noop {} {\bibinfo {title} {{Analytic Continuation Of
  Chern-Simons Theory}}} (\bibinfo {year} {2010}),\ \Eprint
  {https://arxiv.org/abs/1001.2933} {arXiv:1001.2933 [hep-th]} \BibitemShut
  {NoStop}%
\bibitem [{\citenamefont {Cachazo}\ \emph {et~al.}(2004)\citenamefont
  {Cachazo}, \citenamefont {Svr{\v c}ek},\ and\ \citenamefont {Witten}}]{CSW}%
  \BibitemOpen
  \bibfield  {author} {\bibinfo {author} {\bibfnamefont {F.}~\bibnamefont
  {Cachazo}}, \bibinfo {author} {\bibfnamefont {P.}~\bibnamefont {Svr{\v
  c}ek}},\ and\ \bibinfo {author} {\bibfnamefont {E.}~\bibnamefont {Witten}},\
  }\bibfield  {title} {\bibinfo {title} {{MHV vertices and tree amplitudes in
  gauge theory}},\ }\href@noop {} {\bibfield  {journal} {\bibinfo  {journal}
  {JHEP}\ }\textbf {\bibinfo {volume} {09}},\ \bibinfo {pages} {006}},\ \Eprint
  {https://arxiv.org/abs/hep-th/0403047} {arXiv:hep-th/0403047 [hep-th]}
  \BibitemShut {NoStop}%
\bibitem [{\citenamefont {Gukov}\ \emph {et~al.}(2007)\citenamefont {Gukov},
  \citenamefont {Motl},\ and\ \citenamefont {Neitzke}}]{GukovMotlNeitzke}%
  \BibitemOpen
  \bibfield  {author} {\bibinfo {author} {\bibfnamefont {S.}~\bibnamefont
  {Gukov}}, \bibinfo {author} {\bibfnamefont {L.}~\bibnamefont {Motl}},\ and\
  \bibinfo {author} {\bibfnamefont {A.}~\bibnamefont {Neitzke}},\ }\bibfield
  {title} {\bibinfo {title} {{Equivalence of twistor prescriptions for super
  Yang-Mills}},\ }\href@noop {} {\bibfield  {journal} {\bibinfo  {journal}
  {Adv. Theor. Math. Phys.}\ }\textbf {\bibinfo {volume} {11}},\ \bibinfo
  {pages} {199} (\bibinfo {year} {2007})},\ \Eprint
  {https://arxiv.org/abs/hep-th/0404085} {arXiv:hep-th/0404085 [hep-th]}
  \BibitemShut {NoStop}%
\bibitem [{\citenamefont {Guadagnini}\ \emph {et~al.}(1990)\citenamefont
  {Guadagnini}, \citenamefont {Martellini},\ and\ \citenamefont
  {Mintchev}}]{GuadagniniMartelliniMintchev}%
  \BibitemOpen
  \bibfield  {author} {\bibinfo {author} {\bibfnamefont {E.}~\bibnamefont
  {Guadagnini}}, \bibinfo {author} {\bibfnamefont {M.}~\bibnamefont
  {Martellini}},\ and\ \bibinfo {author} {\bibfnamefont {M.}~\bibnamefont
  {Mintchev}},\ }\bibfield  {title} {\bibinfo {title} {{Wilson lines in
  Chern-Simons theory and link invariants}},\ }\href@noop {} {\bibfield
  {journal} {\bibinfo  {journal} {Nucl. Phys. B}\ }\textbf {\bibinfo {volume}
  {330}},\ \bibinfo {pages} {575} (\bibinfo {year} {1990})}\BibitemShut
  {NoStop}%
\bibitem [{\citenamefont {Mari{\~n}o}\ and\ \citenamefont
  {Vafa}(2001)}]{MarinoVafa}%
  \BibitemOpen
  \bibfield  {author} {\bibinfo {author} {\bibfnamefont {M.}~\bibnamefont
  {Mari{\~n}o}}\ and\ \bibinfo {author} {\bibfnamefont {C.}~\bibnamefont
  {Vafa}},\ }\href@noop {} {\bibinfo {title} {{Framed knots at large $N$}}}
  (\bibinfo {year} {2001}),\ \Eprint {https://arxiv.org/abs/hep-th/0108064}
  {arXiv:hep-th/0108064 [hep-th]} \BibitemShut {NoStop}%
\bibitem [{\citenamefont {Gaiotto}\ \emph {et~al.}(2013)\citenamefont
  {Gaiotto}, \citenamefont {Moore},\ and\ \citenamefont
  {Neitzke}}]{GaiottoMooreNeitzke}%
  \BibitemOpen
  \bibfield  {author} {\bibinfo {author} {\bibfnamefont {D.}~\bibnamefont
  {Gaiotto}}, \bibinfo {author} {\bibfnamefont {G.~W.}\ \bibnamefont {Moore}},\
  and\ \bibinfo {author} {\bibfnamefont {A.}~\bibnamefont {Neitzke}},\
  }\bibfield  {title} {\bibinfo {title} {{Spectral networks}},\ }\href@noop {}
  {\bibfield  {journal} {\bibinfo  {journal} {Ann. Henri Poincar\'e}\ }\textbf
  {\bibinfo {volume} {14}},\ \bibinfo {pages} {1643} (\bibinfo {year}
  {2013})},\ \Eprint {https://arxiv.org/abs/1204.4824} {arXiv:1204.4824
  [hep-th]} \BibitemShut {NoStop}%
\bibitem [{\citenamefont {Caron-Huot}\ and\ \citenamefont
  {He}(2013)}]{CaronHuotHe}%
  \BibitemOpen
  \bibfield  {author} {\bibinfo {author} {\bibfnamefont {S.}~\bibnamefont
  {Caron-Huot}}\ and\ \bibinfo {author} {\bibfnamefont {S.}~\bibnamefont
  {He}},\ }\bibfield  {title} {\bibinfo {title} {{Three-loop octagons and
  $n$-gons in maximally supersymmetric Yang--Mills theory}},\ }\href@noop {}
  {\bibfield  {journal} {\bibinfo  {journal} {JHEP}\ }\textbf {\bibinfo
  {volume} {08}},\ \bibinfo {pages} {101}},\ \Eprint
  {https://arxiv.org/abs/1305.2781} {arXiv:1305.2781 [hep-th]} \BibitemShut
  {NoStop}%
\bibitem [{\citenamefont {Costello}(2021)}]{CostelloTwistor}%
  \BibitemOpen
  \bibfield  {author} {\bibinfo {author} {\bibfnamefont {K.~J.}\ \bibnamefont
  {Costello}},\ }\href@noop {} {\bibinfo {title} {{Quantizing local holomorphic
  field theories on twistor space}}} (\bibinfo {year} {2021}),\ \Eprint
  {https://arxiv.org/abs/2111.08879} {arXiv:2111.08879 [hep-th]} \BibitemShut
  {NoStop}%
\bibitem [{\citenamefont {Bullimore}\ and\ \citenamefont
  {Skinner}(2011)}]{BullimoreSkinner}%
  \BibitemOpen
  \bibfield  {author} {\bibinfo {author} {\bibfnamefont {M.}~\bibnamefont
  {Bullimore}}\ and\ \bibinfo {author} {\bibfnamefont {D.}~\bibnamefont
  {Skinner}},\ }\href@noop {} {\bibinfo {title} {{Holomorphic linking, loop
  equations and scattering amplitudes in twistor space}}} (\bibinfo {year}
  {2011}),\ \Eprint {https://arxiv.org/abs/1101.1329} {arXiv:1101.1329
  [hep-th]} \BibitemShut {NoStop}%
\end{thebibliography}%
